\documentclass[twocolumn,aps,epsfig,nofootinbib,preprintnumbers]{revtex4-1}
\usepackage{graphicx}
\usepackage{epstopdf}
\usepackage{latexsym}
\usepackage{amssymb}
\usepackage{amsmath}
\usepackage{color}
\usepackage{float}
\usepackage{mathrsfs}
\usepackage{multirow}
\usepackage{tabularx,booktabs}
\usepackage{esvect}
\usepackage{ulem} 
\newcolumntype{Y}{>{\centering\arraybackslash}X}

\usepackage[center]{subfigure}
\usepackage{makecell}
\usepackage{ulem}
\usepackage[colorlinks=true]{hyperref} 
\setcellgapes{4pt}
\begin{document}

%
%
%
%
%
%




\def\noi{\noindent}
\def\nq{\hspace*{-1em}}
\def\nqq{\hspace*{-2em}}
\def\nhq{\hspace*{-0.5em}}
\def\nhh{\hspace*{-0.3em}}
\def\quad {\hspace*{2em}}
\def\cm{\hspace*{1cm}}
\def\inch{\hspace*{1in}}
\def\wide{\mbox{$\dst\vphantom{\int}$}}
\def\Wide{\mbox{$\dst\vphantom{\intl_a^a}$}}
\def\ten#1{\mbox{$\cdot 10^{#1}$}}

\def\floatpagefraction{.95}
\def\topfraction{.95}
\def\bottomfraction{.95}
\def\textfraction{.05}
\def\dblfloatpagefraction{.95}
\def\dbltopfraction{.95}


\def\Jl#1#2{#1 {\bf #2},\ }
\def\ApJ#1 {\Jl{Astroph. J.}{#1}}
\def\CQG#1 {\Jl{Class. Quantum Grav.}{#1}}
\def\DAN#1 {\Jl{Dokl. AN SSSR}{#1}}
\def\GC#1 {\Jl{Grav. Cosmol.}{#1}}
\def\GRG#1 {\Jl{Gen. Rel. Grav.}{#1}}
\def\JETF#1 {\Jl{Zh. Eksp. Teor. Fiz.}{#1}}
\def\JETP#1 {\Jl{Sov. Phys. JETP}{#1}}
\def\JHEP#1 {\Jl{JHEP}{#1}}
\def\JMP#1 {\Jl{J. Math. Phys.}{#1}}
\def\NPB#1 {\Jl{Nucl. Phys. B}{#1}}
\def\NP#1 {\Jl{Nucl. Phys.}{#1}}
\def\PLA#1 {\Jl{Phys. Lett. A}{#1}}
\def\PLB#1 {\Jl{Phys. Lett. B}{#1}}
\def\PRD#1 {\Jl{Phys. Rev. D}{#1}}
\def\PRL#1 {\Jl{Phys. Rev. Lett.}{#1}}

  \renewcommand\arraystretch{2}
 \newcommand{\bqu}{\begin{equation}}
 \newcommand{\equ}{\end{equation}}
 \newcommand{\bqun}{\begin{eqnarray}}
 \newcommand{\equn}{\end{eqnarray}}
 \newcommand{\nb}{\nonumber}
 \newcommand{\lb}{\label}
 \newcommand{\cb}{\color{blue}}
    \newcommand{\cc}{\color{cyan}}
\newcommand{\rc}{\rho^{\scriptscriptstyle{\mathrm{I}}}_c}
\newcommand{\rd}{\rho^{\scriptscriptstyle{\mathrm{II}}}_c} 
%

%

\def\al{&\nhq}
\def\lal{&&\nqq {}}
\def\eq{ \,}
\def\eqs{ \,}
\def\beq{\begin{equation}}
\def\eeq{\end{equation}}
\def\bear{\begin{eqnarray}}
\def\bearr{\begin{eqnarray} \lal}
\def\ear{\end{eqnarray}}
\def\earn{\nonumber \end{eqnarray}}
\def\besub{\begin{subequations}}
\def\esub{\end{subequations}}

\def\nn{\nonumber\\ {}}
\def\nnv{\nonumber\\[5pt] {}}
\def\nnn{\nonumber\\ \lal }
\def\nnnv{\nonumber\\[5pt] \lal }
\def\yy{\\[5pt] {}}
\def\yyy{\\[5pt] \lal }
\def\eql{\al =\al}
\def\eqv{\al \equiv \al}
\def\sequ#1{\setcounter{equation}{#1}}

\def\dst{\displaystyle}
\def\tst{\textstyle}
\def\fracd#1#2{{\dst\frac{#1}{#2}}}
\def\fract#1#2{{\tst\frac{#1}{#2}}}
\def\Half{{\fracd{1}{2}}}
\def\half{{\fract{1}{2}}}

\def\e{{\,\rm e}}
\def\d{\partial}
\def\re{\mathop{\rm Re}\nolimits}
\def\im{\mathop{\rm Im}\nolimits}
\def\arg{\mathop{\rm arg}\nolimits}
\def\tr{\mathop{\rm tr}\nolimits}
\def\sign{\mathop{\rm sign}\nolimits}
\def\diag{\mathop{\rm diag}\nolimits}
\def\dim{\mathop{\rm dim}\nolimits}
\def\const{{\rm const}}
\def\eps{\varepsilon}
\def\ep{\epsilon}

\def\then{\ \Rightarrow\ }
\newcommand{\toas}{\mathop {\ \longrightarrow\ }\limits }
\newcommand{\aver}[1]{\langle \, #1 \, \rangle \mathstrut}
\newcommand{\vars}[1]{\left\{\begin{array}{ll}#1\end{array}\right.}

\tolerance 4000
\def\red#1{{\color{red} #1}}
\def\blue#1{{\color{blue} #1}}
\def\rf{\eqref}
\def\eqn#1{ \,\rf{#1}}

\def\mn{_{\mu\nu}}
\def\MN{^{\mu\nu}}
\def\mN{_\mu^\nu}
\def\nM{_\nu^\mu}

\def\N{{\mathbb N}}
\def\R{{\mathbb R}}
\def\S{{\mathbb S}}
\def\Z{{\mathbb Z}}
\def\cF{{\cal F}}
\def\cR{{\cal R}}

\def\da{\delta\alpha}
\def\db{\delta\beta}
\def\df{\delta\phi}
\def\dg{\delta\gamma}
\def\Veff{V_{\rm eff}}

\def\wh{wormhole}
\def\whs{wormholes}
\def\bh{black hole}
\def\bhs{black holes}

\def\sph{spherically symmetric}
\def\ssph{static, spherically symmetric}
\def\asflat{asymptotically flat}
\def\Schr{Schr\"o\-din\-ger}
\def\Scz{Schwarz\-schild}
\def\scyl{static, cylindrically symmetric}
\def\cyl{cylindrically symmetric}

\def\off#1{{\it \blue{#1}}}

\def\mn{_{\mu\nu}}
\def\MN{^{\mu\nu}}
\def\mN{_\mu^\nu}
\def\nM{_\nu^\mu}

\def\D{{\mathbb D}}
\def\M{{\mathbb M}}
\def\R{{\mathbb R}}
\def\tK{{\widetilde K}}
\def\tT{{\widetilde T}}

\def\eqv{\al \equiv \al}
\def\kappa{\varkappa}

\def\GR{general relativity}
\def\cy{cylindrical}
\def\cyl{cylindrically symmetric}

\def\wh{wormhole}
\def\whs{wormholes}
\def\asflat{asymptotically flat}



\title{Cylindrical Systems in General Relativity} 

\author{ Kirill A. Bronnikov$^{1, 2, 3}$}
\email{kb20@yandex.ru}

\author{N. O. Santos $^{4, 5}$}
\email{Nilton.Santos@obspm.fr}

\author{ Anzhong Wang $^{6, 7}$}
\email{anzhong$\_$wang@baylor.edu; Corresponding author}

\affiliation{$^1$  Center for Gravitation and Fundamental Metrology, VNIIMS, Ozyornaya ul. 46, Moscow 119361, Russia}
\affiliation{$^2$
Institute of Gravitation and Cosmology, Peoples' Friendship University of Russia (RUDN University), ul. Miklukho-Maklaya 6, Moscow 117198, Russia}
\affiliation{$^3$
National Research Nuclear University ``MEPhI" (Moscow Engineering Physics Institute), Kashirskoe sh. 31, Moscow 115409, Russia}

\affiliation{$^4$ Sorbonne Universit\'e, UPMC Universit\'e Paris 06, LERMA, UMRS8112 CNRS,
Observatoire de Paris-Meudon, 5,
Place Jules Janssen, F-92195 Meudon Cedex, France}
\affiliation{$^5$
School of Mathematical Sciences, Queen Mary, University of London,
Mile End Road, London E1 4NS, UK}
 
\affiliation{$^6$ GCAP-CASPER, Physics Department, Baylor University, Waco, TX 76798-7316, USA}
\affiliation{$^7$
 Institute for Theoretical  Physics $\&$ Cosmology,   
Zhejiang University of Technology, Hangzhou, 310032, China}

\date{\today}

\begin{abstract}

With the arrival of the era of gravitational wave astronomy,   the strong gravitational  field regime will be  explored soon in various aspects. 
In this article, we provide a general review over cylindrical systems in Einstein's theory of general relativity. In particular, we first review the 
general properties, both local and global, of several important solutions of Einstein's field equations, including the Levi-Civita and Lewis 
solutions and their extensions to include the cosmological constant and  matter fields, and pay particular attention to properties that represent 
the generic features of the theory, such as the formation of the observed extragalactic jets and gravitational Faraday rotation. We also review 
studies of cylindrical wormholes, gravitational collapse and Hoop conjecture, and polarizations of gravitational waves.  {In addition, 
by rigorously defining cylindrically symmetric spacetimes, we clarify various (incorrect) claims  existing in the literature, regarding to the generality 
of such spacetimes.}
 
 \end{abstract}

\maketitle

%

 \tableofcontents
 
 \section{Introduction}\label{Introduction}
\renewcommand{\theequation}{1.\arabic{equation}} \setcounter{equation}{0}

The investigations of cylindrically symmetric spacetimes can be traced back as far as to 1919 when Levi-Civita (LC) discovered a class of solutions of Einstein's vacuum field equations, corresponding 
  to static cylindrical spacetimes \cite{Levi}. 
The extension of the LC   spacetimes to  stationary ones was obtained independently by Lanczos in 1924  \cite{Lanczos} and Lewis in 1932 \cite{Lewis}. 
In 1925,    Beck   studied a class of exact solutions and interpreted them as representing the propagation of cylindrical gravitational waves  (GWs) \cite{Beck25}. This class of solutions was later rediscovered  
by Einstein and Rosen  in their seminal work on the studies of the nonlinearity of GWs    in 1937 \cite{ER37}, and in the same year van Stockum solved the problem of a rigidly rotating infinitely long cylinder of dust, and found
explicitly the { {corresponding} metric  \cite{Stockum}.  In 1957, Bonnor \cite{Bonnor57} and Weber and Wheeler \cite{WW57} studied the   Einstein-Rosen waves in great detials, and 
since then, cylindrically symmetric  spacetimes have been extensively investigated  with various motivations     {\cite{StephaniA,GP09}}. 

In particular, to understand some fundamental issues in general relativity (GR), such as the structure of the theory   {\cite{MTW73}}, the nature and formation of spacetime 
singularities \cite{HE73}, the cosmic censorship \cite{Penrose69} and  hoop  \cite{Thorne72} conjectures,  one often  assumes certain symmetries of spacetimes, because    {Einstien's}
field equations are in general so complicated  that it is extremely difficult to study them in their most general   {form}, especially in the times when computers had not been available. In fact,
even now it is still very   {difficult} to study spacetimes with only one Killing vector (both analytically and numerically). Therefore, the next step is naturally to consider spacetimes with two Killing vectors, 
which   include  cylindrically symmetric   spacetimes    \cite{StephaniA,GP09}. 

The studies of   {cylindrically symmetric} spacetimes were further motivated by  topological defects discovered  in 1976  by Kibble  {\cite{Kibble76,Kibble80},} which include cosmic strings, and the latter were 
once believed to provide a mechanism to produce the cosmic microwave background (CMB) and  large-scale structure observed in our universe   \cite{VS00,CK10}.    However, observations of  CMB
have ruled out  cosmic strings, formed in the context of symmetry breaking in grand unified theories (GUT's), as the sources of the cosmological perturbations \cite{Smoot92,Benn96,Spergel07}, 
and led to an upper bound, $G\mu/c^2  \lesssim 10^{-7}$ \cite{Ade16}, where $\mu$ denotes the string's tension, and $G$ and $c$ are, respectively, the Newtonian constant and the speed of 
light in vacuum.  The string tension is intimately related to the energy $E$ of the phase transition,
\bqu
\lb{1.1}
\frac{G\mu}{c^2} = {\cal{O}}\left(\frac{E^2}{M_{pl}^2}\right),
\equ
where $M_{pl} \left(\equiv \sqrt{\hbar c/G}\right)$ denotes the Planck mass.  Currently, the strongest constraint $G\mu/c^2  \lesssim 10^{-11}$ comes from the analysis of stochastic  GW
background, expected from cosmic strings with  the latest pulsar timing   {arrays} \cite{BOS18}. 

Recently, the subject has attracted further attention in the framework of string/M-Theory, the so-called cosmic superstrings  \cite{DV04,CMP04}, 
which were formed before inflation took place and stretched to macroscopic length scales in the inflationary phase.
During the subsequent epochs, a complicated network of various string elements forms \cite{Tye08,CH15,CH15b}. The main phenomenological consequence of a string network is the emission of GWs 
\cite{VV85,Sakel90,DV00,DV01,Dv05,BFKK09,OMS10,BBHS10,RGSM12,Aasi14,Abbott16,SES17,RS17}.  {It can generate} bursts  at cusps, kinks and junctions, as well as a stochastic GW background.  
Low tension strings are natural in string/M-Theory, and can easily satisfy the observational bounds given above \cite{CFW18}. 
 
 With the observations of the 11  GWs \cite{GWs18}, lately the studies of GW phenomena have attracted a great deal of attention, including their nonlinear properties, such as
 memory effects \cite{YZ74,BG85,Chris91,BD92,Thorne92,Harte13,Souriau73,BT87,BP88,BP88b,GP89,BGY17,Zhang2017,Maluf2017,zhang2018a,Andrzejewski2018,Wang18}.
 Such effects might be possibly detected by the Laser Interferometer Space Antenna (LISA)    \cite{Favata10}. 
 
  { From these observations, another important result is that the individual masses of such observed black hole-black hole (BBHs) binaries can be much larger than what were expected previously both theoretically and observationally \cite{Fryer12,Ozel10,Farr11}, and various scenarios have been proposed \cite{Abbott16a,TZWang18}. For such massive binary systems,  their mHz emission of GWs from the early inspiral phase is strong enough to be visible to space-borne detectors, such as LISA \cite{LISA}, TianQin \cite{TianQin} and Taiji \cite{Taiji},  for several years prior to their coalescence \cite{Sesana16}. In fact, because of the poor constraints on the rates and the standard assumption of 10 solar mass objects, BBHs had never been considered as particularly interesting sources of space-borne detectors, and multi-band GW observations had been only proposed in the context of seed  or intermediate mass BBHs \cite{Sesana09,ASS10}. But, now analysis of the BBH population as constrained by current LIGO/Virgo demonstrated that such space-borne detectors can see thousands of such systems, with a variety of profound scientific consequences.  In particular, observations of the same signal in two different detectors provides an efficient independent way to cross check 
 and validate the instruments, which is particularly valuable for a space-based detector. }

  {    More important, multi-band detections will enhance the potential of gravity tests in the strong, dynamical field regime of merging BBHs. Massive systems will be observed by ground based detectors with high merging signal/noise rates, after being tracked for years by space-borne detectors in their inspiral phase. The two portions of signal can be combined to push the search for deviations from GR.  In particular, joint observations of BBHs with a total mass larger than $\sim 60$ solar masses by LIGO/Virgo and space-borne detectors can potentially improve current bounds on dipole emission from BBHs by more than six orders of magnitude \cite{BYC16}, which will impose sever constraints on various theories of gravity \cite{Corda09,Berti15}.
   These include  scalar-tensor theories \cite{Jacobson99,CCD07,Corda07,HB12,YYP16,Zhang17}, screened modified gravity \cite{Liu18,Zhang19a,Zhang19b},  Einstein-dilaton Gauss-Bonnet gravity \cite{YSYT12}, Einstein-aether theory \cite{Gong18,Zhao19,Lin19}, theories with parity violations \cite{Qiao19,Zhao19a,Zhao19b}. For more details, we refer readers to the    review articles \cite{Berti18a,Berti18b}.  }

  {These} studies  recently have gained additional momenta,  due to the close relations between asymptotically symmetric  theorems of soft gravitons and GW memory \cite{HPS16,Strominger16}.
  {Such} studies in general have been carried out in spacetimes with plane symmetry, but to shed some new light on   {the nonlinear} effects, recently cylindrically GWs have been also studied \cite{MT17}. 

Along the above  direction, another interesting phenomenon  related to the nonlinear properties of GWs is the gravitational Faraday rotation, first discovered by Piran, Safier and Stark \cite{PS85,PSS85},
and later generalized to spacetimes with plane symmetry  {\cite{Wang91,Wang91th,Wang19}.} It was found that the polarizations of one GW can be rotated by   the other, due to their nonlinear interaction, an analogue of
the electromagnetic Faraday rotation. Since then, detailed studies of gravitational Faraday rotation have been carried out in various situations, including  Einstein-Maxwell waves \cite{AFH90}, 
rotating cylindrical GWs \cite{PW00}, and more recently, cylindrical gravitational soliton waves \cite{TM14}.

\subsection{Cylindrically symmetric spacetimes}

It is interesting to note that, despite  the extensive studies of cylindrical spacetimes in the past several decades, in the literature   there still exist confusions in the definition of  cylindrically symmetric spacetimes and 
 (wrong/false) claims. 
  {A reason for that is probably connected with different kinds of evolutions that remain cylindrical symmetry:  motions  around and
  along the symmetry axis. In both cases the velocities of these motions may depend on the radial coordinate in an arbitrary manner. }
 
  For example, it has been claimed by various authors that the Kompaneets-Jordan-Ehlers-Kundt (KJEK) metric \cite{Komp58,JEK60},
 \bqu
 \lb{1.2}
 ds^2 =  e^{2(\gamma - \psi)}\left(dt^2 - dr^2\right) - e^{2\psi}\left(dz + \omega d\phi\right)^2 - W^2e^{-2\psi}d\phi^2,
 \equ
is the most general form that describes cylindrically symmetric spacetimes, where $\gamma, \; \psi, \; \omega$ and $W$ are functions of $t$ and $r$ only. However, it is clear that this metric does not include 
 the rotating cylindrical GW spacetimes 
studied by Mashhoon and Quevedo (MQ) \cite{MQ90,MMQ00,QM91}, 
 \bqu
 \lb{1.3}
 ds^2 =  e^{2(\gamma - \psi)}\left(dt^2 - dr^2\right) - W^2e^{-2\psi}\left(d\phi + \omega dt\right)^2 -  e^{2\psi}dz^2.
 \equ
 
 {To clarify these issues, it is necessary first to give a rigorous definition of cylindrically symmetric spacetimes. Unfortunately, this inevitably involves the introduction of  some rather abstract   group theory 
and and  in certain sense complicated geometric conceptions. For the sake of reader's convenience, in the following we shall carry this out in several steps. In doing so, it will become clear 
that both of the above metrics describe spacetimes
with cylindrical symmetry, and that none of them actually represents the most general metric of cylindrical spacetimes.     
In particular,  to obtain metric (\ref{1.2}),  an additional condition that the two-dimensional abelian Lie group $G_2$ acts orthogonally transitively must be imposed \footnote{For the definition of 
a Lie group $G_n$, see, for example,  \cite{StephaniA,JFA96}}, in addition to the general definition of cylindrically symmetric spacetimes, while in metric (\ref{1.3}) such an assumption is left out, but with other further constraints.    }

 {To show the above claims  clearly, let us start with  the    definition of 
   the cyclical symmetry. }

 {\bf Definition 1.} A spacetime $\left({\cal{M}}, g\right)$ is said to have a cyclical symmetry,  if and only if the metric is invariant under the effective smooth action $SO(2) \times {\cal{M}} \rightarrow {\cal{M}}$ 
 of the one-parameter cyclic group $SO(2)$   \cite{Carter70}.

  {\bf Definition 2.} A cyclically symmetric spacetime is said to be axially symmetric, if the set of fixed points (i.e., points that remain invariant under the action of the group) of the isometry is not empty. 
  The set of fixed points is referred to as the symmetry axis \cite{CSV99}. 

Note that Definition 2 implicitly assumes that there exists at least one fixed point  in $\left({\cal{M}}, g\right)$. It can    be shown  that the set of fixed points must be an autoparallel, two-dimensional timelike surface \cite{Carter70,MS93}. 
Furthermore, the infinitesimal generator $\xi^{\alpha}$ of the axial symmetry
is spacelike in a neighborhood of the axis, and that the so-called elementary flatness condition holds  \cite{MS93},  
 \bqu
 \lb{1.4}
 \frac{X_{,\alpha} X_{,\beta} g^{\alpha\beta}}{4X} \rightarrow -1, 
 \equ
 as $r \rightarrow 0^{+}$, which   ensures the standard $2\pi$-periodicity of the axial coordinate near the axis, where $r= 0$ denotes  the location of the axis,  
 \bqu
 \lb{1.5}
X \equiv \left|\xi^{\alpha} \xi_{\alpha}\right|_{r \rightarrow 0^{+}}   \rightarrow 0, 
 \equ
and $(\;)_{,\alpha} \equiv \partial/\partial x^{\alpha}$. Note that in this paper we shall adopt the conversions:   Greek letters  run from $0$ to $3$ ($\mu, \nu,
 ... = 0, 1, 2, 3$);  Latin letters  run from $1$ to $3$ ($i, j,  ... = 1, 2, 3$); and repeated indices represent sum over their individual domains. 

To define the cylindrical symmetry, in addition to $\xi$ we assume that the spactime  $\left({\cal{M}}, g\right)$ contains another spacelike Killing vector $\eta$, which generates together with $\xi$ a $G_2$  group.
Then, one can prove the following:

{\bf Proposition 1.}  In an axially symmetric spacetime, in addition to $\xi$ if there is another Killing vector $\eta$, then both Killing vectors commute,  thus generating an Abelian
$G_2$ group \cite{CSV99}.

 With the above, we are ready to define  cylindrically symmetric spacetimes.

 {\bf Definition 3.} A spacetime $\left({\cal{M}}, g\right)$ is cylindrically symmetric if and only if it admits a $G_2$ on $S_2$  
group of isometries containing an axial symmetry \cite{CSV99}. 
 
 As mentioned above, the assumption on the existence of two-surfaces orthogonal to the group orbits  is not necessary for the definition of cylindrical symmetry nor a consequence of it, but the existence of an axis is
 essential. So, in the above definition, the orthogonal transitivity is   {removed}, and the line-element of cylindrically symmetric spacetimes is provided in the following theorem \cite{Wain81}.
 
 {\bf Theorem 1.} Suppose that a spacetime admits two spacelike commuting Killing vector fields $\xi$ and $\eta$. Let $u$ be a hypersurface-orthogonal  timelike vector field of unit length, which is orthogonal to $\xi$ and $\eta$.
   Then, there exist local coordinates $(t, x, y, z)$, such that $ u = N^{-1}(t, x)\partial_t, \; u_{\alpha}dx^{\alpha} = N(t, x) dt, \; \xi = \partial_y, \; \eta = \partial_z$, and 
  \bqun
 \lb{1.6}
 ds^2 &=&  N^2(t, x) dt^2 - g_{ij}(t,x) dx^i dx^j\nb\\
 &=&  e^{2k}dt^2 - e^{2h}dx^2 - W^2\left[f\left(dy + w_1 dz + w_2 dx\right)^2 \right.\nb\\
 && \left. + f^{-1}\left(dz + w_3 dx\right)^2\right], 
 \equn
where $k, \; h,\; W,\; f$ and $w_i$ are all functions of $t$ and $x$ only. It can be shown that this metric includes both metric (\ref{1.2}) and metric (\ref{1.3}) as its particular cases.  

In particular, when $G_2$ acts orthogonally transitively, we have  $w_2 = w_3 = 0$, and then  (\ref{1.6}) reduces to 
  \bqun
 \lb{1.6a}
 ds^2 &=&    e^{2k}dt^2 - e^{2h}dx^2 - W^2\left[f\left(dz + w_1 d\phi\right)^2 + f^{-1} d\phi^2\right].  \nb\\
 \equn
Note that in writing the above expression, we had made the replacement $(y, z) \rightarrow (z, \phi)$.
Clearly, the two-surfaces of $t,\; x  =$ const.  is gauge-invariant under the coordinate transformations,   
\bqu
\lb{1.6b}
t =f(t', r), \quad x = g(t', r), 
\equ
where $f(t', r)$ and $g(t', r)$ are arbitrary functions of their indicated arguments. Then, using this freedom,
we can always set $g_{t't'} = - g_{rr}$ and $g_{t'r} = 0$, so the resulted metric takes precisely the form of  (\ref{1.2}).
 In this case, there exists a general no-go theorem regarding the existence of black holes \cite{Wang05}.
 
  {\bf Theorem 2.} Let $\left({\cal{M}}, g\right)$ be a four-dimensional Riemannian spacetime obeying    Einstein's field equations, 
    {$R_{\mu\nu} -(R/2) g_{\mu\nu} - \Lambda g_{\mu\nu} =  \kappa T_{\mu\nu}$,} coupled with a cosmological constant $\Lambda$.  
  Assume that the spacetime possesses two commuting spacelike  orthogonally transitive Killing vectors $\xi$ and $\eta$. Then, $\left({\cal{M}}, g\right)$ 
  contains neither outer nor degenerate apparent horizons, if matter satisfies the dominant energy condition and $\Lambda > 0$.

 Note that when $\Lambda = 0$, only the existence of outer apparent horizons is excluded. Thus, in this case only degenerate apparent horizons are allowed. However, when $\Lambda < 0$, 
 both  outer and degenerate apparent horizons can exist. This explains why all the topological black holes found so far are with $\Lambda \le 0$. For more details, we refer readers to \cite{Wang05}, 
  and for a concrete example, see \cite{Wang05b}.

 In addition,  the above theorem does not require any of the orbits of the two  spacelike Killing vectors be closed. As a result, the theorem can be equally applied to plane symmetric spacetimes    
  {\cite{Griffiths16,Wang19}.} 

  When $G_2$ acts non-orthogonally transitively, but one of the two Killing vectors is hypersurface-orthogonal, the corresponding line-element can be obtained from (\ref{1.6}) by setting 
     either $w_1 = w_2 = 0$ or $w_1 = w_3 = 0$, which is referred to as  the A(ii) class of solutions in \cite{Wain81}, while the metric (\ref{1.6a}) is classified as the B(i) class of   solutions. 
   In the current case, without loss of the generality, we only consider  the case $w_1 = w_2 = 0$, with which the metric (\ref{1.6}) reduces to 
  \bqun
 \lb{1.6c}
 ds^2 &=&   e^{2k}dt^2 - e^{2h}dx^2 - W^2\left[fdy^2 + f^{-1}\left(dz + w_3 dx\right)^2\right].   \nb\\
 \equn
 Making the coordinate transformations (\ref{1.6b}) and using the gauge freedom, we can always set $g_{t'r} = g_{zr} = 0$, so the resulted metric takes the form,
   \bqun
 \lb{1.6d}
 ds^2 &=&   e^{2k}dt^2 - e^{2h}dr^2 - W^2\left[fdz^2 \right.\nb\\
 &&  \left. + f^{-1}\left(d\phi + w_3 dt\right)^2\right], 
 \equn
 after the replacement, $(y, z) \rightarrow (z, \phi)$.
  {Clearly,  the above metric reduces to the  MQ metric (\ref{1.3}) when $k = h$ \footnote{Note that in the 2-surfaces of $z, \phi = $ const., we can always
 use the gauge freedom (\ref{1.6b}) to write the reduced metric in the conformally-flat form, $e^{2k}\left(dt^2 - dr^2\right)$. But, now the second part of the metric (\ref{1.6d}) will in general take
   the form, $W^2\left[fdz^2 + f^{-1}\left(d\phi + w_2 dr + w_3 dt\right)^2\right]$.}.  A class of solutions of this metric was studied recently by Chicone, Mashhoon and Santos for 
   gravitomagnetic accelerators \cite{CM11,CM11b, MS00}.}

  When $G_2$ acts orthogonally transitively, and both $\xi$ and $\eta$ are  hypersurface-orthogonal and mutually orthogonal, the corresponding spacetimes are described by metric (\ref{1.6}) with
  $w_1 = w_2 = w_3 = 0$, and with the gauge freedom (\ref{1.6b}), the metric can be further written as
  \bqun
 \lb{1.6e}
 ds^2 &=&   e^{2k}\left(dt^2 -  dr^2\right) - W^2\left(fdz^2 + f^{-1}d\phi^2\right). ~~~
 \equn
 This is referred to as the  B(ii) class of solutions in \cite{Wain81} and represents the Einstein-Rosen (ER) cylindrical GWs \cite{ER37}, which are linearly polarized, that is, only the ``+" polarization exists 
   \cite{Griffiths16}.

It should    {be noted}  that in spacetimes with cylindrical symmetry, closed timelike curves (CTC's) can be easily formed. To guarantee their absence, one usually also imposes that the condition,
\bqun
 \lb{1.7}
 \xi^{\alpha} \xi_{\alpha} < 0,
 \equn
holds in the whole spacetime. 

In addition,   sometimes it is also required that the spacetime be asymptotically flat in the radial direction, especially  for the cases in which the sources are confined 
within a finite region \cite{BCM95}.

Finally, we note that  there are physically realistic situations where an axis may not exist, as it may be singular and so not part of the manifold, or the topology of the manifold may be
such that no axis exists. Barnes noted   \cite{Barnes00} that in this case  the requirement of the existence of an axis in Proposition 1 can be    {removed} and the following holds  (See also \cite{BS84} and the footnote 
given  in \cite{Barnes00}).

{\bf Proposition 2.} Any two-dimensional Lie transformation group that contains a one-dimensional subgroup whose orbits are circles must be Abelian.

\subsection{Einstein's field equations}

The world constructed by Newton to describe his theory of gravitation had much simplicity: Space is always equal to itself where particles move and act upon each other. Furthermore, the gravitational effects propagate with an infinitely large velocity. This concept of space sounded not plausible to Einstein  and he came with the brilliant idea that space is the gravitational field meaning the Newton space itself \cite{Rovelli}. This needed a better formulation of the field equations. Einstein, after a long struggle, found that the Riemann geometry, first proposed by Gauss and then generalized to any dimensions by Riemann, could properly describe the curvature of space produced by different matter distributions. By doing this Einstein managed to express the gravitational field equations in an arbitrary coordinate system.

In Riemann's geometry the parallel transport of a vector is proportional to its curvature which is described by a quantity called the Riemann curvature tensor $R_{\alpha\beta\gamma\delta}$. The justification of calling it curvature lies in the fact that it vanishes if and only if the space is flat.

Einstein \cite{Einstein} derived the field equations in the year 1915 which are given by \footnote{In this review, we shall adopt the conventions of \cite{DIn92}. In particular, the signature of the metric will be $(+, -, -, -)$,   {and when the cosmological constant $\Lambda$ does not vanish, Einstein's equations read,
$R_{\alpha\beta}-\frac{1}{2}Rg_{\alpha\beta} - \Lambda g_{\alpha\beta} =\kappa T_{\alpha\beta}$}.},
\begin{equation}
R_{\alpha\beta}-\frac{1}{2}Rg_{\alpha\beta}=\kappa T_{\alpha\beta}.  \label{1}
\end{equation}
They show how the space curvature, represented by the Ricci tensor $R_{\alpha\beta}={R^{\gamma}}_{\alpha\gamma\beta}$ and its scalar $R=g^{\alpha\beta}R_{\alpha\beta}$, are brought about by the source of curvature, the matter distribution given by the energy momentum tensor $T_{\alpha\beta}$. In fact not only space curves but time as well and for this reason we call spacetime where Einstein field equations dwell. The coupling constant $\kappa$ in normalized units, the velocity of light $c=1$ and the Newton's constant $G=1$, values  {$\kappa\equiv 8\pi G/c^4 = 8\pi$. } This system of equations, called Einstein's field equations, constitutes a set of 10 partial differential equations with respect to the metric $g_{\alpha\beta}$.  {In the vacuum case where $T_{\alpha\beta}=0$, These equations  reduce to  }
\begin{equation}
R_{\alpha\beta}=0. \label{2}
\end{equation}

When spacetime is deprived of sources producing gravitational fields, and thereby producing curvature,    the spacetime is flat and given by the metric $\eta_{\alpha\beta}$, called Minkowski metric,
\begin{equation}
ds^2=\eta_{\alpha\beta}dx^{\alpha}dx^{\beta}=dt^2-dx^2-dy^2-dz^2, \label{2a}
\end{equation}
with $t$  {being}  the time coordinate and $x$, $y$ and $z$ the Cartesian coordinates.
Hence, the Newtonian limit, which one expects to be obtained from Einstein field equations,  is produced when the metric is given by
\begin{equation}
g_{\alpha\beta}=\eta_{\alpha\beta}+f_{\alpha\beta}, \label{2aa}
\end{equation}
with $f_{\alpha\beta}$  {being} a small deviation from the Minkowski spacetime.

In this paper, we shall present an updated review on cylindrically symmetric spacetimes. However, due to the vast scope of the field carried out in the past several decades and the limitation of the  {scope}
of the review, it is impossible to cover all the subjects studied so far. So, one way or another one has to make a choice on which subjects that should be included in a brief review,  like the current one. 
Such a choice clearly contains the reviewers' bias. In addition, in this review we do not intend to exhaust all  the relevant articles even within 
the chosen subjects, as in the information era, one can easily find them, for example, from the list of the citations of relevant articles. With all these in mind, we would like first to offer our sincere thanks 
and apologies  to whom his/her work is not mentioned in this review. 
In addition,  there have already existed two excellent books  to cover the subjects extensively    \cite{StephaniA,GP09} and a brief review on  gravitational collapse in cylindrically symmetric spacetimes \cite{Mena15}.

  The rest of article is organized as follows: In the next four sections, we shall review the general properties, both local and global, and possible sources of the Levi-Civita solution (Sec. 2), coupled with an electromagnetic
field (Sec. 3), with a cosmological constant (Sec. 4), or with a perfect fluid (Sec. 5). Then, in the next two   sections, we shall provide a general review on the Lewis vacuum solution (Sec. 6), and its generalization to include 
a cosmological   constant (Sec. 7). Finally,  In Secs. 8 and 9, we shall provide a review on wormholes, gravitational collapse, polarizations of cylindrical GWs and gravitational Faraday rotations. An appendix is also included, in 
which    Einstein's field equations of static   cylindrical spacetimes are presented.

%
\section{Levi-Civita (LC) vacuum spacetime}   
%
\renewcommand{\theequation}{2.\arabic{equation}} \setcounter{equation}{0}

 Einstein's field equations (\ref{1})
 are highly nonlinear thus imposing  {huge}  difficulties in finding their solutions. Furthermore, once found their solutions, even   {more} difficulties arise in interpreting them. In spite of great efforts to grasp their   {physical meanings}, as is well shown in \cite{Stephani} and   {\cite{GP09}}, the majority or even almost the totality of these solutions are still not well understood given their high nonlinearity. A simple example like the spherical vacuum field for a point mass, presented below, which is perhaps the solution most studied in GR, still bears weird and unexpected properties as we will see in    the following by studying cylindrically symmetric vacuum fields.

Schwarzschild \cite{Schwa} in 1916 obtained the first vacuum solution to   {(\ref{2})}, describing the spherically symmetric vacuum field with the following line element $ds$,
\begin{eqnarray}
ds^2&=&g_{\alpha\beta}dx^{\alpha}dx^{\beta} =\left(1-\frac{2M}{r}\right)dt^2\nb\\
&& -\left(1-\frac{2M}{r}\right)^{-1}dr^2-r^2(d\theta^2+\sin^2\theta d\phi^2), \label{3}
\end{eqnarray}
where the spherical coordinates are numbered $x^0=t$, $x^1=r$, $x^2=\theta$ and $x^3=\phi$. In the Newtonian approximation (\ref{3}) becomes
\begin{equation}
g_{00}=1+2U, \label{4}
\end{equation}
where $U$ is the Newtonian potential and comparing to (\ref{3}) we have
\begin{equation}
U=  {-} \frac{M}{r}, \label{5}
\end{equation}
hence the only parameter stemming from the integration of the field equations for a spherically symmetric vacuum spacetime is the Newtonian mass $M$.

\subsection{ LC solutions}

The second vacuum solution of Einstein field equations   {(\ref{2})} was obtained a few years later,  { in 1919 by LC \cite{Levi},
 corresponding to a cylindrical vacuum spacetime, and can be cast}  in the form   {[Cf. Appendix A]},
\begin{equation}
ds^2=\rho^{4\sigma}dt^2-\rho^{4\sigma(2\sigma-1)}(d\rho^2+dz^2)-\frac{1}{a}\rho^{2(1-2\sigma)}d\phi^2, \label{6}
\end{equation}
where the cylindrical coordinates are numbered $x^0=t$, $x^1=\rho$, $x^2=z$ and $x^3=\phi$. In the Newtonian approximation,   { correspondingly to (\ref{4}) we have, }
\begin{equation}
g_{00}=e^{2U}, \label{7}
\end{equation}
    and compared to (\ref{6})   { for this case} we have the Newtonian potential,
\begin{equation}
U=2\sigma\ln \rho. \label{8}
\end{equation}
Hence, from   {this expression}  we clearly see that for small values of $\sigma$ it is the Newtonian mass per unit length as produced by an infinitely long homogeneous line mass, as observed by LC himself. Ever since much has been written by researchers trying to grasp its physical and geometrical interpretations. However, this endeavor proved to be difficult and uncertain.

Only in 1958 did Marder \cite{Marder}   {established} that the solution (\ref{6}) contains two arbitrary independent parameters,   { $\sigma$ and $a$, }
 differently from the Newtonian fields   {in wich} there is only one independent parameter   {$\sigma$.} We also call the attention  that for the spherical case in the relativistic and Newtonian theories there appears just one parameter. This fact already suggests some   {harder} difficulties in understanding cylindrically symmetric fields. 

In the Newtonian approximation the parameter $a$ is associated with the constant arbitrary potential that exists in the Newtonian solution,  { and can be given any value.
 However,  in GR Bonnor   \cite{BonnorA} observed that $a$ is  dressed  with a relevant global topological meaning, and cannot be removed by coordinate transformations.}

In the   following   we review the main properties and physics that lie behind the LC spacetime, which
 so far have been grasped up to the present time in a large number of papers. We are aware that these results sometimes appear contradictory since some interpretations collide with others. In fact this is one of our main motivations to deepen into Einstein's theory.

\subsection{Nature of the coordinates of the LC solutions}

The LC metric given by (\ref{6}) can be written in the form \cite{HerreraB}
\begin{equation}
ds^2=\rho^{4\sigma}dt^2-\rho^{4\sigma(2\sigma-1)}\left(d\rho^2+\frac{1}{a_m}dm^2\right)-\frac{1}{a_n}\rho^{2(1-2\sigma)}dn^2, \label{9}
\end{equation}
where $-\infty<t<\infty$ is the time and $0\leq\rho<\infty$ the radial coordinates, and $\sigma$, $a_m$ and $a_n$ are constants. The nature of the coordinates $m$ and $n$, so far unspecified, depends upon the behaviour of the metric coefficients. Either $a_m$ or $a_n$ can be transformed away by a scale transformation depending upon the behaviour of the coordinates $m$  and $n$, thus leaving the metric with only two independent parameters. In order to find that behaviour we transform the radius $\rho$ into a proper length radial coordinate $r$ by defining
\begin{equation}
\rho^{2\sigma(2\sigma-1)}d\rho=dr, \label{10}
\end{equation}
thus obtaining
\begin{equation}
\rho=R^{\frac{1}{\Sigma}}, \;\; R=\Sigma r, \;\; \Sigma \equiv 4\sigma^2-2\sigma+1, \label{11}
\end{equation}
and metric (\ref{9}) becomes
\begin{equation}
ds^2=f(r)dt^2-dr^2-h(r)dm^2-l(r)dn^2, \label{12}
\end{equation}
with
\begin{equation}
f(r)=R^{\frac{4\sigma}{\Sigma}}, \;\; h(r)=\frac{1}{a_m}R^{\frac{4\sigma(2\sigma-1)}{\Sigma}}, \;\; l(r)=\frac{1}{a_n}R^{\frac{2(1-2\sigma)}{\Sigma}}. \label{13}
\end{equation}

Consider $0<\sigma<1/2$, which implies for this range $h(r)$ diverging when $r\rightarrow 0$ and $l(0)=0$. Then we can interpret $m$ as the axial coordinate $-\infty<z<\infty$ with $a_m=1$ by rescaling $z$, and $n$ as the angular coordinate $\phi$ with the topological identification of every $\phi$ with $\phi+2\pi$, and the metric (\ref{12}) becomes
\begin{equation}
ds^2=R^{\frac{4\sigma}{\Sigma}}dt^2-dr^2-R^{-\frac{4\sigma(1-2\sigma)}{\Sigma}}dz^2-\frac{1}{a}R^{\frac{2(1-2\sigma)}{\Sigma}}d\phi^2, \label{14}
\end{equation}
where $a_n$ is replaced by $a$.

Consider $1/2<\sigma<\infty$,  which implies  $h(0)=0$ and $l(r)$ diverging when $r\rightarrow 0$. Then we can interpret $m$ as the angular coordinate $\phi$ with topological identification of every $\phi$ with $\phi+2\pi$, and $n$ as an axial coordinate $-\infty<z<\infty$ with $a_n=1$ by again rescaling $z$, and (\ref{12}) becomes
\begin{equation}
ds^2=R^{\frac{4\sigma}{\Sigma}}dt^2-dr^2-R^{-\frac{2(2\sigma-1)}{\Sigma}}dz^2-\frac{1}{a}R^{\frac{4\sigma(2\sigma-1)}{\Sigma}}d\phi^2, \label{15}
\end{equation}
where we replaced  {$a_m$} for $a$.

In both cases (\ref{14}) and (\ref{15}) where $\sigma>0$ we have $g_{00}\rightarrow 0$ as $r\rightarrow 0$, indicating an attractive singularity. While assuming $\sigma <0$, we obtain $g_{00}\rightarrow\infty$ as $r\rightarrow 0$, indicating a repulsive singularity.

The invariant quantity, under coordinate transformations, built out of the Riemann curvature tensor given by ${\mathcal R}\equiv R^{\alpha\beta\gamma\delta}R_{\alpha\beta\gamma\delta}$, called the Kretschmann scalar, is a good indicator of singularities. Calculating $\mathcal R$ for metrics (\ref{14}) and (\ref{15}) we obtain
\begin{equation}
{\mathcal R}=\frac{64\sigma^2(2\sigma-1)^2}{\Sigma^3r^4}. \label{16}
\end{equation}
From (\ref{16}) we see that $\mathcal{R}\rightarrow\infty$ as $r\rightarrow 0$,  and $\mathcal{R}\rightarrow 0$ for $\sigma=0$, $1/2$ and $\infty$.

Metric (\ref{14}) for $\sigma=0$ becomes
\begin{equation}
ds^2=dt^2-dr^2-dz^2-\frac{1}{a}r^2d\phi^2, \label{22}
\end{equation}
representing the Minkowski spacetime when $a=1$ in cylindrical polar coordinates $r$, $z$ and $\phi$. However, if (\ref{22}) has $a>1$ there is an angle deficit $2\pi\delta$ given by $\delta=1-1/\sqrt a$,
 producing flat spacetime everywhere except along the axis $r=0$,  which is interpreted as a cosmic string. The deficit can represent the tension of the string with mass per unit length $\mu$ given by $\mu=\delta/4$. If there is an angle excess $a<1$ it would represent a cosmic string under compression with $\mu<0$. Hence the constant $a$ is directly linked to the gravitational analog of Aharonov-Bohm effect \cite{Dowker}. This effect shows that gravitation depends on the topological structure of spacetime giving rise to an angular deficit $\delta$ as in the electromagnetic Aharonov-Bohm effect, where a (classical) non-observable quantity (the vector potential) becomes observable (part of it) through a quantum non-local effect. Its gravitational analog allows a (Newtonian) non-observable quantity (the additional constant to the Newtonian potential) to become observable in the relativistic theory through the angular deficit in strings. For a review in cosmic strings see \cite{Hindmarsh}.

In the case $\sigma=1/2$ the two metric coefficients $h$ and $l$ in (\ref{13}) are constants, so both $a_m$ and $a_n$ can be  {set to unity}.  Then neither $m$ nor $n$ is entitled to be an angular coordinate, and the three coordinates $r$, $m$ and $n$ are better visualized as Cartesian coordinates   {$x$, $y$ and $z$}. Hence metric (\ref{9}) can be written as
\begin{equation}
ds^2=z^2dt^2-dx^2-dy^2-dz^2, \label{23}
\end{equation}
which is the static plane symmetric vacuum spacetime obtained by Rindler  \cite{GriffithsA,Rindler,Silva1}.

In the next section we calculate the circular geodesics for the above spacetime to try to get further understanding of the properties of  different values that $\sigma$ can attain for $\sigma\geq 0$, since $\sigma<0$ would in some way correspond to negative mass densities.

\subsection{Geodesics}

For the metrics (\ref{14}) and (\ref{15}) the circular geodesics \cite{Silva} have ${\dot r}={\dot z}=0$ and  $g_{tt,r}{\dot t}^2+g_{\phi\phi,r}{\dot\phi}^2=0$, where the dot stands for differentiation with respect to 
 { the affine parameter $s$. The geodesic angular velocity is defined by $\omega={\dot\phi}/{\dot t}$, and the only nonzero component of its velocity  is the  tangential one, given by } $W^{\phi}=\omega/{\sqrt g_{tt}}$ with its modulus defined by
$W^2 \equiv W^{\alpha}W_{\alpha}$ \footnote{This $W$ is different from the one appearing in the metric (\ref{1.2}) or that in (\ref{1.3}).} \cite{Herrera3A}.

In the case $0\leq\sigma<1/2$, from (\ref{14}) we obtain
\begin{eqnarray}
\omega^2&=&\frac{2\sigma}{1-2\sigma}aR^{2(4\sigma-1)/\Sigma}, \label{17}\\
W^2&=&\frac{2\sigma}{1-2\sigma}. \label{18}
\end{eqnarray}
We note that for a given $\sigma$ the  velocity $W$ in (\ref{18}) is the same for all circular geodesics, in agreement with the corresponding Newtonian gravitation. Furthermore, we see that $W$ increases monotonically with $\sigma$, that is from $\sigma=0$ producing $W=0$, to $\sigma=1/4$ attaining $W=1$, the speed of light, and finally $\sigma=1/2$ producing geodesics with $W=\infty$.
For small $\sigma$ and $a=1$ from (\ref{17}) and (\ref{18}) we obtain the Newtonian limit $W=r\omega$.

In the case $1/2<\sigma<\infty$,  from (\ref{15}) we obtain
\begin{eqnarray}
\omega^2&=&\frac{1}{2\sigma-1}aR^{8\sigma(1-\sigma)/\Sigma}, \label{19}\\
W^2&=&\frac{1}{2\sigma-1}. \label{20}
\end{eqnarray}
With $\sigma$ increasing beyond $1/2$, we note from (\ref{20}) that $W$ diminishes, attaining $W=1$ for $\sigma=1$ and $W=0$ for $\sigma=\infty$.

In other words, the circular geodesics are timelike when either $0<\sigma<1/4$ or $\sigma>1$, are lightlike when $\sigma=1/4$ or $1$, and are spacelike when $1/4<\sigma<1$.

If we redefine $\sigma$ by
\begin{equation}
\sigma=\frac{1}{4\bar\sigma}, \label{21}
\end{equation}
then metric (\ref{15}) with a rescaling of its coordinates becomes (\ref{14}), hence (\ref{19}) and (\ref{20}) reduce to (\ref{17}) and (\ref{18}). This means that the parameter range $1/2<\sigma<\infty$ is equivalent to the range $0<\sigma<1/2$ and the coordinates $z$ and $\phi$ switching their nature.

Hence we might have the following picture for the different values of $\sigma$. For small values of $\sigma$ the metric (\ref{14}) with $t$ and $r$ constants describes cylindrical surfaces with $\phi$ a periodic coordinate. As $\sigma$ increases the cylindrical surfaces open out and become infinite planes for $\sigma=1/2$. For values of $\sigma$ bigger that $1/2$ the coordinate $z$ becomes periodic forming new cylindrical surfaces perpendicular to previous ones for $0<\sigma<1/2$.

  {Another interesting geodesic is the one that describes}  the motion of the particle along the axis of symmetry $z$. These geodesics calculated with (\ref{14}) (we restrict the calculation of this metric since (\ref{17}) is equivalent) produce
\begin{equation}
{\ddot z}=\frac{4\sigma(1-2\sigma)}{\Sigma}\frac{\dot r\dot z}{r}. \label{21a}
\end{equation}
It means that particles increase their speed along $z$ when distancing radially from the axis, while diminishing their axial speed when moving radially towards the axis. This result indicates that a force parallel to the $z$ axis appears. In the flat case $\sigma=0$ such an effect vanishes, bringing out its non-Newtonian nature.  We further discuss this   {weird} geodesic property in the section  concerning the Lewis spacetime.

For radial geodesics it has been shown  {\cite{Herrera3A}} that there exist timelike particles approaching $z$ that are reflected at $r=r_{min}$ and move outwards until attaining ${\dot r}=0$ at $r_{max}$ repeating endlessly this trajectory. This motion is called confinement of test particles.

In the next section we see some further limits that LC metric satisfies.

\subsection{LC spacetime as a limiting case of the $\gamma$ spacetime}

In cylindrical coordinates, static axisymmetric solutions of Einstein's vaccuum equations are given by the Weyl metric \cite{Weyl}
\begin{equation}
ds^2=e^{2\lambda}dt^2-e^{-2\lambda}\left[e^{2\mu}(d\rho^2+dz^2)+\rho^2d\phi^2\right], \label{24}
\end{equation}
with $\lambda(\rho,z)$ and $\mu(\rho,z)$ satisfying
\begin{equation}
\lambda_{,\rho\rho}+\frac{1}{\rho}\lambda_{,\rho}+\lambda_{,zz}=0, \label{25}
\end{equation}
and
\begin{eqnarray}
\mu_{,\rho}&=&\rho(\lambda_{,\rho}^2-\lambda_{,z}^2), \label{26}\\
\mu_{,z}&=&2\rho\lambda_{,\rho}\lambda_{,z}, \label{27}
\end{eqnarray}
where the comma stands for partial derivation. Synge writes \cite{Synge}, as the most amazing fact, that (\ref{25}) is just the Laplace equation for $\lambda$ in the Euclidean space.  {
Metric (\ref{24}) with the general solution of (\ref{25}), (\ref{26}) and (\ref{27}) is usually  referred to as the  $\gamma$ metric,  and the corresponding functions  $\lambda$ and $\mu$ are given by}
\begin{eqnarray}
e^{2\lambda}&=&\left(\frac{R_++R_--2m}{R_++R_-+2m}\right)^{\gamma}, \label{28}\\
e^{2\mu}&=&\left[\frac{(R_++R_-+2m)(R_++R_--2m)}{4R_+R_-}\right]^{\gamma^2}, \label{29}
\end{eqnarray}
where
\begin{equation}
R^2_{\pm}=\rho^2+(z\pm m)^2, \label{30}
\end{equation}
and $\gamma$  {and $m$ are two integrations  constants.  These solutions were first found } by Bach and Weyl in 1922 \cite{Bach}. Calculating its Newtonian potential (\ref{7}) we obtain
\begin{equation}
U=\gamma\ln\left(\frac{R_-+z-m}{R_++z+m}\right), \label{31}
\end{equation}
which corresponds to a potential due to a line segment of length $2m$ and mass per unit length $\gamma/2$ symmetrically distributed along the $z$ axis. Hence the total mass $M$ of the line segment is $M=\gamma m$. The particular case $\gamma=1$ corresponds to the Schwarzschild metric. This can be seen by taking spherical Erez-Rosen coordinates \cite{Erez}, defined   by
\begin{equation}
\rho^2=(r^2-2mr)\sin^2\theta, \;\; z=(r-m)\cos\theta, \label{32}
\end{equation}
and the $\gamma$ metric becomes
\begin{equation}
ds^2=Fdt^2-\frac{1}{F}\left[Gdr^2+Hd\theta^2+(r^2-2mr)\sin^2\theta d\phi^2\right], \label{33}
\end{equation}
where
\begin{eqnarray}
F&=&\left(1-\frac{2m}{r}\right)^{\gamma}, \label{34}\\
G&=&\left(\frac{r^2-2mr}{r^2-2mr+m^2\sin^2\theta}\right)^{\gamma^2-1}, \label{35}\\
H&=&\frac{(r^2-2mr)^{\gamma^2}}{(r^2-2mr+m^2\sin^2\theta)^{\gamma^2-1}}. \label{36}
\end{eqnarray}
Now we can easily check that for $\gamma=1$ the metric (\ref{33}) reduces to the Schwarzschild metric (\ref{3}).

If we want to compare the $\gamma$ metric  {to the LC metric,  in the limit when its length segment $m\rightarrow\infty$, } one notices that by taking directly this limit in (\ref{28}) and (\ref{29}) the metric diverges. For this reason 
 {one can}  use the Cartan scalars approach to obtain a finite limit. These scalars are the components of the Riemann tensor and its covariant derivatives calculated in a constant frame. Two metrics are equivalent if and only if there exist coordinate and Lorentz transformations which transform the Cartan scalars of one of the metrics into the Cartan scalars of the other. Although the Cartan scalars provide a local characterization of the spacetime, global properties such as topological defects do not appear in them. By doing all this procedure one can prove that locally in the limit $m\rightarrow\infty$ the $\gamma$ metric is the same as the LC metric. Details of these long calculations are given in  {\cite{Herrera1A}}. Here we come to an interesting and, so far, weird result showing how apparently unexpected results can stem from long known results like the Schwarzschild and LC solutions. When the   density per unit length of the rod $\sigma=\gamma/2$ has the value $\gamma=1$, or the mass density per unit length $\sigma=1/2$, it becomes the Schwarzschild spherically symmetric spacetime, and  { in the limit $m\rightarrow\infty$,  it } becomes the Rindler static plane symmetric vacuum spacetime. This is a remarkable result.

For all the different limiting metrics that LC spacetime can undergo see  {\cite{Herrera1A}}. The limits for the circular geodesics of the $\gamma$ spacetime  to the LC spacetime are well studied in  {\cite{Herrera3A}}.

In the next section we make a brief review of possible sources to the LC spacetime.

\subsection{Sources producing LC spacetime}

The LC spacetime, as we saw, contains two essential constants denoted by $a$ and $\sigma$. The constant $a$ refers to an angle deficit or excess that produces cosmological strings. However, it is $\sigma$ that presents the most serious obstacles to its interpretation. Indeed, for small values $0<\sigma< 1/4$, LC describes the spacetime generated by an infinite line mass, with $\sigma$ as mass per unit coordinate length. When $\sigma=0$ the spacetime is flat. However, circular timelike geodesics only exist for $0<\sigma<1/4$, becoming null for $\sigma=1/4$ and being spacelike for $\sigma>1/4$. Furthermore, as the value of $\sigma$ increases from $1/4$ to $1/2$  the corresponding Kretschmann scalar (\ref{16}) diminishes monotonically and vanishes when attaining $\sigma=1/2$ implying that the spacetime is flat.

Thus, the question is what does the LC spacetime represent outside the range $0\leq\sigma\leq 1/4$ ?

First,  we observe that there are known physical satisfactory fluid sources for the LC spacetime satisfying boundary conditions for both ranges of $\sigma$ (see for example \cite{Stephani,Herrera2}). On the other hand, the fact that the Kretschmann scalar decreases with increasing $\sigma$ may not be associated with the strength of the gravitational field. Instead,  it could be associated with the acceleration of a test particle that would measure more suitably the strength of the gravitational field, which is the case for $1/4<\sigma<1/2$.

A possible interpretation of the LC solution  is a spacetime generated by a cylinder whose radius increases with $\sigma$ and tends to infinity as $\sigma$ approaches $1/2$. This interpretation suggests that when $\sigma=1/2$ the cylinder becomes a plane.

It might be instructive to consider a static cylinder filled with anisotropic perfect fluid and calculate its mass per unit length by using the junction conditions to its exterior LC spacetime. The Whittaker formula \cite{Whittaker} for the active gravitational mass per unit length $\nu$ of a static distribution of perfect fluid with energy density $\mu$ and principal stresses $P_r$, $P_z$ and $P_\phi$ inside a cylinder of surface $S$ is
\begin{equation}
\nu=2\pi\int_0^{r_S}(\mu+P_r+P_z+P_\phi)\sqrt{-g}dr.  \label{37}
\end{equation}
Considering a static cylindrically symmetric metric
\begin{equation}
ds^2=-Adt^2+d\rho^2+Cdz^2+D\rho^2d\phi^2, \label{38}
\end{equation}
in which $A$, $C$ and $D$ are only functions of $\rho$,  from Einstein's fields equations we obtain \cite{Griffiths1}
\begin{equation}
\frac{A_{,rr}}{A}-\frac{1}{2}\frac{A_{,r}}{A}\left(\frac{A_{,r}}{A}-\frac{2}{\rho}-\frac{C_{,r}}{C}-\frac{D_{,r}}{D}\right)=\kappa\left(\mu+P_{\rho}+P_z+P_{\phi}\right). \label{39}
\end{equation}
Substituting (\ref{39}) into (\ref{37}) we have the simple expression
\begin{equation}
\nu=\frac{1}{4}\left(\frac{A_{,r}}{A}\sqrt{-g}\right)_S , \label{40}
\end{equation}
where regularity conditions on the axis of symmetry \cite{Herrera2} have been assumed. Now taking  {the LC metric (\ref{14}) for the exterior spacetime of the cylindrical surface $S$, 
 and imposing that it satisfies Darmois' junction conditions \cite{Herrera2,Darmois} with its interior spacetime described by  (\ref{38}), it amounts to impose the continuity of the metric functions and its derivatives on $S$. By doing so,
 from (\ref{40})  we obtain  }
\begin{equation}
\nu=\frac{\sigma}{\sqrt a}, \label{41}
\end{equation}
 {where $a$ is the constant defined in (\ref{14})}. From (\ref{41}) we have that when $a>1$, there is a topological angle defficit, then $\nu<\sigma$, and if $a<1$ there is an angle excess, producing $\nu>\sigma$.
However, when there is no topological defect, $a=1$, then $\nu$ can in fact be interpreted as mass per unit length of its source. Furthermore, since with cylindrical sources no black holes are formed, one might conclude that the minimum mass per unit length to form a black hole satisfies the constraint $\nu>1/2$. This result would fulfill the present knowledge of black holes that a lower mass limit is required for its formation. We are aware that $\nu$ is model-dependent and cannot be given a general meaning, but nonetheless it fulfills some of the expected properties.

A last comment on sources for the LC spacetime is the fact that conformally flat static sources do not admit Darmois'  matching conditions satisfied for an exterior LC spacetime. This result is proved in
 {\cite{Herrera1}}. Conformal flatness of a metric means that its corresponding Weyl tensor vanishes. The interpretation of the Weyl tensor is that it describes the purely gravitational  {characteristics} of the source. This interesting result means that a static cylindrical source deprived of its purely gravitational character cannot be smoothly matched to the exterior LC spacetime. It is conspicuous that for spherical symmetry this result does not stand, since there are conformally flat static spherical sources matched to the Schwarzschild spacetime  {\cite{HPOF01}}.


%
\section{Static electrovacuum spacetimes} 
%
%
\renewcommand{\theequation}{3.\arabic{equation}} \setcounter{equation}{0}


\subsection{Electromagnetic fields in cylindrical spacetimes}

 In this section we discuss the gravitational fields whose only source is a free 
 electromagnetic field described by Maxwell's theory. This field in curved spacetime is 
 characterized by the antisymmetric tensor   {$F\mn = \d_\nu A_\mu -\d_\mu A_\nu$}
 ($A_\mu$ is the vector 4-potential) satisfying the Maxwell equations
 
\beq                     \label{Max}
		\nabla_\alpha F^{\alpha\mu}   { = 4\pi j^\mu}, \quad  \nabla_\alpha {}^*F^{\alpha\mu}=0,
\eeq 
 where $\nabla_\alpha$ is a covariant derivative, $j^\mu$ is the electric 4-current, and 
 $^*F^{\alpha\beta} = -\eps^{\alpha\beta\mu\nu}F\mn/(2\sqrt{-g})$ 
 is the axial tensor dual to $F\mn$ ($\eps^{\alpha\beta\mu\nu}$ is the LC completely
 antisymmetric unit object). In a vacuum region of a spacetime we have $j^\mu =0$.

 Nonzero components $F\mn$, compatible with the geometry (\ref{ds-cy}), 
 depending on the radial coordinate $u$ only can have three different directions:
\begin{itemize}  
\item
	 Radial (R) fields:  electric, $F_{01} (u)$ (so that $E^2 =F_{01}F^{10}$), and
      magnetic, $F_{23} (u)$ (so that $B^2 = F_{23}F^{23}$).
\item
	 Longitudinal (L) fields: electric, $F_{02}(u)$  ($E^2 =F_{02}F^{20}$), and 
      magnetic, $F_{13}(u)$ ($B^2 = F_{13}F^{13}$).
\item
	 Azimuthal (A) fields: electric, $F_{03} (u)$ ($E^2 =F_{03}F^{30}$),
      and magnetic, $F_{12}(u)$ ($B^2 = F_{12} F^{12}$).
\end{itemize} 
  Here $E$ and $B$ are the absolute values of the electric field strength and magnetic induction,
  respectively. Self-gravitating static, cylindrically symmetric configurations of such electromagnetic
  fields have been considered in   {\cite{em1,em2,em3,em4,MSWS01}} ( {in which see}  also references to earlier
  work, and more historical details can be found in \cite{StephaniA}). In our presentation we follow 
  the lines of \cite{em3,em4}. 

  The Maxwell equations follow from the least action principle with the Lagrangian 
\beq 				\label{L_em}
  	L_{\rm em} = - \frac{F\mn F\MN}{16\pi} \equiv \frac{E^2 - B^2}{8\pi}.
\eeq
  The same Lagrangian leads to the following well-known form of   {the stress-energy tensor (SET)}
  to be used as a source of gravity in Einstein's equations:
\beq            		      				\label{SET}
	T\mN = \frac{1}{16\pi}\left(-4F_{\mu\alpha}F^{\nu\alpha} 
					+ \delta\mN F_{\alpha\beta}F^{\alpha\beta}\right). 
\eeq

  This expression leads to an important ( { result: } it turns out that the metric \rf{ds-cy} admits only
  a single sort of electromagnetic field  {(R, L, or A, defined above).} Any simultaneous existence of fields of different
  directions is forbidden due to the emergence of off-diagonal components of $T\mN$,  whereas 
  the left-hand side of   Einstein's equations \rf{1} ( {has only diagonal components. For example, a nonzero electric
  R-field ($F_{01} \ne 0$) together with a magnetic L-field ($F_{13}\ne 0$) leads} to $T_0^3 \ne 0$.

  However, simultaneous existence of electric and magnetic fields of the same direction is 
  admissible, and due to the well-known electric-magnetic duality, such fields lead to similar
  contributions to $T\mN$ (in the general case  there appears the sum $E^2 + B^2$). Therefore in
  what follows we consider each sort of fields (R, L, A) separately,  and for definiteness, speak 
  of electric R-fields ($F_{01} = -F_{10}$), magnetic L-fields ($F_{13} = -F_{31}$) and magnetic 
  A-fields ($F_{12} = -F_{21}$), bearing in mind that in each case mixtures of electric and
   magnetic fields of the same direction can be obtained by duality rotations without changing
  the SET and therefore the metric. With this ansatz the Maxwell equations
  reduce to the first part of \eqs \rf{Max}. In vacuum ($j^\mu =0$) they lead to
\beq                                                                   \label{q_i}
			\frac d{du} \left(\sqrt{-g}F^{\mu 1}\right) = 0 \;\; \then \;\;
						F^{\mu 1} = \frac{q_\mu}{\sqrt{-g}}, \;\; q_\mu = \const,  
\eeq
  where $\mu = 0, 3, 2$ for R-, L-, A-fields, respectively. 

  To solve   Einstein's equations, we choose the harmonic radial coordinate $u$ 
  \rf{harm} in the metric \rf{ds-cy}, such that $\alpha(u) = \beta(u) + \gamma(u) + \mu(u)$.  
  Then $\sqrt{-g} = e^{\alpha + \beta + \gamma + \mu} = e^{2\alpha}$, and 
  the solutions \rf{q_i} can be written as
\beq                                                        \label{q_a}
                 \left\{F^{01}, F^{31}, F^{21}\right\} = e^{-2\alpha} \left\{q_R, q_L, q_A\right\},
\eeq
  where in each curly bracket one should consider a single term for each sort of field, and the
  constants $q_{N}$ have been renamed accordingly. In what follows, describing each of these
  fields separately, we denote by simply $q$ any of these ``charges'' without a risk of confusions.  

  Since the tensor $T\mN$ has zero trace, $T = T_\alpha^\alpha =0$,   { Einstein's} equations 
  can be used in the form   {$R\mN = 8\pi T\mN$}, with the expressions \rf{Ric} for $R\mN$.
  They are easily solved by the same method in all three cases, but the resulting geometries are
  substantially different. 

\subsection{Geometry with a radial electric field}

  In this case, as one easily finds, the SET \rf{SET} takes the form   
\beq                                     \label{SET-R}
		T\mN = \frac {q^2}{8\pi} e^{-2\mu -2\beta} \diag (1,\ 1,\ -1,\ -1),
\eeq
  where $q = q_R$ may be interpreted as a linear electric charge along the $z$ axis. 
  Due to high symmetry of the SET \rf{SET-R}, independent components of 
    {Einstein's  equations $R\mN = 8\pi T\mN$ }
  reduce to
\bearr                                        \label{R1}
	\beta'' = \mu'' = -\gamma'',
\yyy								\label{R2}
	\gamma'' = q^2 e^{2\gamma},
\yyy  							\label{R3}
	\beta'\gamma'+\beta'\mu' + \gamma'\mu' = q^2 e^{2\gamma}. 
\ear
  Equations \rf{R1} are trivial while \rf{R2} is a Liouville equation which is easily solved, giving
\bearr                                           \label{R4} 
	  \mu(u) = au - \gamma(u), \quad  \beta(u) = bu - \gamma(u),
\yyy								\label{R5}
	  e^{-\gamma(u)} = s(k, u) \equiv \vars{(1/k) \sinh ku, &  k>0,\\
										u, &  k =0,\\
										(1/k) \sin ku, &  k <0,}
\ear
  where $a, b, k$ are integration   {constants},  which can take any real values, and three more 
  constants have been removed by rescaling $t$ and $z$ and by choosing the zero point 
  of the coordinate $u$. 
  Furthermore, \rf {R3} is a first
  integral of \rf{R1} and \rf{R2}, which, after substitution of \rf{R4} and \rf{R5} yields a 
  simple relation between the integration constants
\beq 					\label{R6}
	    ab = k^2 \sign (k).
\eeq
  Thus we have a solution with three significant integration constants $q$, $a$, $b$, which 
  splits into three subfamilies according to the sign of $k$.

\medskip\noi
  {\bf 1.} $k >0$, so that $ab = k^2$. The metric takes the form
\bqun                          \label{ds-R+}
		ds^2 &=& \frac{k^2 dt^2}{q^2\sinh^2 ku} - \frac{q^2}{k^2}\sinh^2 ku 
			\Big[e^{2(a+b)u}du^2   \nb\\
			&& + e^{2au}dz^2 +e^{2bu} d\phi^2\Big]. 
\equn
  Without loss of generality we may take $u \in \R_+ = (0,\infty)$ as the range of $u$,
  and evidently $u=0$ corresponds to the axis since the cylindrical radius $r \equiv e^\beta$
  vanishes there. The electric field strength diverges there as $E\approx |q|/u^2$, therefore 
  $u=0$ can be interpreted as the location of a stretched electric charge. Since there
  $g_{00} = e^{2\gamma} \sim 1/u^2 \to \infty$, it is a repulsive singularity for 
  small test bodies. 

  The behavior of the metric at the other end, $u\to \infty$, depends on the interplay of 
  the constants $a$ and $b$: while in all cases $e^{2\gamma}\to 0$, the radius $e^\beta$
  and the longitudinal scale factor $e^\mu$ may tend to zero, infinity or finite values, as is
  apparent in \rf{ds-R+}. In the cases where $r\to 0$ as $u\to\infty$, we have to conclude
  that the  spacetime has there one more axis besides $u=0$, where again $E\to \infty$,
  but since it is the other end of the electric lines of force, it is a location of a charge of 
  opposite sign to that existing at $u=0$.
\begin{figure}
\centering
\includegraphics[width=8.5cm]{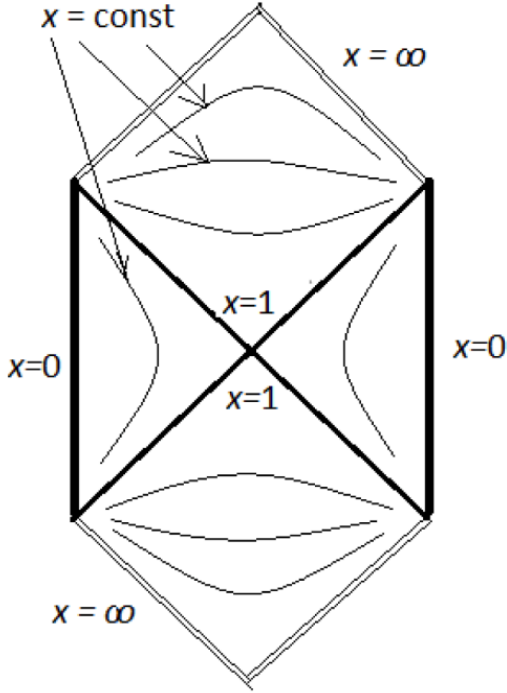}
\caption{\small
	Carter-Penrose diagram for the inverted \bh\ spacetime  The lines $x=0$ are 
	repulsive singularities, the intersecting lines $x=1$ form a Killing horizon, and $x=\infty$ is
	a temporal infinity.} 
\label{fig-invBH}
\end{figure}

  An exceptional case is $a=b=-k$, such that both $e^\mu$ and $e^\beta$ are finite 
  at $u=\infty$, and this cylinder turns out to be a Killing horizon. This is confirmed by a
  transition to the new radial coordinate $x = 1 - e^{-2ku}$, after which the metric \rf{ds-R+}    
  takes the form
\beq 					\label{ds-R1}
	ds^2 = \frac{4(1-x)k^2}{q^2 x^2} dt^2 - \frac{q^2 x^2}{16 k^4}\frac{dx^2}{1-x}
				- \frac{q^2 x^2}{4k^2} \left(dz^2 + d\phi^2\right).
\eeq
  Here we have the axis $x=0$ that coincides with $u=0$; the cylinder $x=1$ (at which $u=\infty$)
  is a horizon, beyond which, at $x>1$, there is a homogeneous anisotropic (Bianchi type I)
  cosmology where $x$ is a temporal coordinate, and $x\to\infty$ is a highly anisotropic
  temporal infinity where two scale factors, $e^\mu$ and $e^\beta$, tend to infinity 
  while the third one, $e^\gamma$, turns to zero. Since, unlike the Schwarzschild and other
  black hole  spacetimes, a static region is here inside the horizon, this solution was named
  ``inverted black hole'' \cite{em4} (Fig.\,\ref{fig-invBH}). 

 \medskip\noi
{\bf 2.} $k=0$, so that by \rf{R6} either $a$ or $b$ is zero (or both). If we choose, for 
 definiteness, $b \ne 0$, then the metric takes the form 
\beq                           \label{ds-R0}
		ds^2 = \frac{dt^2}{q^2 u^2} - q^2 u^2 
				\Big(e^{2bu} du^2 +  dz^2 + e^{2bu} d\phi^2\Big). 
\eeq
  Again $u\in \R_+$, the axis $u=0$ is a repulsive singularity pertaining to a charged thread, 
  while the ``far end'' $u \to \infty$ is either a spatial infinity ($b \geq 0$, $e^\beta \to \infty$) 
  or the second axis  ($b < 0$, $e^\beta \to 0$). The latter is now an attracting singularity 
  where $e^\gamma \to 0$  but $E \sim e^{-\mu-\beta} \to \infty$. 

  The case $a \ne 0,\ b=0$ differs from the previous one where $a=b=0$ by a changing 
  scale along the $z$ axis, and consequently at spatial infinity ($u\to \infty$) the electric field 
  strength $E$ may grow (if $a<0$) or vanish (if $a > 0$). 

\medskip\noi
 {\bf 3.} $k < 0$, so that $ab = -k^2$. The metric is
\beq                           \label{ds-R-}
		ds^2 = \frac{k^2\ dt^2}{q^2\sin^2 ku} - \frac{q^2}{k^2}\sin^2 ku 
			\Big[e^{2(a+b)u}du^2 + e^{2au}dz^2 +e^{2bu} d\phi^2 \Big]. 
\eeq
  Without loss of generality the range of $u$ can be chosen as $0 < u < \pi/|k|$, and its 
  both ends correspond to repulsive (since $e^\gamma \to \infty$) axial (since $e^\beta \to 0$)
  singularities with $E\to \infty$, to be ascribed to axial electric charges of opposite signs. The 
  constants $a$ and $b$ do not affect the qualitative nature of the configuration,
  being only involved in exponentials which are finite and smooth in this range of $u$.

  In all cases with two axes, the 2D subspaces $t = \const$, $z=\const$ can be imagined as 
  warped spheres with singularities at their poles.

\subsection{Geometry with a longitudinal magnetic field}

 In this case, as one easily finds, the SET \rf{SET} takes the form   
\beq                                     \label{SET-L}
		T\mN = \frac {q^2}{8\pi} e^{-2\mu -2\beta} \diag (1,\ -1,\ 1,\ -1),
\eeq
  where $q = q_L$ can be ascribed to an electric current in the azimuthal  ($\phi$) direction,
  like that in a solenoid. Due to high symmetry of the SET \rf{SET-R}, independent components
  of     {Einstein's equations}   in full similarity with \rf{R1}--\rf{R3} reduce to
\bqun                                   \label{L1}
	&&  -\mu'' =  -\gamma'' = \beta'' = - q^2 e^{2\beta},\nb\\
	&& \beta'\gamma'+\beta'\mu' + \gamma'\mu' = q^2 e^{2\beta}, 
\equn
  so excluding the insignificant integration constants by rescaling the $t$ and $z$ axes and 
  choosing the zero point of $u$, we can write the solution as 
\bearr 						\label{sol-A}
		\gamma = -\beta + au, \quad   \mu = -\beta + bu,  \quad
		\e^{2\beta} = \frac{k^2}{q^2 \cosh^2 ku},
\nnn
		a, b, k = \const, \quad  k > 0, \quad    ab = k^2.
\ear
  Thus the metric can be written as
\bqun                                  \label{ds-A}
             ds^2 &=&  \frac{q^2 \cosh^2 ku}{k^2}\Big[e^{2au} dt^2 - e^{2(a+b)u} du^2 
					- e^{2bu} dz^2 \Big]\nb\\
					&&  - \frac{k^2 d\phi^2}{q^2 \cosh^2 ku}. 
\equn
  The range of $u$ is $u \in \R$; moreover, without loss of generality we can assume $a>0$ and
  $b >0$ (making them negative is equivalent to changing $u \to -u$).
  
  The circular radius $e^\beta \to 0$ as $u \to \pm \infty$, that is, we have two axes at two ends
  of the $u$ range. The end $u \to \infty$ is, for any $k, a, b$ infinitely distant since the proper 
  length integral $l = \int \e^{\alpha} du$ diverges. The coefficients $e^\mu$ and $e^\gamma$ also
  diverge there, the latter meaning that the gravitational field is repulsive from this remote axis.
  The magnetic induction B vanishes there.

   As to $u\to -\infty$, the axis there  is  {at finite proper distance from any point with $u < + \infty$,  since}    $\int \e^{\alpha} du$ converges.
   The metric behavior is more diverse there: there is repulsive (if $a < k$) or attractive (if $a>k$)
   singularity according to the behavior of $e^\gamma$, while the scale along $z$ behaves 
   oppositely to $e^\gamma$: it vanishes if $a<k$ and blows up if $a > k$. In addition,
   $B \to \infty$ as $u\to -\infty$ in all cases except $a=b=k$.
   
   But the most interesting  geometry is observed if $a=b=k$: the magnetic field  {there} is  finite, 
   and there is no curvature singularity at $u = \infty$. In this case the substitution 
\beq 				\label{trans-L}
			e^{ku} = \rho = r\frac {|q|}{2k},  \;\;\;   
			 t = \bar t \frac {|q|}{2k^2}, \;\;\;   z = \bar z \frac {|q|}{2k^2}, 
\eeq          
   transforms the metric to
\beq                \label {ds-melvin}
			ds^2 = \frac{q^2}{k^2} (1+\rho^2)^2 \left(d {\bar t}^2 - dr^2 - d{\bar z}^2 \right)
						- \frac {r^2}{\left(1+\rho^2\right)^2} d\phi^2. 
\eeq      
  This metric is regular on the axis $r=0$ if $q^2 = 4k^3$ and coincides with the metric of Melvin's 
  ``geon'' \cite{melvin}. Otherwise there is a conical singularity at $r=0$, and this axis 
   behaves as a cosmic string.  
   
   In the limit $r\to\infty$ ($u\to \infty$), as can be easily verified, all curvature invariants vanish. Thus 
   the metric \rf{ds-melvin} has no curvature singularities. It describes a magnetic (or electric) field 
   configuration supported by its own gravitational field (``a magnetic universe''), with or without
   a cosmic string on its axis.  

\subsection{Geometry with an azimuthal magnetic field}

   For an azimuthal magnetic field, the SET \rf{SET} takes the form   
\beq                                     \label{SET-A}
		T\mN = \frac {q^2}{8\pi} e^{-2\beta -2\gamma} \diag (1,\ -1,\ -1,\ 1),
\eeq
  where $q = q_A$ may be interpreted as a steady electric current along the $z$ axis. 
  The set of equations is the same as for the L-field up to the replacement 
  $\phi \leftrightarrow z$, $\beta \leftrightarrow \mu$, and the solution may be written as
\bqun						\label{sol-A}
		 && \gamma = -\mu + au, \quad  \beta = -\mu + bu,  \quad  
		\e^{2\mu} = \frac{k^2}{q^2 \cosh^2 ku},\nb\\
		&& a, b, k = \const, \quad  k > 0, \quad    ab = k^2,\\
							          \label{ds-A1}
             && ds^2 =  \frac{q^2 \cosh^2 ku}{k^2}\Big[e^{2au} dt^2 - e^{2(a+b)u} du^2 
					- e^{2bu} d\phi^2 \Big] \nb\\
					&& ~~~~ - \frac{k^2 dz^2}{q^2 \cosh^2 ku}. 
\equn
  As before, $u \in \R$; and without loss of generality we assume $a > 0$ and $b > 0$.
  
  Despite the similarity   between \rf{ds-A} and  \rf{ds-A1}, their interpretations are quite different due to
  our assumptions of the topology of $z$ and $\phi$ axes. 
  
   In the metric \rf{ds-A}, ``on the right end'' $u \to +\infty$ we have $e^\gamma \to \infty$ 
   (repulsion), $e^\beta \to \infty$ (spatial infinity in the sense of growing circular radius),
   $l = \int e^\alpha du \to \infty$ (an infinite radial distance), and the magnetic induction 
   $B \to 0$. Meanwhile, $e^\mu \to 0$ (shrinking longitudinal scale) on both ends.  
   
   On the other end, $u\to -\infty$, $l < \infty$, but it is a repulsive singular axis ($e^\beta \to 0$, 
   $e^\gamma \to \infty$) only if $a< k$. If $a>k$, we have, on the contrary, 
   $e^\gamma \to 0$ (attraction) but $e^\beta \to \infty$, and the whole configuration is  
   wormhole-like \cite{wh25} though with a singular end at $u = -\infty$.
   As in the case of L-fields, $B \to \infty$ as $u\to -\infty$ in all cases except $a=b=k$.
      
   If $a=b=k$, then at $u = -\infty$ both  $e^\gamma$ and $e^\beta$ are regular, and to clarify 
   the metric properties it is helpful to substitute $e^{2ku} = y$, after which the metric reads
\bqun
	ds^2 &=& \frac{(1+y)^2}{K} \left(dt^2 -  {\frac{4k^2 dy^2}{ y}} - d\phi^2\right) - \frac{K y}{(1+y)^2}dz^2, \nb\\
	&&
	\;\;   K = \frac{4k^2}{q^2}. 
\equn       
  Now, let us further transform the $x$ coordinate to make the metric
  of  {$(y, z)$} 2-surfaces explicitly conformally flat:
\beq
 	  dl^2_{y,z} =  \frac{K y}{(1+y)^2}(dF^2 + dz^2), \quad    {F = \int \frac{2k(1+y)^2}{Ky} dy}.   
\eeq  
  To regularize it at $y=0$ we can use the notion of conformal mappings on the complex plane
  using the function $W =\ln Z$ for $W(\zeta) = F + {\rm i}z$ with $\zeta = \xi + {\rm i}\eta$.
  More precisely, in terms of $F$ and $z$, we substitute
\bqun
	&& F = \frac {1}{2K} \ln (\xi^2 + \eta^2), \quad  z = \arctan\frac{\eta}{\xi} 
	\ \then \nb\\
	&& dF^2 + dz^2 = \frac {d\xi^2 + d\eta^2}{K(\xi^2 + \eta^2)},  
\equn   
  and since $\xi^2 + \eta^2 \sim y$ as $y\to 0$, the metric is nonsingular at $y=0$.
  However, under this transformation the range of $z$ is $\pi/2 < z < \pi/2$, and the whole 
  $(\xi, \eta)$ plane  $\R^2$ maps in terms of $(y, z)$ to this interval times $y\in \R_+$. 
  To extend the solution to $z \in \R$, we have to consider a Riemannian surface $(\xi, \eta)$
  with in infinite number of sheets and $\xi=\eta=0$ as a branching point. 
  
  Thus in this case we have a self-supported magnetic distribution in a spacetime with 
  a branching-point singularity. 
  
  {From the above analysis it can be seen}  that static cylindrically symmetric Einstein-Maxwell fields are rather diverse, 
  and the most interesting and nontrivial configurations exist in special cases where there is a kind 
  of equilibrium between gravity and electromagnetism: a horizon in an R-field, Melvin's nonsingular
  universe with an L-field, and a system with a branching point singularity in an A-field.   
  
  { To understand the solutions further, sources producing such spacetimes were considered in
  \cite{MSWS01}. It was shown that,  in contrast to the vacuum case,   when the 
  electromagnetic field is present, the situation changes dramatically. In particular, in the case of a purely magnetic
field, all the solutions can be produced by physically acceptable cylindrical thin
shells, while in the case of a purely electric field, no such shells are found for
any choice of the free parameters involved in the solutions.}

%
 \section{LC solutions coupled with  cosmological constant} 
 %
%
\renewcommand{\theequation}{4.\arabic{equation}} \setcounter{equation}{0}

\subsection{Linet-Tian (LT) solutions}
 
The  Linet-Tian (LT) solutions can be cast in the form \cite{Linet86,Tian86,SWPS00},
\bqun
\lb{4.2}
ds^2 &=& Q^{2/3}(r)\Big[P(r)^{-2(4\sigma^2-8\sigma + 1)/(3\Sigma)}dt^2 \nb\\
&& - P(r)^{2(8\sigma^2-4\sigma - 1)/(3\Sigma)}dz^2 \nb\\
&& - C^{-2}P(r)^{-4(2\sigma^2+2\sigma - 1)/(3\Sigma)}d\varphi^2\Big] - dr^2,
\equn
where $\Sigma$ is given by  (\ref{11}),   
\bqu
\lb{4.3}
P(r) \equiv \frac{1}{\beta}\tan\left(\beta r\right), \quad Q(r) \equiv \frac{1}{2\beta}\sin\left(2\beta r\right),
\equ
 $\beta \equiv {\sqrt{3\Lambda}}/{2}$ and $C$ is an arbitrary real integration constant. Without loss of the generality, we assumed $C > 0$. It is easy to show that the above solutions reduce to the LC solutions given by  (\ref{12}), when 
    {$\Lambda = 0$}, so in the following we do not consider this case any more.   When $\Lambda < 0$, the trigonometric functions become hyperbolic, but the  {resulting} solutions are still  well-defined and real.
 All the solutions with $\Lambda \not= 0$ are Petrov type I according to the classifications given in \cite{StephaniA}, except for
 the particular cases, $\sigma = -1/2, \; 0, \; 1/4,\; 1/2,\; 1$,  which are all Petrov type D, as shown explicitly in \cite{SWPS00}.

  In addition, it was also found that the solutions with $\sigma = 1/4 + \tau$ can be obtained from the ones with $\sigma = 1/4 - \tau$, together with the replacement \cite{SWPS00}, 
 \bqu
\lb{4.17}
 t = i C^{-1}\varphi', \quad  \varphi = i C t'.  
\equ
Yet, the solutions also remain the same if we replace $\sigma$ with $1/(4\sigma)$ and exchange the $z$- and $\varphi$-coordinates \cite{ZB08}, similarly to the vacuum case.

 When $\sigma = 0$,  (\ref{4.2}) reduces to 
 \bqu
\lb{4.4}
\left. ds^2\right|_{\sigma = 0} = \cos^{4/3}\left(\beta r\right) \left(dt^2 - dz^2\right) - \frac{\sin^2\left(\beta r\right)}{\beta^2 C^2 \cos^{2/3}\left(\beta \; r\right)} d\varphi^2 - dr^2.
\equ
Clearly, in addition to the usual three Killing vectors $\xi_{(t)} = \partial_t, \; \xi_{(z)} = \partial_z, \; \xi_{(\varphi)} = \partial_{\varphi}$,  the solutions now admit  one more Killing vector $\xi_{(0)} = t\partial_z + z \partial_t$, which 
corresponds to a Lorentz boost in the ($t, z$)-plane.  As $r \rightarrow 0$, 
 we find that
\bqu
\lb{4.5}
\left. ds^2\right|_{\sigma = 0} \simeq dt^2 - dz^2 - \frac{r^2}{C^2}d\varphi^2 - dr^2, 
\equ
which represents the external spacetime of a cosmic string located on the symmetry axis $r = 0$ with the angular defect $\Delta\varphi \equiv 4\mu = 1 - C^{-1}$, 
where $\mu$ denotes the mass per unit length of the string. Note that the metric (\ref{4.4}) also becomes singular at $r = r_g \equiv \pi/(2\beta)$ for $\Lambda > 0$. As shown below, this singularity is a curvature one, which makes
the interpretation of the solutions as representing a cosmic string spacetime coupled with a positive cosmological constant unclear. However, in the case $\Lambda < 0$, the trigonometric functions become hyperbolic ones, and
  this singularity does not exist any more. So, in the latter the corresponding spacetime  {indeed} can be interpreted as representing a cosmic string embedded in an asymptotically anti-de Sitter spacetime.

To study the LT solutions further,    { let } us consider the cases $\Lambda > 0$ and $\Lambda < 0$, separately. 

\subsection{Main properties of  LT solutions with $\Lambda > 0$}

As  $r \rightarrow 0$, we find that $P(r) \rightarrow r$ and $Q(r)  \rightarrow r$, so that  (\ref{4.2}) approaches to the vacuum LC solutions. Then, the singularity behavior of the LT solutions at $r = 0$ are similar to that of the LC solutions. 
In particular, all the solutions  have a curvature singularity, except for the cases $\sigma = 0$ and $\sigma = 1/2$ \cite{WSS97}.  

When $\sigma = 1/2$, similarly to the case $\sigma = 0$, the corresponding solution  also possesses an additional Killing vector, $\xi_{(1/2)} = C^{-1}\varphi \partial_z - C z \partial_\varphi$. This  corresponds to a rotation in the ($z, \varphi$)-plane.   
Since in the latter case the  curvatures of the two spacelike surfaces $ t , r = $ Constant are identically zero, it is difficult to consider this spacetime as having cylindrical symmetry. Instead, one may extend the $\varphi$ coordinate from the range [$0, 2\pi$] 
to the range ($-\infty, +\infty$), so the resulted spacetime has a plane symmetry, as in the vacuum case \cite{SWPS00,WSSW03}. 

However, the metric (\ref{4.2}) shows that the solutions usually are also singular on the hypersurface $r_g \equiv \pi/(2\beta)$. To see the nature of the singularities, let us first note that  $Q(r) \rightarrow r - r_g$ and $P(r) \rightarrow 
(r - r_g)^{-1}$, as $r \rightarrow r_g$. Then, we find that the metric (\ref{4.2}) takes the asymptotic form,
\bqun
\lb{4.6}
ds^2 &\simeq& R^{4(4\sigma^2 -5\sigma +1)/(3\Sigma)}dt^2 - R^{-4(2\sigma^2 -\sigma -1)/(3\Sigma)}dz^2\nb\\
&&  - C^{-2}R^{2(8\sigma^2 + 2\sigma -1)/(3\Sigma)}d\varphi^2 - dr^2,
\equn
as $r \rightarrow r_g$, where $R \equiv r -r_g$. Then, the corresponding Kretschmann scalar is given by,
\bqu
\lb{4.7}
{\cal{R}} \equiv R^{\alpha\beta\gamma\delta} R_{\alpha\beta\gamma\delta}  \simeq \frac{64(\sigma-1)^2 (2\sigma+1)^2 (4\sigma -1)^2}{27 \Sigma^3 R^4}, \; (r \simeq r_g). 
\equ
Therefore, the spacetime is always singular at $r = r_g$, except for the cases $\sigma = -1/2, \; 1/4, \; 1$. It can be shown that all 14 independent scalars, built from
the Riemann tensor in 4-dimensional Riemann spacetimes  \cite{CW77}, have the same properties. 
Therefore, in the cases  $\sigma = -1/2, \; 1/4, \; 1$ the singularities on the hypersurface
$r = r_g$, are  coordinate ones, and to have geodesically complete spacetimes, the solutions need to
be extended beyond this surface.  Note that, in addition to the usually three Killing
vectors, they also have a fourth Killing vector, given 
by $\xi_{(-1/2)} = C^{-1}\varphi \partial_z - Cz \partial_{\varphi}$,  $\xi_{(1/4)} =  C^{-1}\varphi \partial_t + Ct \partial_{\varphi}$, and $\xi_{(1)} = z \partial_t + t \partial_z$,  respectively. 
Using the same arguments as those given for the solution with $\sigma = 1/2$,  the solution with $\sigma = - 1/2$ can be also considered as representing plane symmetry, and the coordinate $\varphi$
can be extended to the whole axis $- \infty < \varphi < +\infty$.

Therefore,   except for the particular cases, $\sigma = -1/2$, 
$0$, $1/4$, $1/2$, $1$, the solutions are singular at both $r = 0$ and $r = r_g$. The physics of these singularities is unclear, and some of them (if not all) can be considered as representing matter sources \cite{Bonnor92,BGM94}. 

When $\Lambda > 0$, in addition to (\ref{4.17}), extra  symmetry exists \cite{GP10},
\bqun
\lb{4.18}
r &=& \frac{\pi}{\sqrt{3\Lambda}} - r', \; t = \left(\frac{4}{3\Lambda}\right)^{\frac{4\sigma^2 - 8\sigma +1}{3\Sigma}} t', \nb\\
\varphi &=& \left(\frac{4}{3\Lambda}\right)^{\frac{2\sigma^2 + 2\sigma -1}{3\Sigma}} z',\;
z = \left(\frac{4}{3\Lambda}\right)^{-\frac{8\sigma^2 -4\sigma -1}{3\Sigma}} \varphi', 
\equn
where $\sigma \equiv {(1-4\sigma')}/{[4(1- \sigma')]}$. Because of this symmetry, it was argued  that $z$ and $\varphi$ should have the same character, and in particular  should be all periodic \cite{GP10}. 

Due to the particular symmetry and singular behavior of the cases $\sigma = \pm 1/2$, in the following let us consider them separately. For more details, interested readers are referred to
\cite{SWPS00,WSSW03}.   

\subsubsection{Global structure of the solution with $\sigma =  \frac{1}{2}$}

When $\sigma = 1/2$, making the coordinate transformations \cite{SWPS00},
\bqu
\lb{4.8}
T = \frac{2}{3} t, \; X = \cos^{2/3}(\beta r), \; Y = \frac{2\beta}{3C}\varphi, \;Z = \frac{2\beta}{3} z, 
\equ
the corresponding metric takes the form,
\bqu
\lb{4.9}
\left. ds^2\right|_{\sigma = \frac{1}{2}} = \frac{9}{4\beta^2}\Bigg[f(X)dT^2 - \frac{dX^2}{f(X)} - X^2\Big(dY^2 + dZ^2\Big)\Bigg],   
\equ
where 
\bqu
\lb{4.10}
f(x) \equiv \frac{1}{X} - X^2.
\equ
From  (\ref{4.8}) we can see that the region $0\le r\le r_g$ is mapped to the region $0\le X\le 1$, and the point $r = r_g$  is mapped  to the point $X = 0$, which is singular. Extending
$X$ to the range $X \in (-\infty, +\infty)$,  we find that in the extended spacetime two new regions $X > 1$   and $X < 0$   are obtained.
However, the curvature singularity at $X = 0$   divides the whole spacetime into two unconnected regions $X \ge 0$ and $X \le 0$.   

In the region $X\le 0$, the function $f(X)$  is always negative and
the $X$ coordinate is timelike, while $T$ is spacelike. Then, the spacetime is essentially
time-dependent and the singularity at $X = 0$ is spacelike.  As $X \rightarrow - \infty$,  the metric becomes asymptotically de Sitter, 
\bqu
\lb{4.11}
\left. ds^2\right|_{\sigma = \frac{1}{2}} \simeq d\tilde{T}^2 - e^{4\beta \tilde{T}/3}\Big(dX^2 + dY^2 + dZ^2\Big),      \; (X \rightarrow - \infty),
\equ
where $T = e^{2\beta \tilde{T}/3}$, and $X, \; Y, \; Z$ had been rescaled.  The corresponding Carter-Penrose diagram is  
given by Fig.\ref{Fig4.1a}. 

\begin{figure}
\centering
\includegraphics[width=8.cm]{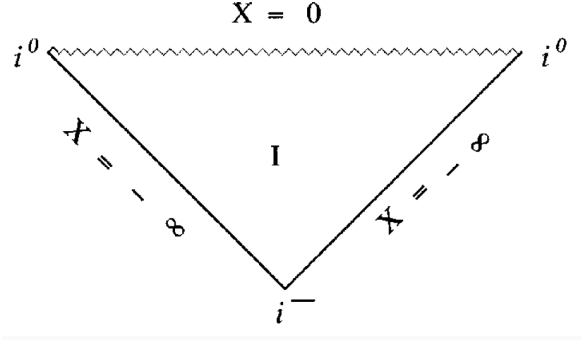}
\caption{
	Carter-Penrose diagram for the LT solutions with $\sigma = \frac{1}{2}$ and $\Lambda > 0$ in the region $X \le 0$, given by  (\ref{4.10}) and (\ref{4.11}). 
	The horizontal   line ($X = 0$) represent a spacetime singularity \cite{SWPS00}.   } 
\label{Fig4.1a}
\end{figure}

In the region  $X\ge 0$,  the function $f(X)$ is greater than zero for $0 \le X <1$ and
less than zero for $X >1$; that is, $X$ is spacelike in the region $0 \le X <1$ and timelike in the region $X >1$. 
On the hypersurface $X = 1$,   it becomes null, which represents a horizon. Since the spacetime
singularity at $X = 0$ now is timelike, the horizon is actually a Cauchy horizon \cite{HE73}. 
As $X \rightarrow + \infty$,  the spacetime is also asymptotically de Sitter and approaches the same form as
that given by  (\ref{4.11}). The corresponding Carter-Penrose diagram is given by Fig.\ref{Fig4.1}. 

\begin{figure}
\centering
\includegraphics[width=6.cm]{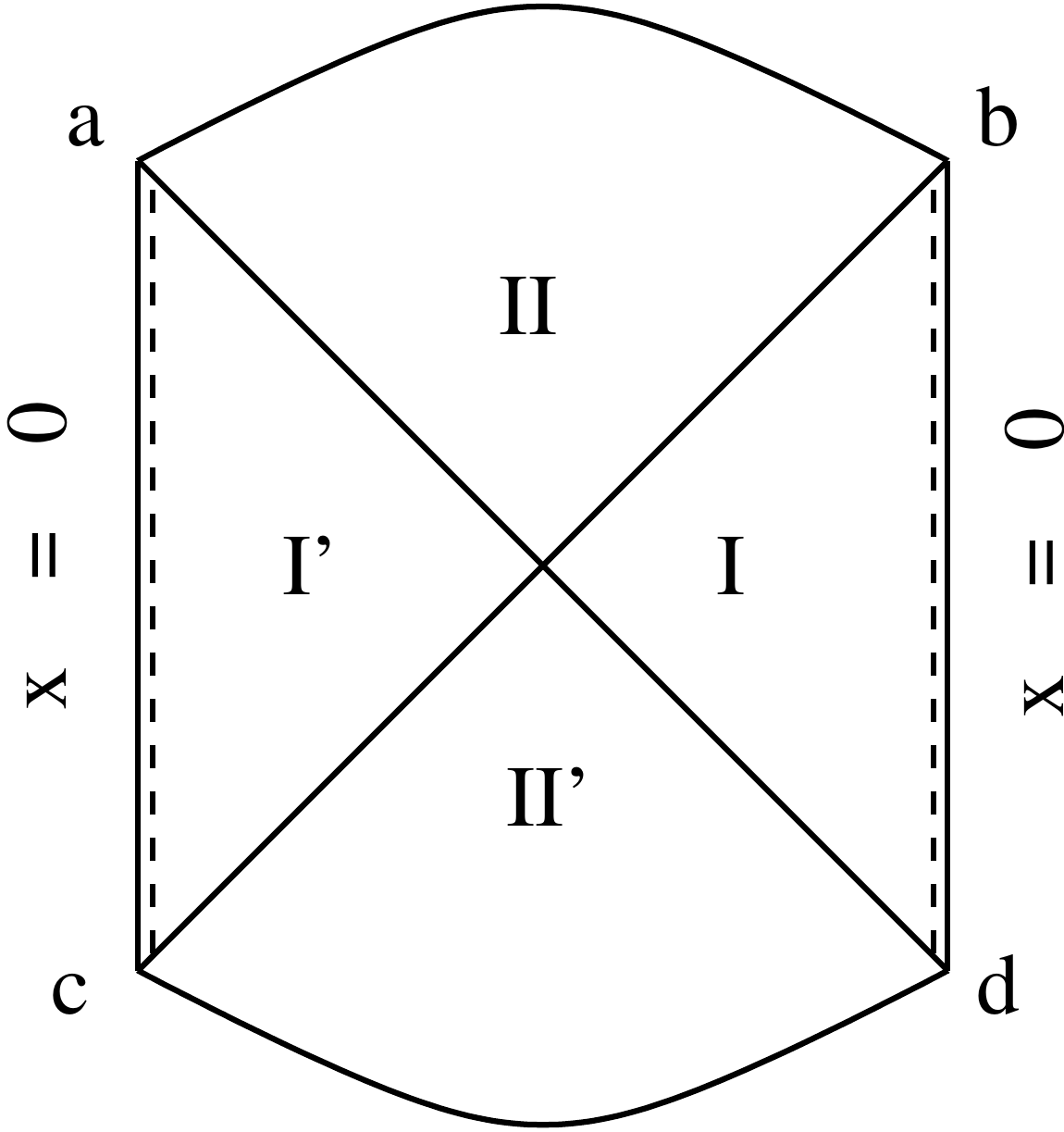}
\caption{
	Carter-Penrose diagram for the LT solutions with $\sigma = \frac{1}{2}$ and $\Lambda > 0$ in the region $X \ge 0$, given by  (\ref{4.10}) and (\ref{4.11}). 
	The vertical  lines ($X = 0$) represent spacetime singularities, while
	the ones $ad$ and $bc$ ($X = 1$) represent Cauchy horizons. As $X \rightarrow \infty$ (represented by the curves $ab$ and $cd$), the spacetime is asymptotically de Sitter  \cite{WSSW03}. } 
\label{Fig4.1}
\end{figure}

\subsubsection{Global structure of the solution with  $\sigma = - \frac{1}{2}$}

When $\sigma = -1/2$,  the spacetime is singular at $r = 0$, and is free of curvature singularity at $r = r_g$. Thus, to have a geodesically complete spacetime, we
need to extend the solution beyond the hypersurface  $r = r_g$. To make such an extension, we can introduce a new coordinate
$X$ by $X = \sin^{2/3}(\beta r)$,  and then rescale the coordinates $t, \; z$, and $\varphi$, we will find that the corresponding metric takes
the same form as that given by  (\ref{4.11}).  Thus, the solution with $\sigma = 1/2$ is actually the same as that with $\sigma = -1/2$. As first noticed in  \cite{SWPS00}, this is not expected!
As a matter of fact,   in the limit $\Lambda \rightarrow 0$,   the solution with $\sigma = 1/2$   approaches the Rindler space \cite{Rindler}, which represents a uniform
gravitational field and is free of any kind of spacetime curvature singularities, while the one with $\sigma = -1/2$ is
the static Taub solution with plane symmetry \cite{Taub51},  and is singular on the hypersurface $X = 0$. The total mass of the Taub
spacetime is negative, while the one of Rindler is not \cite{Silva1}.   However, the presence of the cosmological constant makes
up these differences and turns the two spacetimes to be identical!  We would also like to call the attention to the comments on conformally flat sources given at the end of section 2.5.

It is also interesting to note that none of these two solutions represents black holes. In fact, there is a general theorem for spacetimes with a positive cosmological constant and 
two spacelike Killing vectors  \cite{Wang05}, which shows clearly that in such cases no black holes
exist, as long as matter fields satisfy the  dominant energy condition \cite{HE73}.

 \subsection{Main properties of  LT solutions with $\Lambda < 0$}

Near $r = 0$, the solutions with $\Lambda < 0$ have the same asymptotic properties as those with $\Lambda > 0$, and all approach to the LC vacuum solutions. 
Therefore, all the LT solutions with $\Lambda < 0$ are singular at $r = 0$, except for the two particular cases, $\sigma = 0, \; 1/2$. On the other hand, from  (\ref{4.3}) we can see that both $P(r)$ and
$Q(r)$ are monotonically increasing functions of $r$ now, so the LT solutions with $\Lambda < 0$ are well-defined in the whole range $r \in (0, +\infty)$. In particular, as 
$r \rightarrow \infty$, we find that 
\bqu
\lb{4.12}
\left. ds^2\right|_{r \rightarrow \infty} \simeq C^2_0e^{4\beta r/3}\Bigg[dt^2 - dz^2 - \frac{d\varphi^2}{C^2}\Bigg] - dr^2,      
\equ
where $C_0$ is a real constant. As first noted in \cite{SWPS00}, this is  the anti-de Sitter spacetime, but written in the horospherical coordinates  \cite{CGS93}.
Therefore, all the LT solutions with  a negative cosmological constant are asymptotically anti-de Sitter.  

\subsubsection{Global structure of the solution with $\sigma = 0$}

   {When} $\sigma = 0$, the solutions are given  by  (\ref{4.4}), i.e.,
 \bqun
\lb{4.13}
\left. ds^2\right|_{\sigma = 0} &=& \cosh^{4/3}\left(|\beta| r\right) \left(dt^2 - dz^2\right) \nb\\
&& - \frac{\sinh^2\left(|\beta| r\right)}{|\beta|^2 C^2 \cosh^{2/3}\left(|\beta| \; r\right)} d\varphi^2  - dr^2,
\equn
which can be considered as representing a cosmic string located on the symmetry axis $r = 0$ embedded in the anti-de Sitter spacetime with its mass per unit length given by 
$\mu = \left(1 - C^{-1}\right)/4$. Clearly, when $C = 1$, the cosmic string disappears, and a regular axis is obtained. Then, the whole spacetime is free of any kind of singularities, 
including a conic one, and has a well-defined symmetry axis at $r = 0$. In fact,  the spacetime is locally flat at $r = 0$. So, we have an anti-de Sitter spacetime with cylindrical symmetry,
and this solution is different from the anti-de Sitter spacetime,
 \bqun
\lb{4.14}
ds^2_{AdS} &=& \left(1 + \frac{|\Lambda| r^2}{3}\right) dt^2  - \left(1 + \frac{|\Lambda| r^2}{3}\right)^{-1} dr^2\nb\\
&& - r^2\left(d\theta^2 + \sin\phi^2 d\phi^2\right),
\equn
as first noted by Bonnor in \cite{Bonnor08}, and dubbed  {\em the non-uniform (NoU) AdS universe}. The difference between these metrics can be seen clearly by calculating their  Kretschmann scalars,
 \bqun
\lb{4.15}
K &\equiv& R^{\alpha\beta\gamma\delta} R_{\alpha\beta\gamma\delta}  \nb\\
&= &
\begin{cases}
\frac{4\Lambda^2}{3}\Big[3 - 2\tanh^2(|\beta| r) + \tanh^4(|\beta| r)\Big], & {\mbox{NoU-AdS}},\cr
\frac{8\Lambda^2}{3},  & {\mbox{AdS}}. \cr
\end{cases}\nb\\
\equn
In fact,   the AdS spacetime given by  (\ref{4.14}) is conformally flat $\left(C^{{\mbox{AdS}}}_{\mu\nu\alpha\beta} = 0\right)$, while the one given by  (\ref{4.12}) is not $\left(C^{\mbox{Non-U-AdS}}_{\mu\nu\alpha\beta} \not= 0\right)$,
where $C_{\mu\nu\alpha\beta}$ denotes the Weyl tensor.

\subsubsection{Global structure of the solution with $\sigma = \frac{1}{2}$}

When $\sigma = 1/2$, the corresponding solution also possesses an additional Killing vector, $\xi_{(1/2)} = C^{-1}\varphi \partial_z - Cz \partial_{\varphi}$, which represents a rotation in the ($z, \varphi$)-plane, and implies that this plane is
flat. Thus, one can also consider that the solution has a plane symmetry by simply extending $\varphi$ to the range $\varphi \in (- \infty, \; +\infty)$. Then, we should also need to extend $r$ to the full range $r \in (- \infty, \; +\infty)$. However, the metric
has a coordinate singularity at $r = 0$, and extension beyond it is needed, in order to obtain a geodesically complete spacetime. The extension is quite similar to the case $\sigma = 1/2,\; \Lambda > 0$. In fact, setting
\bqu
\lb{4.8a}
T = \frac{2}{3} t, \; X = \cosh^{2/3}(|\beta| r), \; Y = \frac{2|\beta|}{3C}\varphi, \; Z = \frac{2|\beta|}{3} z, 
\equ
the corresponding metric takes the form,
\bqu
\lb{4.9a}
\left. ds^2\right|_{\sigma = \frac{1}{2}} = \frac{9}{4\beta^2}\Bigg[- f(X)dT^2 + \frac{dX^2}{f(X)} - X^2\Big(dY^2 + dZ^2\Big)\Bigg],    \equ
where $f(X)$ is given by  (\ref{4.10}). Comparing it with  (\ref{4.9}) we can see that we can obtain one from the other by simply replacing $f(X)$ with $-f(X)$. From the expression of $X$ we
can see that the region $0 \le r < \infty$  is mapped to the region $1\le X < \infty$,  and  region $X < 1$ is an extended region. After
the extension, a spacetime curvature singularity appears at $X = 0$, which divides the whole $X$-axis into two parts $X\le 0$    {and}
 $X \ge 0$. It can be shown that, unlike the case $\Lambda > 0$, now the spacetime is static in the region $X\le 0$ and the curvature
singularity at $X = 0$ is timelike and naked. As $X \rightarrow -\infty$,  the
spacetime is asymptotically anti-de    {Sitter},
\bqu
\lb{4.16}
\left. ds^2\right|_{\sigma = \frac{1}{2}} \simeq \frac{9}{4|\beta|^2 \tilde{X}^2}\Big(dT^2 - d\tilde{X}^2   - dY^2 - dZ^2\Big),    
\equ
where $\tilde{X} = X^{-1}$.  The corresponding Carter-Penrose diagram is given by Fig.\ref{Fig4.2}.   

\begin{figure}
\centering
\includegraphics[width=6.cm]{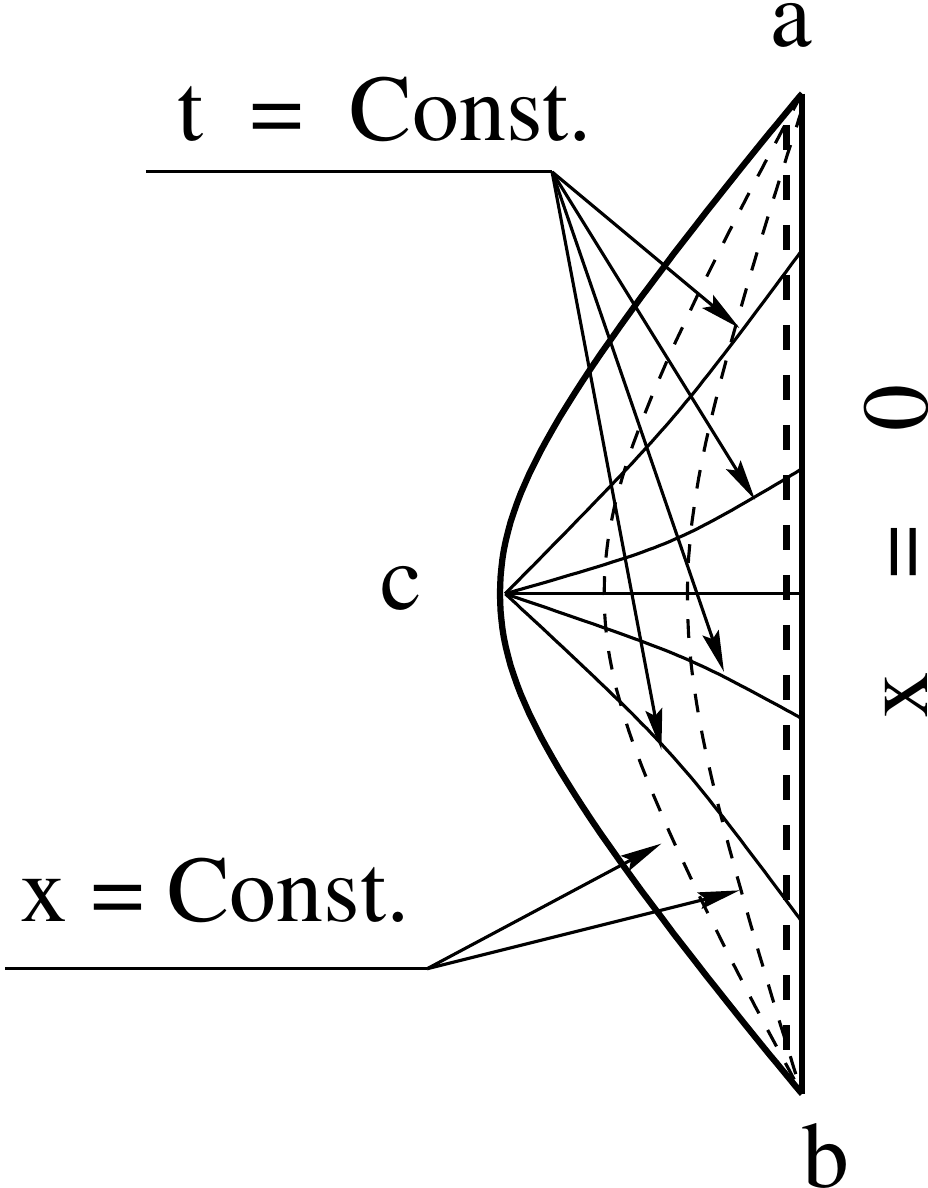}
\caption{\small
	Carter-Penrose diagram for the LT solutions with $\sigma = \frac{1}{2}$ and $\Lambda < 0$ in the region $X \le 0$. The vertical double  line  ($X = 0$) represents a spacetime singularity, which is timelike. 
	The curve $bca$ corresponds to $X = -\infty$, which is also timelike   \cite{WSSW03}. } 
\label{Fig4.2}
\end{figure}

In the region $X\ge 0$, the spacetime singularity at $X = 0$ becomes
spacelike. Except for this curvature singularity, there
is a coordinate one located at $X = 1$. This coordinate singularity
actually represents an event horizon. As shown in the
last section, the spacetime is asymptotically anti-de Sitter
 as $X \rightarrow + \infty$. The corresponding Carter-Penrose diagram is given by
Fig.\ref{Fig4.3}.  This is a black hole solution with plane symmetry
 and vanishing electromagnetic charge first found in \cite{CZ96}, and then  studied in detail in \cite{WSSW03}. It should be noted that this does not contradict the theorem given in \cite{Wang05},
as now we have $\Lambda < 0$.

\begin{figure}
\centering
\includegraphics[width=6.cm]{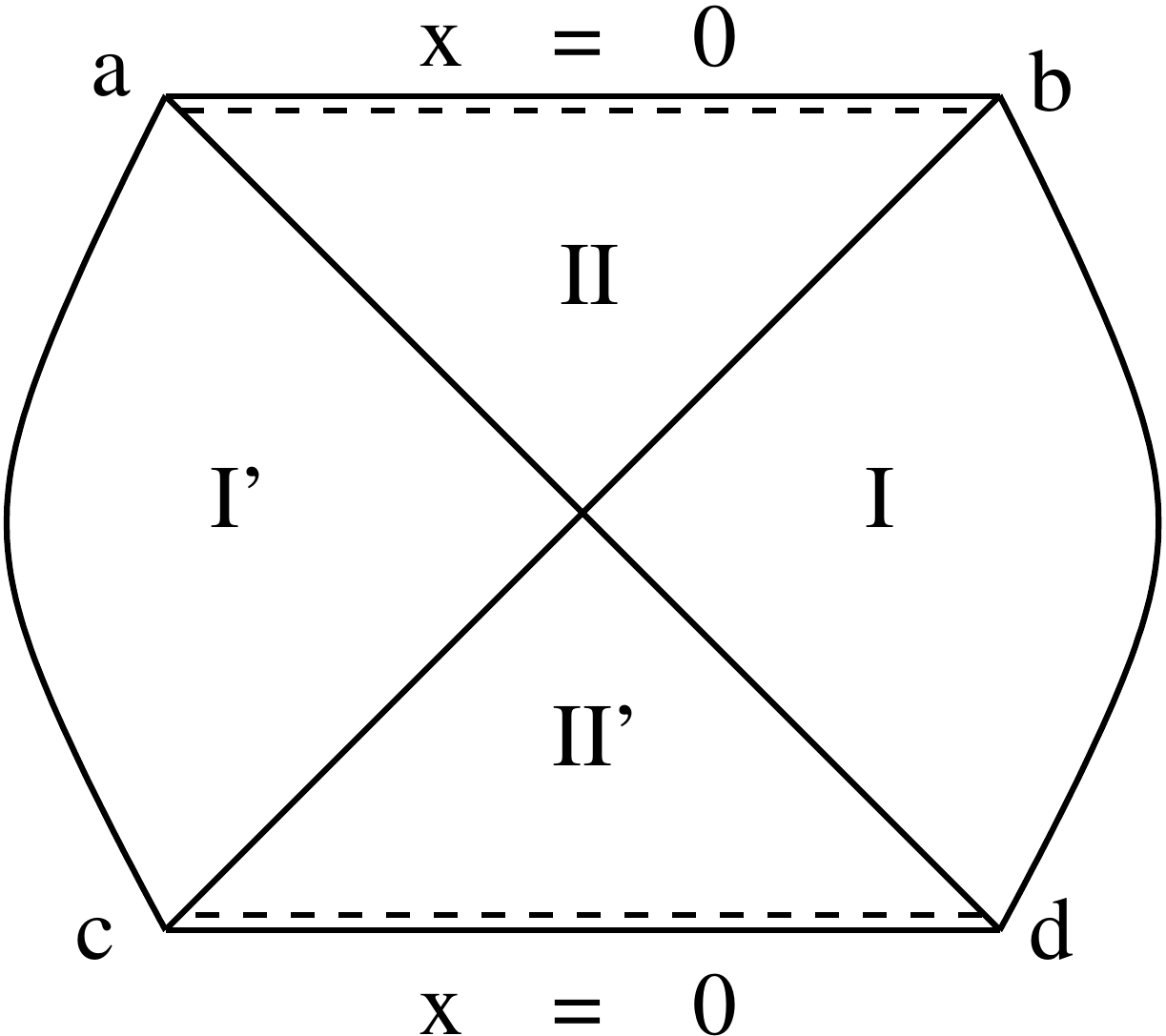}
\caption{\small
	Carter-Penrose diagram for the LT solutions with $\sigma = \frac{1}{2}$ and $\Lambda < 0$ in the region $X \ge 0$. The horizontal  lines ($X = 0$) represent spacetime singularities, while
	the ones  $X = 1$, represented by the straight lines $ad$ and $bc$,  correspond to event horizons. As $X \rightarrow \infty$ (represented by the curves $ac$ and $bd$), the spacetime is asymptotically anti-de Sitter  \cite{WSSW03}. } 
\label{Fig4.3}
\end{figure}

 \subsubsection{Global structure of the solution with $\sigma = - \frac{1}{2}$}
 
 Note that in this case an additional Killing vector $\xi_{(-1/2)} = C^{-1}\varphi \partial_z - Cz \partial_{\varphi}$ also exists, and one may consider that the spacetime also has a plane symmetry. Hence,  the range of $r$ should be
taken as $r \in (-\infty, \; +\infty)$.   Then one may ask, what is the physical interpretation of the spacetime in the region $r \le 0$? To answer this question, let us first introduce a new coordinate $X = - 
\sinh^{2/3}(|\beta| r)$,  and then  rescale the other three,  we  find that the metric takes the same form as that given by  (\ref{4.9a}). 
From the expression for $X$ we can see that the region $0 \le r < +\infty$  now is mapped to the region $ - \infty < X \le 0$, while  
 the region $- \infty < r \le 0$, is mapped   {to}  the region $0 \le X < +\infty$.   In the region $0 \le r < \infty$, the solution represents a
static spacetime with a naked singularity located at $X = 0$. The spacetime is asymptotically anti-de Sitter, and the corresponding
Carter-Penrose diagram is given by Fig.\ref{Fig4.2}. In the region $- \infty < r \le 0$,   the solution represents a black hole with
 plane symmetry, and the corresponding carter-Penrose diagram is given by Fig.\ref{Fig4.3}. 
Thus, similarly to the case with $\Lambda > 0$, now the solution with $\sigma = 1/2$ and the one  with $\sigma = -1/2$ actually describes the same spacetime.

\subsection{Geodesics of LT solutions}    
 
 Geodesics describe the orbits of light rays and material particles, and an important physical aspect is the influence of the cosmological constant on the stability of the geodesics' orbits in an otherwise vacuum spacetime with $\Lambda=0$. In this context spherical spacetime with $\Lambda$ has been studied in \cite{Lebedev}. Here we summarize, in an unified way, the results about the impact of $\Lambda$ on the dynamics of the geodesics in the LT spacetime. We recall the LT metric   (\ref{4.2}) with $P$ and $Q$  given by
 \begin{equation}
 P=\frac{2}{\sqrt {3|\Lambda|}}T(R), \; Q=\frac{1}{\sqrt {3|\Lambda|}}S(2R), \; R=\frac{\sqrt {3|\Lambda|}}{2}r, \label{x1a}
 \end{equation}
 where $T$ and $S$ are the tangent and sine functions if $\Lambda>0$, or the corresponding hyperbolic functions if $\Lambda<0$.

 We follow closely the results obtained for $\Lambda>0$ in \cite{Brito1} and $\Lambda<0$ in \cite{Brito}. We recall too that the range of the radial coordinate $r$ is   $r\in(0,\pi/\sqrt{3\Lambda})$ and there are two singularities for $\Lambda>0$,
 and  the range of the radial coordinate $r$ is   $r\in(0,\infty)$ and there is only one singularity for $\Lambda<0$. If $\Lambda=0$, one simply gets $P=Q=r$.

 The geodesics for the LT metric are given by
 \begin{eqnarray}
 {\dot t}&=&EQ^{-2/3}P^{2(1-8\sigma+4\sigma^2)/(3\Sigma)}, \label{x1}\\
 {\dot r}^2&=&Q^{-2/3}P^{2(1-8\sigma+4\sigma^2)/(3\Sigma)}[E^2-V(r)], \label{x2}\\
 {\dot z}&=&P_zQ^{-2/3}P^{2(1+4\sigma-8\sigma^2)/(3\Sigma)}, \label{x3}\\
 {\dot\phi}&=&C^2L_zQ^{-2/3}P^{-4(1-2\sigma-2\sigma^2)/(3\Sigma)}, \label{x4}
\end{eqnarray}
 where the dot stands for differentiation with respect to an affine parameter $\lambda$; the constants $E$, $P_z$ and $L_z$ represent, respectively, the total energy of a test particle,
 its momentum along the $z$ axis and the angular momentum about the $z$ axis; and the effective potential $V(r)$ is given by
 \bqun
V(r)&=& \epsilon Q^{2/3}P^{-2(1-8\sigma+4\sigma^2)/(3\Sigma)}+P_z^2P^{8\sigma(1-\sigma)/\Sigma}\nb\\
&& +C^2L_z^2P^{-2(1-4\sigma)/\Sigma}, \label{x5}
\equn
where $\epsilon =0$, $1$ or $-1$ if the geodesics are, respectively, null, timelike or spacelike.

We note that in (\ref{x2}) $E^2=V(r)$ gives ${\dot r}^2=0$ and corresponds to circular helices along cylindrical surfaces, or to planar circles if $P_z=0$. We also note that unbounded
geodesics for $\Lambda=0$ remain unbounded after the introduction of any $\Lambda\neq 0$ so our aim here will focus on stability issues for bounded geodesics, where we also consider
$L_z\neq 0$.

\subsubsection{Planar geodesics ${\dot z}=0$}

In this case, let us consider   the null and timelike geodesics,  separately. 

\vspace{.3cm}

{\bf $\bullet$ Null geodesics $\epsilon=0$:} For the null geodesics, when  $\sigma\leq 1/4$,  if $E^2>V_{\infty}=(CL_z)^2P^{-2(1-4\sigma)/\Sigma}$, then outgoing particles escape to infinity for $\Lambda<0$, while they reach the outer singularity for $\Lambda>0$. Incoming geodesics hit the axis for $\sigma=1/4$, but for $\sigma<1/4$, they always bounce at a minimum distance $r_{min}$ from the axis, where
$(\sqrt{3|\Lambda|}/2)r_{LCmin}=(\sqrt{3\vert\Lambda\vert}r_{min}/2)T$ and $r_{LCmin}=(CL_z/E)^{\Sigma/(1-4\sigma)}$.

When $\sigma>1/4$, using $V(r\to 0)\to 0$ and $V(r\to\pi/{\sqrt{3\Lambda}})\to +\infty$, for $\Lambda>0$, and $V(r\to\infty)\to V_{\infty}=(2/{\sqrt {3|\Lambda|}})^{-2(1-4\sigma)/\Sigma}(CL_z)^2$ and
$V(r\to 0)\to 0$, for $\Lambda<0$ (and $E^2\leq V_{\infty}$), one concludes that incoming particles approach the axis with infinite speed. In turn, outgoing particles move with decreasing speed and negative acceleration, attaining a maximum distance $r_{max}$ from the axis, where
$({\sqrt {3|\Lambda|}}/2)r_{LCmax}=({\sqrt{3|\Lambda|}}r_{max}/2)T$ with $r_{LCmax}=(E/(CL_z)^{\Sigma/(4\sigma-1)}$ being the value of $r_{LCmax}$ for $\Lambda=0$. This shows, in a particularly simple way, that $\Lambda>0$ decreases the maximum distance to the axis reached by the null particle, while $\Lambda<0$ increases it. In   {the} latter case, now with
$E^2>V_{\infty}$, outgoing null particles escape to infinity.


{\bf Proposition 1.} {\it Consider null geodesics with $\Lambda=0$, $P_z=0$ and $L_z\neq 0$. Then, bounded geodesics are: (i) unstable, if $E^2>V_{\infty}$, with the introduction of $\Lambda<0$; (ii) stable, if $E^2\leq V_{\infty}$, with the introduction of $\Lambda<0$; (iii) stable, with the introduction of $\Lambda>0$.}

\vspace{.3cm}

{\bf $\bullet$ Timelike geodesics $\epsilon=1$:} In this case, when $\sigma<1/4$, for $\Lambda=0$, since $V(r)=\epsilon Q^{2/3}P^{-2(1-8\sigma+4\sigma^2)/(3\Sigma)}+(CL_z)^2P^{-2(1-4\sigma)/\Sigma}$ always has a minimum, the geodesics is always confined between two nonzero radii and this is also the case for $\Lambda<0$. On the contrary, if $\Lambda>0$, then $dV/dr=0$ does not always have a solution. In fact, for sufficiently large $\Lambda$,
bounded geodesics become unbounded and converge to the $r=\pi/{\sqrt{3\Lambda}}$ singularity.

When $\sigma\geq 1/4$, by similar methods, we prove that outgoing geodesics reach a maximum radial value before bouncing back inwards to the axial singularity and incoming particles hit the axis.


{\bf Proposition 2.} {\it Consider timelike geodesics with $\Lambda=0$ and $P_z=0$. Then, bounded geodesics are: (i) stable, for a sufficiently small $\Lambda>0$; (ii) unstable, for sufficiently large $\Lambda>0$; (iii) stable, against any value of $\Lambda<0$.}

\subsubsection{Non-planar geodesics ${\dot z}\neq 0$}

\vspace{.3cm}

{\bf $\bullet$ Null geodesics $\epsilon=0$:} In this case, when $\sigma<1/4$, for $\Lambda=0$, the geodesics are always confined between two nonzero radii and the same happens for $\Lambda>0$, since $V(r)$ has a minimum and the equation $E^2=V(r)$ has two real roots. For $\Lambda<0$, if $E^2\geq V_{\infty}$, where
$V_{\infty}=P^2_z(2/{\sqrt{3|\Lambda|}})^{8\sigma(1-\sigma)/\Sigma}+(CL_z)^2(2/{\sqrt{3|\Lambda|}})^{-2(1-4\sigma)/\Sigma}$, a null particle approaches the axis with decreasing negative acceleration and increasing speed, until it arrives at its minimum distance from the axis at $E^2=V(r)$, where it has vanishing speed. From there on, the null particle is reflected escaping to infinity. If $E^2<V_{\infty}$, the equation $E^2=V(r)$ has two roots and the geodesics are confined between two nonzero radii.

When $\sigma\geq 1/4$, for $\Lambda=0$, the geodesics are always confined between two nonzero radii. If $\Lambda<0$ and $E^2\geq V_{\infty}$, the null geodesics become unbounded. Otherwise, the geodesics reach a maximum.

\vspace{.3cm}

{\bf $\bullet$ Timelike geodesics $\epsilon=1$:} In this case, there is always geodesic confinement in the radial direction for $\Lambda\leq 0$. If $\Lambda>0$, the geodesics are only confined for $\sigma<1/4$, in which case
$V(r\to 0)\to +\infty$ and
$V(r\to\pi/{\sqrt{3\Lambda}})\to +\infty $. If $\sigma\geq 1/4$, then $V(r\to 0)$ is finite, therefore, incoming particles hit the axis and outgoing particles reach a maximum   {distance before} turning back to the axis.
We summarize the results of this subsection below.

\vspace{.3cm}

{\bf Proposition 3.} {\it Consider geodesics with $P_z\neq 0$. Then, radially bounded geodesics with $\Lambda=0$ are: (i)stable with the introduction of any $\Lambda>0$; (ii) stable
with the introduction of any $\Lambda<0$ in the case of timelike geodesics; (iii) unstable, if $E^2\geq V_{\infty}$, for $\Lambda<0$ in the case of null geodesics.}

\subsection{Instability of LT solutions}

The stability of the solutions with $\Lambda < 0$ was recently studied in \cite{Gleiser17}. Consider the kind of perturbations,
\bqu
\lb{4.19}
g_{\mu\nu} =  {g_{\mu\nu}^{(0)}(x) + \epsilon h_{\mu\nu}},
\equ
where $g_{\mu\nu}^{(0)}(x)$ is the LT solutions, and $h_{\mu\nu} = h_{\mu\nu}(t, r, z)$ with $\epsilon \ll 1$ and $x \equiv \sinh(|\beta| r)$.  Since the background depends only on $x$, one finds that $ h_{\mu\nu}(t, r, z)$ takes the form,
\bqu
\lb{4.20}
 h_{\mu\nu}(t, x, z) = \int{e^{-(\Omega t - kz)}{\cal{C}}(\Omega, k) H_{\mu\nu}\left(\Omega, k, x\right)d\Omega dk},
\equ
where ${\cal{C}}(\Omega, k)$ is determined by the initial data. It is well-known that the spectrum of perturbations (sensitively) depends on the boundary conditions, and the choice of them is subtle here with the LT solutions. First, they are singular
at the symmetry axis $r = 0$, except for some particular cases, as shown above. These singularities are timelike, and there is no general rule on how to impose boundary conditions at such singular points. Following the studies carried out for the vacuum LC solutions \cite{Gleiser15}, Gleiser found that if one imposes certain physically acceptable restrictions as the boundary conditions at $x = 0$, it is possible to define unique evolutions for arbitrary perturbations that satisfy such imposed 
boundary conditions. Second, the spacetime is asymptotically anti-de Sitter as   {$x \rightarrow \infty$}, as shown by  (\ref{4.12}), and this boundary is timelike. As a result, the hypersurfaces $t = $ Constant are not Cauchy surfaces, and  there always exist null geodesics that remain to the future of any given constant $t$ hypersurface, and never intersect the hypersurface [cf. Fig.\ref{Fig4.2}]. Then, the future evolution of any perturbation may be arbitrarily modified by information coming from $x = \infty$. To overcome the second problem, following \cite{IW06}, Gleiser found that under certain circumstances some boundary conditions can lead to a well-defined evolution of appropriate initial perturbation data. With such boundary conditions, he was able to show that unstable modes always exist. Then, he argued that this should be generically present in the  evolution of arbitrary initial data, and concluded that the LT spacetimes with a negative cosmological constant are generically unstable under the linear gravitational perturbations. 

It should be noted that the above analysis cannot be applied to the cases $\sigma = \pm 1/2$, as in these cases the solutions actually represent black holes, as shown in Fig.\ref{Fig4.3}, and the boundary conditions should be now at the horizons $X = 1$,
 instead   {of}  the symmetry axis $r = 0$, while the ones at $r = \infty$ should be similar to what Gleiser considered.  So, it would be very interesting to see if such black holes are stable or not.
 
 Recently, the above studies were generalized to the   {case $\Lambda > 0$}   \cite{Gleiser18}. However, due to the complexities of the perturbation equations, a similar conclusion for the general case is absent, but in all the cases   analyzed  
 unstable modes were found, which strongly suggests that all the   {LT spacetimes with $\Lambda > 0$} are  linearly unstable under gravitational perturbations, although  the problem of determining the
time evolution of arbitrary initial data  still remains open.
   
  \subsection{Sources producing LT spacetimes}
  
  To study the LT solutions further, it is important to study the possible sources that could produce such spacetimes. In particular, the spacetime singularities are normally considered as some kind of indications that in reality they should be replaced by matter sources that produce such spacetimes \cite{Bonnor92,BGM94}. 
  
  Along this vein, in the case $\Lambda > 0$ the solutions are usually singular at both $r = 0$ and $r = \pi/(2\beta)$. Then, one can see that at least two matter sources should exist in this case, and the LT solutions are only valid in between these two sources.
  Then,  it is not clear what are the physical meanings of such spacetimes, if there is any.  In \cite{GP10}, because of the symmetry given by  (\ref{4.18}) it was argued that the coordinate $z$ is more like an angular coordinate than an infinitely long axis. So, replacing $z$ by $\psi$ which has the range $\psi \in [0, 2\pi)$, the LT solutions with $\Lambda > 0$ and $\sigma \in [0, 1/4]$ were matched with the static Einstein universe,
 \bqu
 \lb{4.21}
 ds_{ES}^2 = A_1^2dt^2 - \frac{B_1^2}{\Lambda}\cos^2\Delta d\psi^2 - \frac{C_1^2}{\Lambda}\sin^2\Delta d\varphi^2 - dr^2,
 \equ
 along the hypersurface $r = r_1$, where $\Delta \equiv \sqrt{\lambda}\left(r - r_0\right)$, $A_1, \; B_1, C_1$ and $r_0$ are real constants and are uniquely determined by the junction conditions along $r = r_1$. 
However, as noted in  \cite{GP10}, the static Einstein universe can replace only one of the two singularities, and to have a singularity-free spacetime, one needs to replace the remained singularity, too. 

With the  static Einstein universe as an example, one may also consider $r$ as an angular coordinate, so that the hypersurfaces $r = 0$ and $r = \pi/(2\beta)$ are identical, and the two singularities become one. In fact, setting 
\bqu
\lb{4.23}
\sin\left(\frac{r-r_0}{a}\right) = \sin\chi\sin\theta, \quad \tan(B_1\psi) = \tan\chi\cos\theta, 
\equ
the metric (\ref{4.21}) will take the form,
\bqu
\lb{4.24}
ds_{ES}^2 = dT^2 - a^2\Big[d\chi^2 + \sin^2\chi\left(d\theta^2 + \sin^2\theta d\Phi^2\right)\Big],
\equ
where $a$ is a constant (the radius of the static Einstein universe), and $T \equiv A_1 t$ and $\Phi \equiv C_1\varphi$, with $T \in (-\infty, \infty)$, $\chi \in [0, \pi]$, $\theta \in [0, \pi]$, and $\Phi \in [0, 2\pi)$. It is clear that  (\ref{4.24}) represents a static universe with $R\times S^3$ topology. One can apply  a similar coordinate transformation to the LT solutions with the identifications of $r = 0$ and $r = \pi/(2\beta)$. Then, one can consider the Einstein static universe as the internal spacetime of a ball with radius, say, $a_1$, while the spacetime outside of the spherically symmetric ball is given by the LT solutions. 

In addition, in \cite{BCMV12}, the authors considered the matching of an elastic matter to the LT solutions and found that such a matching  is possible for both $\Lambda >0$ and $\Lambda < 0$, but it is impossible for 
$\Lambda = 0$.  On the other hand,   in \cite{BSMS13} the matching, across cylindrical surfaces, of static cylindrically symmetric conformally flat spacetimes with a cosmological constant, satisfying
regularity conditions at the axis, to an exterior LT spacetime was studied, and it was found that   for $\Lambda \le 0$, such matching is impossible, while  the one with $\Lambda > 0$ is possible.
   
 In the case $\Lambda < 0$, for the black holes solutions (with $\sigma = \pm 1/2$) it was  first shown \cite{WSSW03} that they can be produced by gravitational collapse of type-II fluid \cite{HE73}. In \cite{ZB08}, the junctions of the LT solutions with
 $(\sigma_{-}, \Lambda_{-}, C_{-})$ with the ones with $(\sigma_{+}, \Lambda_{+}, C_{+})$ were considered with a matter shell appearing on the matching surface.    
 
 It should be noted that the matching given in \cite{BCMV12,ZB08} for $\Lambda > 0$ faces the problem that a spacetime singularity still exists at $r = \pi/(2\beta)$ in the LT spacetime.

%
\section{Static cylindrical spacetimes with perfect fluids} 
%
%
\renewcommand{\theequation}{5.\arabic{equation}} \setcounter{equation}{0}

  Static perfect fluid distributions with cylindrical (as well as planar) symmetry are rather
  widely discussed in the literature. This allows for approximately considering the fields 
  of bodies whose shape is drastically nonspherical (e.g., like rods) avoiding substantial
  mathematical difficulties of more general geometries.  In a number of papers 
  (see \cite{Marder, flu2, flu3} and references therein) 
  solutions of this kind were obtained for special choices of matter equations of 
  state (EoS) or for some restricted forms of the metric. In \cite{flu4} the problem of finding 
  such solutions for an unspecified EoS was reduced  to a single second-order ordinary 
  differential equation with two unknown functions, and  for the most frequently used EoS,
  $p = w\rho$, $w=\const$, to a first-order equation integrable only numerically. 
  In \cite{flu5} the problem was solved completely for disordered radiation ($w=1/3$). 
  In a more general form the problem was solved in \cite{flu6, flu7} both for fluids with an 
  arbitrary EoS and for $p = w\rho$. 

  Later there appeared a number of studies (e.g., \cite{flu8, flu9, Herrera2}) discussing global 
  properties of \scyl\ perfect fluid configurations. This actually already began in \cite{flu7},
  considering systems with both cylindrical and (pseudo)planar symmetries. 

  In the recent years, owing to new discoveries in astronomy, fluids with unusual EoS 
  attracted much attention, including those with negative pressures and even those with 
  $w = p/\rho <-1$, the so-called phantom ones. Such \cyl\ distributions were considered 
  in \cite{Marder1}. In this section  we describe the main results of these studies mostly 
  following \cite{flu7, Marder1}. However, some of the relationships and observations are new.

\subsection{General consideration}

   We write the \scyl\ metric in the form \rf{ds-cy}
\beq      \label{flu-1}
      ds^2 = \e^{2\gamma} dt^2 - \e^{2\alpha} dx^2 - \e^{2\mu} dz^2 - \e^{2\beta}d\varphi^2 ,
\eeq
  and choose the harmonic radial coordinate such that \rf{harm}
\beq               \label{flu-2}
      \alpha(x) = \gamma(x) + \beta(x) + \mu(x).
\eeq
  Einstein's equations    {(\ref{1})}
  for the metric (\ref{flu-1}) then take the following simple form:
\bear                \label{flu-3}
  	\beta'' + \mu'' - \mu'\beta' - \mu'\gamma' - \beta'\gamma'
  				&=& -  \kappa T_0^{0}\e^{2 \alpha},
\yy         \label{flu-4}
  	\mu' \beta' + \mu' \gamma' + \beta' \gamma'
				&=& -  \kappa T_1 ^{1}\e^{2 \alpha},
\yy           \label{flu-5}
  	\mu'' + \gamma'' - \mu' \beta' - \mu' \gamma' - \beta' \gamma'
				&=& - \kappa T_{2}^2\e^{2 \alpha},
\yy              \label{flu-6}
  	\gamma'' + \beta'' - \mu' \beta' - \mu' \gamma' - \beta' \gamma'
				&=& - \kappa T_{3}^{3} \e^{2 \alpha},
\ear
  where the prime denotes $d/dx$, and we take $T\mN$ as the stress-energy tensor 
  of a perfect fluid in its comoving reference frame, so that the 4-velocity is 
  $u^\mu = (\e^{-\gamma}, 0, 0, 0)$:   
\beq		\label{flu-7}
     T\mN = (p+\rho) u_\mu u^\nu -\delta\mN p  = \diag (\rho, -p, -p, -p).      
\eeq
   The conservation law $\nabla_\mu T\mN =0$ implies
\beq  							\label{flu-8}
		p' + \gamma' ( p + \rho ) = 0.
\eeq

   For a \cyl\ matter distribution, a natural boundary condition is regularity on the symmetry
   axis where $r^2 \equiv \e^{2\beta} = - g_{33} \to 0$ (the coordinate circles shrink to a point). 
   In terms of the metric \rf{flu-1}, this condition reads
\beq            								\label{flu-9}
		\gamma= \gamma_0 + O(r^2), \;                   
		\mu= \mu_0 + O(r^2), \;
		\e^{-2\alpha}r'{}^2 = 1 + O(r^2),
\eeq      
  where the last condition is the requirement of a correct circumference to radius ratio ($2\pi)$
  for small coordinate circles near the axis. As to the other end of the range of $x$, there can 
  be a matching condition with some  external solution or some requirement to the behavior 
  of variables as $r\to \infty$. It is necessary to note that if the external solution is the vacuum 
  (LC) one, then matching on the boundary requires $p=0$, which restricts the 
  choice of possible EoS. On the other hand, one could require asymptotic flatness at 
  infinity, but this is hard to achieve since the total mass of a \cyl\ configuration is, in general,
  infinite due to its being infinitely extended in the $z$ direction. Even the LC metric is 
  \asflat\ only if it is completely flat. If cylindrical symmetry is approximate and valid only 
  in a certain neighborhood of an extended body, then the asymptotic flatness requirement
  applies to regions far outside this neighborhood.
    
  Before considering particular solutions, let us make some general observations. 
  
  If the EoS $p = p(\rho)$ is not unspecified, there are four Einstein equations for five 
  unknowns $\beta, \gamma, \mu$, $\rho$ and $p$, or, equivalently, three
  independent Einstein equations and the hydrostatic equilibrium condition \rf{flu-8},
  and one unknown function may be chosen arbitrarily. This makes possible obtaining 
  a formal general solution with an arbitrary function.  
   
  Indeed, the difference of  \rf{flu-5} and \rf{flu-6} yields 
\beq  				\label{flu-10}
	\beta'' = \mu'' \ \then \ \mu = \beta + a_1 x + a_2, \;   a_1, a_2 = \const.
\eeq      
   Furthermore, the difference of  \rf{flu-4} and \rf{flu-5} can be written as
\beq                             \label{flu-11}
		\eta'' - 2a_1 \eta' - 2 \eta'^2 + 2 \gamma'^2 =0, \; \eta \equiv \beta+\gamma,		
\eeq      
   and $\gamma (x)$ is easily found by integration if we take $\eta(x)$ as an arbitrary function.
   Thus the metric is completely determined. The quantities $\rho(x)$ and $p(x)$ are then 
   obtained directly from Einstein's equations, and the EoS is thus found in a parametric 
   form.
   
   On a regular axis it should be finite $\mu$ while $\beta \to -\infty$, hence $a_1\ne 0$, 
   and the axis is located at $x = +\infty$ if $a_1 > 0$ and at $x = -\infty$ if $a_1 < 0$.   
   Examples of solutions with a regular axis will be presented below.

\subsection{Solutions for $p = w\rho$, the general case}

   With the EoS $p=w \rho$, $w = \const$, the condition \rf{flu-8} leads to
\beq                            \label{flu-12}
    \rho = \rho_0  \e^{-(w+1)\gamma(x)/w},\;
     p = w \rho_0 \e^{-(w+1)\gamma(x)/w},
\eeq
  for $w\neq 0$, where $\rho_0 = \const> 0$. At $w=0$ the tensor (\ref{flu-7}) would correspond to dust which
  cannot be in static equilibrium under pure gravitational forces, so we assume $w \ne 0$.

  From \rf{flu-5}, \rf{flu-6} we have, as before, \rf{flu-10}, and we take it in the form
\beq                                                \label{flu-13}
  	\mu(x) = \beta(x) + a_1 x - \ln r_0,
\eeq  	 
 where the arbitrary constant $r_0$ with the dimension of length will determine a length 
 scale.  Another linear combination of \rf{flu-3}--\rf{flu-6} then leads to
\beq                                                \label{flu-14}
    	  (1+3w) \beta'' + (1-w) \gamma'' =0.
\eeq
  The simpler cases $1+3w=0$ and $1-w =0$ will be considered separately. Assuming that 
  these quantities are nonzero, we can write
\beq   		                           \label{flu-15}
        \beta(x) = \frac{(w-1)}{(3w+1)} \gamma(x) + b_1 x + b_2 + \ln r_0,
	\;  b_1,  b_2 = \const ,
\eeq
  and put $b_2=0$ by properly choosing the zero point of $x$. With \rf{flu-13}, \rf{flu-15} and
  \rf{flu-2}, all metric functions are expressed in terms of $\gamma(x)$.
  The latter can be determined using the first-order equation \rf{flu-4} after proper substitutions.

  Let us denote 
\bearr                                                   \label{flu-16}
      A = 7w^2 - 6w -1 \cm  B = \frac {A}{w(3w+1)},    
\yyy 						           \label{flu-17}
      2 \eta(x) =  B \gamma(x)  + 2ax.
\ear
  Then \rf{flu-4} leads to the equation 
\beq                                                  \label{flu-18}
        4\eta'{}^2(x) = \frac{A}{w^2} \bigg[  \kappa w \rho_0 r_0^2
					\e^{2\eta(x)} + \frac {4w^2}{A}(a_1+2b_1)^2  - b_1 (a_1 + b_1) \bigg],
\eeq
   which is easily solved by quadratures to finally yield all unknown functions. 
  Thus under the conditions $w \neq 0, 1, -1/3, -1/7$ ($A \neq 0, B$ is finite) the solution
   is completely obtained analytically, though its particular form depends on the values of 
   $w$ and the integration constants $a_1, b_1$. 
    
  One can show that perfect fluid distributions with $p = w\rho$ cannot behave as 
  isolated systems in space, such as threads, tubes or cosmic strings. To verify that, 
  let us write the corresponding conditions at spatial infinity, $x\to x_{\infty}$, such that
\beq                          \label{flu-19}
           r(x) \equiv \e^{\beta(x)} \Big|_{x=x_{\infty}} = \infty.
\eeq
  The metric \rf{flu-1} has a flat or string spatial infinity as  $x \to x_{\infty}$ if
\beq                                  \label{flu-20}			
	   |\gamma| < \infty,\;   |\mu| < \infty, \; 
	    \e^{2\beta-2\alpha}\beta'{}^2 \Big|_{x \to x_{\infty}}  \to 1-\xi, 
\eeq
 where $\xi  =\const$.  If  $\xi =0$, we have asymptotic flatness, at $0< \xi < 1$ there is a deficit of the angle
  $\varphi$ (the surfaces $z=\const$ asymptotically behave as cones rather than planes,
  it is a feature of cosmic string behavior), while at $\xi < 0$ there is an angular excess: 
  this   {is } a cosmic string with negative linear density.
  
  In our solutions, if $w \ne -1/3$, from \rf{flu-15} it follows that $\gamma$ can be finite 
  while $\beta \to \infty$ only if  $b_1 x \to \infty$. Assume without loss of the generality
  $b_1 >0$, so $x_{\infty}= \infty$. Then (\ref{flu-13}) shows that $|\mu | < \infty$  requires
  $a_1 = - b_1$, and as  $ x\to \infty$, $\eta(x) \approx b_1 x$. Then the left-hand side 
  of \rf{flu-18} tends to a constant while the right-hand side infinitely grows. This 
  contradiction shows that our assumption was wrong, and a flat or  {string-like} asymptotic 
  behavior of our solutions is impossible.  

\subsubsection{Solutions with a regular axis}

  Let us find explicitly, without fixing $w$, perfect fluid configurations with a regular 
  symmetry axis. On such an axis we should simultaneously have $\beta\to -\infty$ 
  and finite $\mu$ and $\gamma$, which is only possible if  $b_1 x \to -\infty$ and 
  $a_1 + b_1 =0$. Assuming without loss of generality $b_1 = -a_1 = a > 0$ (so that
  the axis corresponds to $x \to -\infty$), we can rewrite  \rf{flu-18} in the form
\beq       						\label{flu-21}
		\eta'{}^2 = a^2 - K \e^{2\eta}, \cm    K := - \frac{\kappa \rho A r_0^2}{4w}.
\eeq
  Let us further assume that the fluid density is positive, hence $\rho_0 >0$, then 
  $K >0$ for the values of $w$
\beq        						\label{flu-22}
		0 < w < 1, \quad {\rm or}  \quad  w < -1/7.
\eeq    
  Then \rf{flu-21} is integrated to give
\beq      						\label{flu-23}
	     \e^{\eta(x)} = \frac {a}{\sqrt{K}\cosh (ax)},\;
	     \e^{\gamma(x)} = \big(1 + \e^{2ax}\big)^{-2/B} a^{\tfrac{3w+1}{4w}},
\eeq    
   where we have suppressed the emerging integration constant by choosing the zero point of 
   $x$. 
   
   The next step is to apply the third regularity condition \rf{flu-9}, taking into account that 
   $\gamma'(-\infty) =0$ and $\beta(-\infty) = a$. We obtain  
\beq  						\label{flu-24}
		\frac{4a^2}{K} = a^{A/(4w^2)} \ \then\  
				a = \bigg(\frac K4\bigg) ^{\tfrac{4w^2}{w^2 + 6w +1}}
				    = \bigg(\frac {-A\kappa \rho_0 r_0^2}{16 w}\bigg) ^{\tfrac{4w^2}{w^2 + 6w +1}},
\eeq      
   where $K$ is defined in \rf{flu-21}. This completes the solution. So $\gamma(x)$ is given
   in \rf{flu-21}, while the other metric functions are
\bearr  						\label{flu-25}
	       \mu (x) = \frac{(w-1)}{(3w+1)} \gamma(x),  \;
		\beta(x) = \frac{(w-1)}{(3w+1)} \gamma(x) + ax + \ln r_0,  \;
\nnnv		
		\alpha(x) = \frac{(5w-1)}{(3w+1)} \gamma(x) + ax + \ln r_0.
\ear      
    
   The metric takes a simpler form in terms of the new radial coordinate $y = \e^{ax}$, such that 
   $y =0$ is the axis and $y=\infty$ is spatial infinity. Rescaling the time coordinate so that  
   $g_{00} = \e^{2\gamma} =1$ on the axis, we obtain
\bqun 						\label{flu-26}
		ds^2 &=& \big(1+y^2\big)^{s_0} dt^2 - a^{(w-1)/(2w)}
		\Big[ \big(1+y^2\big)^{s_1} r_0^2 dy^2 \nb\\
		&& 
		- \big(1+y^2\big)^{s_2} (dz^2 + r_0^2 y^2 d\varphi^2) \Big],\nb\\
	s_0 &=& -\frac 4B =  - \frac{4w(3w+1)}{(7w+1)(w-1)},\;
	s_1 =  - \frac{4w(5w-1)}{(7w+1)(w-1)}, \nb\\
	s_2 &=& - \frac{4w}{7w+1},		
\equn
  with $a$ given in \rf{flu-24}, and the density $\rho$ and pressure $p=w\rho$ are determined 
  by \rf{flu-12}. The solution is valid for all $w \in (0,1)$ and $w < -1/7$. Its properties strongly 
  depend on $w$, so in what follows we discuss this dependence as well as some particular 
  values of $w$. 
  
  One should notice that \rf{flu-24} is meaningless if $w^2 + 6w +1=0 \then w = -3\pm\sqrt{8}$,
  so the constant $a$ cannot be determined and remains arbitrary. It means that for these 
  values of $w$ (approximately $-0.16$ and $-5.84$) the axis with finite $\gamma$ and $\mu$
  cannot be regular and is inevitably a conical-type singularity.   
   
  If $w \in (0, 1)$, then both the circular radius $r = \sqrt{-g_{33}}$ and the temporal metric
  component  $g_{00}(y)$ are increasing functions, so the gravitational field attracts 
  test particles towards the axis,  while both $\rho$ and $p$ are positive and decrease
  with growing $r$.
  
  However, with negative pressures ($w <0$) things are much more diverse at large $x$ (or  
  $y = \e^{ax}$). Let us present for different $w$ the large $x$ behavior of the metric coefficients
  $g_{00}=\e^{2\gamma}$ (describing an attractive or repulsive nature of gravity at large $x$) 
  and $r^2 \equiv \e^{2\beta}$ (squared circular radius), the fluid density, and the convergence or
  divergence of some integral characteristics of the solution: 
 
 \medskip\noi  
  $L = \int e^\alpha dx$ (the total length along the radial direction), 

  \medskip\noi  
  $V = 2\pi \int \e^{\alpha+\mu+\beta} dx$ (the total volume per unit length along the 
   symmetry axis),  and  
 
 \medskip\noi   
   $E = 2\pi \int \rho(x) \e^{\alpha+\mu+\beta} dx$ (the total  fluid energy per unit length along the
    symmetry axis).
 
 \medskip      
  Note that, up to a positive constant factor,
\bearr                              \label{flu-27}
	\e^{\alpha+\mu+\beta} \sim \exp \bigg[ \frac {2(1 + 7 w^2)}{(1- w) (1 + 7 w)}\,ax\bigg],
\nnn
	\rho\e^{\alpha+\mu+\beta} \sim \exp \bigg[ -\frac {2 (1 + 8 w - w^2)}{(1- w) (1 + 7 w)}\,ax\bigg].	
\ear       

The large $x$ behavior of these parameters is presented in the Table II. 

\begin{table}
\label{Table2}
\caption{ The large $x$ behavior for different $w$, where $F$ means finite.}
\begin{center}
\begin{tabular}{|c|c|c|c|c|c|c|}
\hline
      \vphantom{$\Big($} 
      {\rm  Parameters:} &   $\e^{2\gamma}$ &  $ r^2$  & $\rho$    & $L$      & $V$  & $E$ 
\\    \hline          
\wide
       $ 0 < w < 1 $               &  $\infty$             & $\infty$    &   0     & $\infty$& $\infty$ & F    
\yy         
    $-\frac{1}{3} < w <  -\frac{1}{7}$          & $\infty$              & 0              &$\infty$ & F       &  F      & F  
\yy     
     $-1 < w <  -\frac{1}{3} $           & 0                        & $\infty$    & 0          & F       &F      &F  
\yy
     $w = -1  $                      & 0                        & $\infty$    & F       & F       &F     &F    
\yy
     $w < -1   $                     & 0			  & $\infty$    & $\infty$ &F     &F      &F    
\\[5pt]
\hline
\end{tabular}
\end{center}
\end{table}

   We see that in the range $-1/3 < w < -1/7$ we obtain a closed type geometry with one
   regular axis and another singular one. All solutions with negative pressure ($w<0$) describe 
   geometries with a finite volume of space per unit length along the axis, and in all cases 
   under consideration the corresponding total fluid energy is finite. In the phantom range 
   $w < -1$ the density grows with growing radius, much like the phantom energy behavior 
   in cosmology.    
   
\subsection{Solutions for particular values of $w = p/\rho$}

\subsubsection{Stiff matter, $w=1$}

  In this case, $A = (7w+1)(w-1)=0 $, and the above scheme does not work, but 
  \rf{flu-12}--\rf{flu-14} are valid and lead to
\bearr                                                            \label{flu-28}
        \beta(x) = b_1 x + \ln r_0,\cm  \mu(x)=(a_1 + b_1 )x, 
\nnnv        
         \alpha(x)  = \gamma(x) + ax + \ln r_0,\;\;\;   a:= a_1 + 2b_1,
\ear
   where, as before, $r_0 = \const$ specifies the length scale,  while from (\ref{flu-4}) we get
    the equation $ a\gamma'(x) + (a_1+b_1) b_1  =\kappa \rho_0 \, \e^{2a x}$  for $ \gamma(x)$, 
    whence (provided $a \ne 0)$
\bearr                                                         \label{flu-29}
        \gamma(x)=\frac{\kappa \rho_0 }{2a^2} \e^{2ax} - \frac{b_1(a_1+b_1)}{a} x, \;\;
        \alpha(x)=\gamma(x)+ax + \ln r_0.\nb\\
\ear
  If $a=0$ (that is, $a_1= -2b_1$), then \rf {flu-4} reduces to a relation between constants,
  $-2 b_1^2 = \kappa \rho_0 r_0^2$, incompatible with $\rho >0$, not to be considered. 

  If $b_1 >0$, a symmetry axis ($\e^\beta \to 0$) corresponds to $ x \to -\infty $. 
  As in the general case, to have finite $ \gamma(x) $ and $ \mu(x) $ as $x \to -\infty$, 
  we need  $a_1 + b_1 =0$, so $b_1=a >0$, and
\bearr                         		\label{flu-30}
        \mu(x) \equiv 0,\quad    \beta(x) = ax + \ln r_0, \quad 
\nnnv       
	\gamma(x)=\frac{\kappa \rho_0 }{2a^2} \e^{2a x}, \;
	\alpha(x)= \frac{\kappa \rho_0 }{2a^2} \e^{2a x} + a x + \ln r_0.
\ear

  From the third axis regularity condition \rf{flu-9} it follows $a = 1$, and the solution with a regular
  axis even more simplifies. It is  conveniently written in terms of the coordinate $r = r_0 e^x$:
\bearr                          		\label{flu-31}
	ds^2 = \e^{\kappa\rho_0 r^2} (dt^2 - dr^2) - dz^2 - r^2 d\varphi^2,  
\yyy                              		\label{flu-32}
	\rho = p = \rho_0 \e^{-\kappa \rho_1 r^2}.
\ear		  
  
  Consider the fluid energy distribution in stiff matter for the solution \rf{flu-25}, \rf{flu-26}.
  The density is $T_0^0 =\rho = \rho_0$ at $r =0$, and it quickly decays as $r\to\infty$, 
  being thus localized near the axis. Let us find the energy per unit length along the $z$ axis:
\beq                                                         \label{flu-33}
        E=2\pi \int_0^{\infty} T_0 ^0  \sqrt{-^{3}g}\, dr = \frac{2\pi}{\kappa} = \frac{1}{4 G},
\eeq
  where ${}^3 g = - \e^{2\alpha + 2\beta + 2\mu}$ is the determinant of the spatial metric, 
  and $G$ is the Newtonian gravitational constant. Thus the total energy per unit length 
  along $z$ does not depend on $\rho_0$ and is of the order of Planck linear density.

\subsubsection{Disordered radiation, $w= 1/3$}

  It is a special case of the above general solution: in particular,  $A = -20/9$, $B = -10/3$. 
  From \rf{flu-13}--\rf{flu-18} we find
\bqun                                                          \label{flu-34}
       \beta(x)&=& -\frac{1}{3} \gamma(x)+b_1 x + \ln r_0,  \;\;    
       \mu(x) = -\frac{1}{3} \gamma(x) + (a_1 + b_1) x, \nb\\       
       \alpha(x) &=& \frac{1}{3} \gamma(x) + a x + \ln r_0, \;\;\; a = a_1 + 2b_1
\equn
   where $\e^{\gamma(x)}$ is found using (\ref{flu-18}) as:
\bqun                                    \label{flu-35}
        \e^{\gamma(x)}&=&\bigg[ \frac{\kappa \rho_0 r_0^2}{3h^2}	\cosh^2(\sqrt{5}hx) \bigg]^{3/10} 
        		\exp\Big(\frac{3}{5}ax\Big),\nb\\
	h &\equiv& \sqrt{\frac{a^2}{5} + b_1(a_1+b_1)}.
\equn

  A regular axis at $x = -\infty$ is obtained with $a_1+b_1 =0$, $a=b_1 > 0$, and
  the third regularity condition \rf{flu-9} requires a special value of $a$. Indeed, as $x\to -\infty$, 
\bqun                               \label{flu-36}
    \e^{-2\alpha} r'^2 &=& \e^{2\beta-2\alpha} \beta'{}^2 \to 
   		\bigg( \frac{12 a^2}{5\kappa \rho_0 r_0^2}\bigg)^{2/5} = 1,\nb\\
    	& \then&   a = \bigg(\frac{5 \kappa \rho_0 r_0^2}{12}\bigg)^{1/7}.	
\equn
  In terms of $y = \e^{ax}$, so that $0 < y < \infty$, we obtain the metric \rf{flu-26} with
\beq                                          \label{flu-37}
			s_0 = 6/5, \quad   s_1= 2/5, \quad    s_2 = -2/5,
\eeq    
  so that 
\bqun                               \label{flu-38}
               ds^2 &=& (1+y^2)^{6/5}dt^2 - a^{-1} \left[(1+y^2)^{2/5}r_0^2 dy^2\right. \nb\\
               &&
				               			\left. - (1+y^2)^{-2/5}(dz^2 + r_0^2 y^2 d\varphi^2)\right],\nb\\
							            			\label{flu-39}
	       \rho &=& 3p = \rho_0 (1+y^2)^{-12/5}.  
\equn            
  This solution coincides with the one presented in \cite{flu5, flu6, flu7}.
   
  The fluid energy density $\rho$ is finite on the regular axis and vanishes as $y \to +\infty$. 
  The total fluid energy per unit coordinate length $\Delta z$ along the $z$ axis is finite:
\bqun                     \label{flu-40}
        E&=& \frac{2\pi r_0^2}{a^3} \int_0^{\infty} \rho (1+y^2)^{-1/5} y dy\nb\\
	&=& \frac{2\pi \rho_0 r_0^2}{a^3} \int_0^{\infty} (1+ y^2)^{-13/5} y\, dy = \frac {5\pi \rho_0 r_0^2}{8a^3}.
\equn	
  However, since $\e^\mu = (1+y^2)^{-1/5}$ vanishes at large $y$, this finite energy 
  belongs to a vanishing asymptotic length $\e^\mu \Delta z$, and it looks meaningless 
  to speak of its localization.
  
\subsubsection{A gas of cosmic strings, $w=-1/3$}

  It is a special case excluded in \rf{flu-15}--\rf{flu-18}, but we can use \rf{flu-12}--\rf{flu-14} and 
  write (choosing proper scales along the $z$ and $t$ axes)
\beq                           \label{flu-41}
	    \mu = \beta + a_1 x - \ln r_0, \;\; \gamma = b_1 x, \;\;   a_1, b_1, r_0 = \const,
\eeq    
  where, as before, $r_0$ is an arbitrary length scale.
  Substituting this into \rf{flu-4}, we obtain an equation for $\beta(x)$:
\beq                           \label{flu-42}
            \beta'{}^2 + a \beta'+ a_1 b_1 	= -\frac{\kappa \rho_0 }{3r_0^2} \e^{4\beta + 2ax},
\eeq
  preserving the previous notation $a = a_1 + 2b_1$.
  Already from \rf{flu-36} we can conclude that a regular axis is impossible in this system.
  Indeed, since on such an axis $|\gamma| < \infty$, it should correspond to finite $x$.
  But then it is impossible to have simultaneously $|\mu| < \infty$ and $\beta \to -\infty$.
  
  Equation \rf{flu-37} can be rewritten as 
\beq                                            \label{flu-43}
            \eta'{}^2 = a_1^2 + 4b_1^2 - \frac {4\kappa \rho_0}{3r_0^2} \e^{2 \eta(x)},   
            \;\;    \eta(x) := 2\beta(x) + ax. 
\eeq
  It is easily solved, leading to 
\beq 						\label{flu-44}
      \e^{2\beta(x)} = \sqrt{\frac{3r_0^2}{4\kappa \rho_0 }}  \frac{h \e^{-ax}}{\cosh (hx)},
      \quad    h = \sqrt{a_1^2 + 4b_1^2}.
\eeq
     This completes the solution. The range of $x$ is $x\in \R$. It is easy to verify that the 
     nature of the geometry described depends on the product $a_1 b_1$:
     
      (a) If $a_1 b_1 >0$, then $h < |a|$, hence there is a (singular) axis ($e^\beta \to 0$)
      on one end of the range of $x$ and a spatial infinity ($e^\beta \to \infty$) on the other.
      
      (b) If $a_1 b_1  < 0$, then $h > |a|$, hence there are two singular axes at $x \to \pm \infty$.
      
      (c) If $a_1 b_1 = 0$, then $h = |a| >0$, hence there is a singular axis on one end and 
      a finite radius on the other, $r = \e^\beta \to \const$, where, however, either $\mu(x)$ or
      $\gamma(x)$ tends to infinity, depending on which of the constants $a_1, b_1$ is
      nonzero and which sign it has.
      
    Thus a gas of disordered cosmic strings can create different geometries depending 
    on the parameter values, which actually express different boundary conditions.
    However, none of these geometries is nonsingular.
    
\subsubsection{A cosmological constant, $w=-1$}

  In this case, $\rho = -p = \Lambda/\kappa = \const$.
  It is again a special case of the general solution, such that 
  $A=12$, $B= 6 $, and $\eta = 3 \gamma + ax$. We have
\bearr 					\label{flu-45}
	\beta(x)= \gamma(x)+b_1 x + \ln r_0,  \quad\       
       \mu(x) = \gamma(x) + (a_1 + b_1) x, \quad\       
\nnnv       
       \alpha(x) = 3 \gamma(x) + a x + \ln r_0, \quad\ a = a_1 + 2b_1.	
\ear    
  The function $\gamma(x)$ is found from \rf{flu-18} which now reads
\beq        					\label{flu-46}
		\eta'{}^2 = -3 \Lambda r_0^2 \e^{2\eta} + a_1^2 + 3b_1^2 + 3a_1 b_1. 
\eeq   
  Its solution takes different forms depending on the sign of $\Lambda$. 
  This class of solutions is already discussed in detail in other coordinates in section  4, so let 
  us here focus on $\Lambda >0$ and the solution with a regular axis, which corresponds to 
  $a_1+b_1 =0$ and (assuming $b_1 = a >0$) $x \to -\infty$. We have 
\bearr  					\label{flu-47}
	a = \frac{4}{3\Lambda r_0^2},\quad\
        \e^{\gamma(x)} = \e^{\mu(x)} = \frac{\sqrt{a}}{(1+ \e^{2a x})^{1/3}},
 \nnn
         \e^{\beta(x)}  = \frac{\sqrt{a} r_0\e^{ax}}{(1+ \e^{2a x})^{1/3}},
 \quad\ 
         \e^{\alpha(x)}  = \frac{a^{3/2} r_0\e^{ax}}{1+ \e^{2a x}},
\ear
  In terms of the coordinate $y = \e^{ax}$, ranging from zero to infinity, the metric reads
\bqun      				\label{flu-48}
              ds^2 &=& \frac{4}{3\Lambda r_0^2}\Bigg[\frac{dt^2}{(1+y^2)^{2/3}}
              			- \frac{r_0^2 dy^2}{(1+y^2)^2} \nb\\
			&& - \frac 1 {(1+y^2)^{2/3}}(dz^2 + r_0^2y^2 d\varphi^2)           
              							\Bigg].
\equn    
  
  One could also easily describe examples of solutions with phantom matter: $w < -1$, but
  this will hardly add anything significant to the description of the more general solution 
  \rf{flu-23}--\rf{flu-26}.

\subsection{Some further results}

  In this subsection we mention some results on perfect fluids in \scyl\ spacetimes, related to
  the content of the present section but go beyond it.

  Much effort has been devoted to finding the conditions of matching cylindrical perfect fluid 
  distributions with external LC spacetime and an analysis of global properties of the obtained 
  configuration. This problem was already discussed in section  2, but here we find it appropriate 
  to add some discussion.   
  
  Thus, in particular, in \cite{flu8} it is shown that matching is possible for $-1/2 < \sigma < 1/2$,
  ($\sigma$ is the LC solution parameter) but $\sigma < 0$ corresponds to a negative fluid
  density. Global properties of cylindrical spacetimes are discussed in \cite{flu9}, where 
  analytical and numerical solutions have been obtained for an incompressible fluid. 
  The existence and uniqueness of global solutions are shown for rather general fluid EoS
  admitting nonvanishing density at zero pressure. In \cite{Herrera2}, it is shown that conformally flat 
  internal solutions cannot be matched to an LC exterior.  

   An interesting result is presented in \cite{Marder2}: assuming that there is a regular axis, that
  the fluid has a nonnegative density and that the integral $\int dp/(\rho+p)$ from zero to the 
  pressure value on the axis converges, it is proven that at some finite radius the pressure is zero,
  hence the fluid distribution is finite in the radial direction and can be matched with the LC vacuum 
  solution. It generalizes a theorem from \cite{flu9}. Note that for fluid distributions with $w >0$
  described above this integral diverges, and this matching is impossible. On the other hand, 
  as we saw, fluids with $w < -1/7$ create spaces of finite 3-volume, fill the whole space, 
  and the total fluid energy per unit length along the axis is finite.  
   
  A number of papers discuss \cyl\ perfect fluid distributions, which can be charged or neutral,
  in the presence of scalar or/and electromagnetic fields \cite{flu6, flu7, Marder3}.
  In \cite{flu6, flu7}, it is shown how to obtain perfect fluid solutions with elecrtomagnetic fields 
  of three possible directions admitted by the symmetry (see section  3)  as well as scalar fields
  using the inverse problem method, starting with certain combinations of metric functions 
  which can be chosen by hand. This freedom is related to arbitrariness of the EoS and electric 
  charge or current or scalar charge distributions. The regular axis conditions are also discussed 
  there. In \cite{Marder3} different kinds of electrically and scalarly charged \cyl\ perfect fluid 
  distributions are discussed  both in static and stationary spacetimes, a number of exact 
  solutions are found and analyzed.   
  
  It is known that dust ($p=0$) without charges cannot maintain static equilibrium in its own
  gravitational field. However, for electrically or scalarly charged dust it is possible, and the 
  corresponding solutions, in particular, \cyl\ ones, are known \cite{Marder4, Marder5, Marder6, Marder7}. 
  According to \cite{Marder4, Marder5}, such static equilibrium configurations of arbitrary shape can
  only exist if the mass density is everywhere equal (in proper units) to the absolute value of 
  the electric charge density: in this way, the electric repulsion exactly balances the gravitational 
  attraction.  However,  in the presence of a scalar field with its charge density,
  the situation is not so strictly determined, and an equilibrium is possible with different 
  relationships between the charge densities \cite{Marder6, Marder7}. Also, in the framework of dilaton gravity 
  (general relativity coupled to dust, scalar and electromagnetic fields, and the latter two interact 
  in a gauge-invariant way),  it has been shown that cylindrical black holes are possible, 
  but cannot exist without negative energy density somewhere in space  {\cite{Marder6, Marder7}}.

  These are some important results concerning \scyl\ perfect fluid distributions in general relativity.

%
\section{Lewis vacuum spacetime} 
%
%
\renewcommand{\theequation}{6.\arabic{equation}} \setcounter{equation}{0}

\subsection{Stationary cylindrical vacuum spacetime}

The extension of the LC static cylindrically symmetric vacuum spacetime to a stationary cylindrically symmetric vacuum spacetime was obtained independently by Lanczos in 1924  \cite{Lanczos} and Lewis in 1932 \cite{Lewis}. 
We  {consider the spacetime described} by the  cylindrically symmetric stationary metric
\begin{equation}
ds^2=fdt^2-2kdtd\phi-e^{\mu}(dr^2+dz^2)-ld\phi^2, \label{L1}
\end{equation}
where $f$, $k$, $\mu$ and $l$ are functions only of $r$. The ranges of the coordinates are $-\infty<t<\infty$ for the time coordinate, $0\leq r<\infty$ for the radial coordinate, $-\infty<z<\infty$ for the axial coordinate and
$0\leq\phi\leq 2\pi$ for the angular coordinate with the hypersurfaces $\phi=0$ and $\phi=2\pi$ being identified. The coordinates are numbered $x^0=t$, $x^1=r$, $x^2=z$ and $x^3=\phi$. The general vacuum solution $R_{\alpha\beta}=0$ for the metric (\ref{L1}), in the notation given by \cite{Silva} and \cite{Silva1}, is
\begin{eqnarray}
\label{L2}
f&=&ar^{-n+1}-\frac{c^2r^{n+1}}{n^2a}, \;\; 
k=-Af, \;\;
e^{\mu}=r^{(n^2-1)/2}, \label{L4},\nb\\
l&=& \frac{r^2}{f}-A^2f, \;\; 
A=\frac{cr^{n+1}}{naf}+b. 
\end{eqnarray}
The constants $n$, $a$, $b$ and $c$ can be either real or complex, and the corresponding solutions belong to the Weyl class or Lewis class, respectively. For the Lewis class these constants are given by
\begin{eqnarray}
\label{L7}
n&=& im, \;\; 
a=\frac{1}{2}(a_1+b_1)^2, \;\; 
b=\frac{a_2+ib_2}{a_1+ib_1}, \nb\\
c&=& \frac{m}{2}(a_1^2+b_1^2), \label{L10}
\end{eqnarray}
where $m$, $a_1$, $a_2$, $b_1$ and $b_2$ are real constants satisfying
\begin{equation}
a_1b_2-a_2b_1=1. \label{L11}
\end{equation}
The metric coefficients (\ref{L2})  with (\ref{L7})  become \cite{Lewis,Silva1}
\begin{eqnarray}
\label{L12} 
f&=& r(a_1^2-b_1^2)\cos(m\ln r)+2ra_1b_1\sin(m\ln r), \nb\\ 
k&=&-r(a_1a_2-b_1b_2)\cos(m\ln r)\nb\\
&& -r(a_1b_2+a_2b_1)\sin(m\ln r), \nb\\ 
l&=&-r(a_2^2-b_2^2)\cos(m\ln r)-2ra_2b_2\sin(m\ln r),\nb\\
e^{\mu}&=&r^{-(m^2+1)/2}. 
\end{eqnarray}

A simple deduction of the Lewis metric, where one does not need to consider complex parameters to obtain the Lewis class, is given in \cite{Gariel}.  {There a  physical  interpretation of the field equations
is also provided,  which}  permits to get some understanding of the four parameters appearing in the Lewis solution (\ref{L2}). Another derivation of the Lewis metric is given in \cite{MacCallum} where it is extended to Einstein's spaces by including the cosmological constant. It is found that three parameters are essential, of which one characterizes the local gravitational field, while the remaining two give information about the topological identification made to produce cylindrical symmetry.

In Newtonian physics the potential due to a cylindrical matter source, being static or stationary,   {has the same dependence, that is, it depends  only on}  one parameter, the mass per unit length. For the static vacuum cylindrical field in GR the solution is the LC metric, the one we  {studied} in section  2, revealing two essential parameters, while for the stationary cylindrical rotating source it has, in its usual form, four parameters reducible to three essential parameters \cite{MacCallum}.

In the next subsection we study the meaning of the parameters appearing in (\ref{L2}).

\subsection{The parameters of the Weyl class}

The transformation  {\cite{BonnorC}}
\begin{eqnarray}
d\tau=\sqrt{a}(dt+bd\phi), \label{L16} \\
d\bar{\phi}=-\frac{1}{n}[cdt-(n-bc)d\phi], \label{L17}
\end{eqnarray}
casts the metric (\ref{L1}) with (\ref{L2}) into
\begin{equation}
ds^2=r^{1-n}d\tau^2-r^{(n^2-1)/2}(dr^2+dz^2)-\frac{r^{n+1}}{a}d\bar{\phi}^2. \label{L18}
\end{equation}
This is locally the LC metric. Nevertheless, since $\phi=0$ and $\phi=2\pi$ are identified, $\tau$ defined in (\ref{L16}) attains a periodic nature unless $b=0$  {\cite{BonnorC}}. On the other hand, the new coordinate $\bar\phi$ ranges from $-\infty$ to $\infty$. A more detailed account of this subject can be found in \cite{Stachel}. In order to globally transform the Weyl class of the Lewis metric into the static LC metric we have to make $b=0$ and $c=0$. Note that in this case, from the transformations (\ref{L16}) and (\ref{L17}), $\tau$ and $\bar\phi$ become respectively true time and angular coordinates.

Hence we can say that $b$ and $c$ are responsible for the non-staticity of this family of   {solutions} of the Lewis metric.

 {As mentioned previously, the  Cartan scalars provide the local characteristics } of a metric. They are obtained through the components of the Riemann tensor and its covariant derivatives calculated in a constant frame. Two metrics are equivalent if and only if there exist coordinate and Lorentz transformations which transform  the Cartan scalars of one of the metrics into the Cartan scalars of the other. By performing these calculations for both metrics, the LC metric and the Weyl class of the Lewis metric, we obtain that both are equivalent locally and indistinguishable,  which confirms the coordinate analysis made at the beginning of this subsection. Furthermore, we showed that only the parameter $n$ curves spacetime for both, static and stationary Weyl class, metrics. However, we shall see that the two metrics possess very different topological behaviour.

Details of the calculations of the Cartan scalars are given in \cite{Silva}.

Considering a cylindrical matter source for the Weyl class metric consisting of a rigidly rotating anisotropic fluid,  one of Einstein's field equations can be integrated. This integration produces
\begin{equation}
fk_{,r}-kf_{,r}=\xi r, \label{L19}
\end{equation}
where the commas stand for differentiation with respect to $r$ and $\xi$ is an integration constant. Calculating the rotation of the source as given in \cite{Stephani,Silva} produces its rotation magnitude given by $\xi/(2fe^{\mu/2})$. Now using the matching conditions on the surface of the source cylinder as given by Darmois \cite{Darmois} we find,   
\begin{equation}
c\equiv -\frac{\xi}{2}. \label{L20}
\end{equation}
 {Note that this constant $c$ is different from the  speed of light used in other places of this article. In fact,  now it measures the rotation of the cylindrical source, as can be seen from  (\ref{L20}). }

In the Newtonian limit, the velocity term is negligible, then from (\ref{L20}) $c\approx 0$, and recalling (\ref{L2}), we find that 
\begin{equation}
f=e^{2U}. \label{L21}
\end{equation}
Then,  the Newtonian potential is
\begin{equation}
U=2\sigma\ln r+\frac{1}{2}\ln a, \label{L22}
\end{equation}
where $\sigma$ is given by
\begin{equation}
\sigma=\frac{1}{4}(1-n). \label{L23}
\end{equation}
In Newtonian theory, (\ref{L22}) is the gravitational potential of an infinite uniform line mass with mass per unit length $\sigma$. The constant $(\ln a)/2$ represents the constant arbitrary potential that exists in the Newtonian solution.
The metric (\ref{L2}) has infinite curvature, according to its Cartan scalars, only at $r=0$ for all $n$ except $n=\pm 1$, i.e., $\sigma=0$ and $1/2$. Thus the Weyl class metric has a singularity along the axis $r=0$, then we can say that this spacetime is generated by an infinite uniform line source for $0<\sigma<1/2$.

Considering the static limit for the Weyl class metric, when $n=1$ ($\sigma=0$) and $b=c=0$, we have from (\ref{L2})
\begin{equation}
ds^2=d\tau^2-dr^2-dz^2-\frac{r^2}{a}d\phi^2, \label{L24}
\end{equation}
which is the limit of the LC metric when $\sigma=0$. In Section  2 it has been pointed that it generates strings when $a>1$ with mass per unit length $\mu=\delta/4$ and how $a$ is directly linked to the gravitational analog of the Aharonov-Bohm effect \cite{Dowker}.

Considering $c=0$ and $n=1$ ($\sigma=0$) in (\ref{L2}) we have
\begin{equation}
ds^2=d\tau^2+2b\sqrt ad\tau d\phi-dr^2-dz^2-\left(\frac{r^2}{a}-b^2a\right)d\phi^2, \label{L25}
\end{equation}
producing a locally flat spacetime. In this case (\ref{L25}) represents the exterior spacetime of a spinning string along the axis of symmetry \cite{Jensen} with the same mass per unit length $\mu=\delta/4$ but with angular momentum $J$ given by
\begin{equation}
J=-\frac{b\sqrt a}{4}, \label{L26}
\end{equation}
for $a>1$.

It has been shown \cite{Jensen} that a quantum scalar particle moving around a spinning cosmic string as given by (\ref{L25}), exhibits a phase factor proportional to $J$, in its angular momentum. It is a reminiscence of the Aharonov-Bohm effect. It is also worth mentioning that even if $b=0$, an Aharonov-Bohm like effect (of a different kind) appears (as commented in the static case), since the angular momentum spectrum differs from the usual one, if only $a>1$.

\subsection{Geodesics for the Weyl class}

 {The Weyl class of solutions is given by (\ref{L1}) with (\ref{L2}) and  all  parameters being real, as shown above. Then,   for circular geodesics  {\cite{HerreraB}} ${\dot r}={\dot z}=0$, we find  that 
$f_{,r}{\dot t}^2-2k_{,r}{\dot t}{\dot\phi}-l_{,r}{\dot\phi}^2=0$, where the dot stands for differentiation with respect to an affine parameter $s$.  The geodesic angular velocity is defined by $\omega\equiv {\dot\phi}/{\dot t}$,
 and the velocity of the test particle
has only two nonzero components, } $W^t=k\omega/(f^{3/2}-\sqrt fk\omega)$ and $W^{\phi}=\sqrt f\omega/(f-k\omega)$. For $\omega$ and $W$ we obtain
\begin{eqnarray}
\label{LL27}
\omega=\frac{c\pm n\omega_0}{n-b(c\pm n\omega_0)}, \; 
W=\frac{\left(\frac{cr^n}{na}\pm W_0\right)}{\left(1\pm\frac{cr^n}{na}W_0\right)}, ~~
\end{eqnarray}
where $\omega_0$ is the LC angular velocity and $W_0$ is the LC tangential velocity,
\begin{eqnarray}
\label{LL28}
\omega_0^2=\left(\frac{1-n}{1+n}\right)\frac{a^2}{r^{2n}}, \;\;\; 
W_0=\frac{1-n}{1+n}. 
\end{eqnarray}
We note that $\omega$ and $W$ vanish for $\omega_0=\mp c/n$ and $W_0=\mp cr^n/na$, respectively,  {which are equivalent to say that  the free particle in the present case  is simply static. }
This could come about if the ``centrifugal repulsion" balances the gravitational attraction.

The geodetic motion of a particle along the axis of symmetry $z$ for metric (\ref{L1}) produces
\begin{equation}
{\ddot z}=\frac{1-n^2}{2}\frac{{\dot r}{\dot z}}{r}. \label{LL29}
\end{equation}
It is interesting to note that for this geodesic the parameters $b$ and $c$, due to the stationarity of spacetime, do not appear in (\ref{LL29}) and in fact it is indistinguishable from its static limit,  the LC spacetime. There is a force that tends to damp the motion along the axis, $\ddot z<0$, whenever the particle approaches the axis, $\dot r<0$, and reverses this tendency in the opposite case. In the flat case, $n=1$ or $\sigma=0$, such an effect vanishes, exposing its non-Newtonian nature.

For quasi-spherical objects it has been shown  {\cite{Herrera1B}} that positive radial acceleration can be produced along its axis of symmetry.

It is also worth noticing that non-Newtonian forces, parallel to the $z$ axis, also appear in the field of axially symmetric rotating bodies   \cite{Bonnor1B}. However, the force parallel to $z$ in   \cite{Bonnor1B}, 
unlike  the current case,} is directly related to the spin of the source. For the Kerr black hole,  { it is shown that particles, produced by a Penrose process, can be ejected from the ergosphere surface, covering the black hole through  repulsive gravitational fields. } In this case too, unlike in the Lewis spacetime, the gravitational repulsion is created by the spin of the Kerr black hole. These ejected particles are highly collimated and might be a mechanism for the observed extragalactic jets \cite{Pacheco,Gariel1,Gariel2,Gariel4,Gariel3}.

\subsection{The parameters of the Lewis class}

Using the transformation
\begin{equation}
d\phi=d{\bar\phi}+\omega dt, \;\; \omega=-\frac{k}{l}, \label{L27}
\end{equation}
the metric (\ref{L12}) can be diagonalized. In order to have an integral coordinate transformation $\omega$ must be constant; therefore, from (\ref{L12}), $m=0$. This implies, from (\ref{L7}), that $n=0$ and $c=0$. Thus the line element becomes
\begin{equation}
ds^2=-\frac{r}{a_2^2-b_2^2}dt^2-\frac{dr^2+dz^2}{\sqrt r}-r(a_2^2-b_2^2)d{\bar\phi}^2. \label{L28}
\end{equation}
This is a particular case of the static LC metric with the energy density per unit length $\sigma$ given by (\ref{L23})  {being equal to} $1/4$. Nevertheless, the transformation (\ref{L27}) is not global, since the new coordinate $\bar\phi$ ranges from $-\infty$ to $\infty$, instead of ranging from $0$ to $2\pi$ \cite{Silva,Stachel}.

Considering,  {as in the case of the Weyl class,}  a rigidly rotating anisotropic fluid,  one of the Einstein field equations can be integrated producing (\ref{L19}). With (\ref{L12}) the matching conditions given by Darmois \cite{Darmois} yield
\begin{equation}
\xi=-m(a_1^2+b_1^2), \label{L29}
\end{equation}
and so
\begin{equation}
c=-\frac{\xi}{2}. \label{L30}
\end{equation}
Hence, in order to have the rotation equal to zero, i.e. $\xi=0$, we need $m=0$ since $a_1^2+b_1^2\neq 0$.

Observe the difference, at this point, between the Weyl class and the Lewis class. In the latter the vanishing of the rotation yields a locally LC spacetime, whereas in the former the vanishing of the rotation does not necessarily imply that the metric can be reduced  {either globally  or locally to a static spacetime.}

For the Cartan scalars that produce the local characteristics of a metric, we have the following results for the Lewis class.
 {In the Lewis class as in the Weyl class, } only the constant $n$ appears in the Cartan scalars. Nevertheless, now $n$ must be substituted by its complex value (\ref{L7}) $im$. However, contrary to the Weyl class, the Cartan scalars for the Lewis class are distinguishable from the LC metric, except for $m=0$. Furthermore, there is no value of $m$ for which the Cartan scalars are all zero, implying at once that the Lewis class does not include the Minkowski spacetime as a special case. This fact implies too that there must be a lower limit to the energy per unit length of its source. The Cartan scalars impose a  { upper  bound } on the parameter $m$, given by
\begin{equation}
m\leq\sqrt 3, \label{L31}
\end{equation}
since for larger values of $m$ than this, the singularity is at $r=\infty$ and not at $r=0$.

\subsection{Sources producing Lewis spacetime}

In a fine paper, well ahead of his time, as observed by Bill Bonnor  {\cite{BonnorB}},  in 1937 van Stockum \cite{Stockum} completely solved the problem of a rigidly rotating infinitely long cylinder of dust, including the application of adequate boundary conditions. The solution is a remarkable one. The metric for the interior is simple and unique, depending on one parameter, $w$ in our notation. But for the vacuum exterior, $r>R$, where $R$ is the coordinate radius of the cylinder, there are three cases depending on the mass per unit length of the interior. For the metric (\ref{L1}) we have the following results.

Case $wR<1/2$:
\begin{eqnarray}
\label{L32}
f&=&-r\left[2\beta\cosh(2N\ln r)+\frac{\alpha^2+\beta^2}{\alpha}\sinh(2N\ln r)\right], \nb\\ 
k&=&-r\left[\cosh(2N\ln r)+\frac{\beta}{\alpha}\sinh(2N\ln r)\right],\nb\\ 
e^{\mu}&=& \lambda\left(\frac{r}{R}\right)^{(4N^2-1)/2}, \;\; 
l= \frac{r}{\alpha}\sinh(2N\ln r), 
\end{eqnarray}
with,
\begin{eqnarray}
\label{L36}
N&=& \frac{1}{2}(1-4w^2R^2)^{1/2}, \;\; 
\alpha=\frac{(1-4w^2R^2)^{1/2}}{2w^3R^4}, \nb\\ 
\beta&=& -\frac{1-2w^2R^2}{2w^3R^4}, \;\; 
\lambda=e^{-w^2R^2}. 
\end{eqnarray}

Case $wR>1/2$:
\begin{eqnarray}
\label{L40}
f &=& r\left[2\beta\sin(2N\ln r)+\frac{\alpha^2-\beta^2}{\alpha}\cos(2N\ln r)\right], \nb\\ 
k &=& r\left[\sin(2N\ln r)-\frac{\beta}{\alpha}\cos(2N\ln r)\right], \nb\\ 
e^{\mu} &=& \lambda\left(\frac{r}{R}\right)^{-(4N^2+1)/2}, \;\; 
l= \frac{r}{\alpha}\cos(2N\ln r), 
\end{eqnarray}
with,
\begin{eqnarray}
\label{L44}
N&=& \frac{1}{2}(4w^2R^2-1)^{1/2},\;\; 
\alpha=\frac{(4w^2R^2-1)^{1/2}}{2w^3R^4},\nb\\ 
\beta&=& \frac{2w^2R^2-1}{2w^3R^4}, \;\; 
\lambda=e^{-w^2R^2}. 
\end{eqnarray}

For the case $wR=1/2$ one obtains the relations  from the limits either from $wR<1/2$ or $wR>1/2$ which are equal. The solution for $wR<1/2$ belongs to the Weyl class and its real parameters $n$, $a$, $b$ and $c$ assume the values
\begin{eqnarray}
\label{L48}
n&=&(1-4w^2R^2)^{1/2},\;\; 
a=\frac{(\alpha-\beta)^2}{2\alpha}, \nb\\ 
b&=& \pm\frac{1}{\alpha-\beta},\;\; 
c=\frac{\alpha^2-\beta^2}{\alpha}N. 
\end{eqnarray}
The solution $wR>1/2$ belongs to the Lewis class and its real parameters $m$, $a_1$, $a_2$, $b_1$ and $b_2$ assume the values
\begin{eqnarray}
\label{L52}
m&=& (4w^2R^2-1)^{1/2}, \; \; 
a_1=\frac{\beta}{b_1},\;\; 
a_2=-\frac{1}{b_1}, \nb\\ 
b_1&=& (-\alpha)^{1/2}, \;\; 
b_2=0. 
\end{eqnarray}

For the Weyl class we have the Newtonian mass per unit length given by $\sigma=(1-n)/4$, which implies, for case $wR<1/2$, from (\ref{L48}),
\begin{equation}
\sigma=\frac{1}{4}\left[1-(1-4w^2R^2)^{1/2}\right]. \label{L57}
\end{equation}
Hence (\ref{L57}) establishes a lower limit for $\sigma$ in the Lewis class and being $\sigma=1/4$. This value $\sigma=1/4$ is the frontier between the Weyl class metric and the Lewis class metric, at least for the rotating dust solution obtained by van Stockum \cite{Stockum}.

For the Lewis class metric the Cartan scalars, as it was remarked, do not admit the Minkowski spacetime. This is in accordance with the existence of a lower limit for $\sigma$ in the van Stockum solution $wR>1/2$, since with a lower limit the source cannot be made a vacuum, and therefore the exterior solution cannot be Minkowski.

The Cartan scalars also impose a  {upper bound }  on the parameter $m$, given by
\begin{equation}
m\leq\sqrt 3, \label{L58}
\end{equation}
since for values of $m$ larger than this, the singularity is at $r=\infty$ and not in $r=0$. When we substitute this value in (\ref{L52}), considering the equality, we have $wR=1$.

Van Stockum   {solution} is studied {at length} in  {\cite{BonnorB}}, its properties concerning gravitoelectric and gravitomagnetic fields in  {\cite{Bonnor1B}}, its confinement properties in \cite{Opher}, its extension to nonrigid rotation in \cite{Bonnor2}. A range of stationary cylindrically symmetric perfect fluids sources are presented in \cite{Stephani},   a rigidly rotating dust coupled with a thin disk is found in \cite{dAW00}, and an  anisotropic cylindrical stationary source can be found in \cite{Debbasch}.

  Another source that consists of an infinitely rotating thin shell was constructed in \cite{PW00}, in which the spacetime inside the shell is a constantly rotating Minkowski spacetime so the symmetry axis is well defined, and the shell satisfies all the energy conditions, the weak, strong and dominant \cite{HE73}. Later, this problem was studied in details in \cite{DHSW02}, and shown explicitly that the
parameters $a$ appearing in (\ref{L2}) and $\sigma$ defined in (\ref{L23})  must be restricted to the ranges   $0 \le \sigma \le 1/4$ for $a > 0$ or $ 1/4 \le \sigma \le 1/2$ for $a < 0$, in order to have no CTC's outside of the shell and meanwhile all the energy conditions hold.

\subsection{Rotation and translation of cylinders}

In GR  we show that the vacuum field produced by a rotating mass cylinder is mathematically closely related to the field produced by a translating mass cylinder along its axis of symmetry. Nonetheless, its physical and geometrical properties differ significantly since the relativistic frame dragging for rotation and translation physically differ considerably.

We assume the general cylindrically symmetric metric with its source translating parallel to its axis of symmetry given by
\begin{equation}
ds^2=Adt^2-2Kdtdz-Bd\rho^2-Cdz^2-B\rho^2d\phi^2, \label{L59}
\end{equation}
with the usual properties of its coordinates and $A$, $K$, $B$ and $C$ functions only of $\rho$. Making $\rho=e^r$ and after rescaling, the metric (\ref{L59}) can be written as
\begin{equation}
ds^2=fdt^2-2kdtdz-e^{\mu}(dr^2+d\phi^2)-ldz^2. \label{L60}
\end{equation}
The general vacuum solution  for (\ref{L60}), with the $z$ and $\phi$ interchanged is the stationary Lewis metric. So, the vacuum solution corresponding to (\ref{L59}) is simply the Lewis solution with the coordinates $z$ and $\phi$ interchanged. Hence the metric coefficients in (\ref{L60}) are the same as in (\ref{L2}).

In spite of the mathematical similarity between the vacuum solutions for the fields produced by rotating and translating cylinder filled with perfect fluid they differ substantially. Unlike the rotating case, the translating cylinder cannot be filled with pressure free dust, as there is nothing here equivalent to a centrifugal force that would prevent the matter collapsing to the axis. The pressure must therefore be nonzero. Furthermore, unlike the rotating case where matter can be rigidly rotating, 
 which means with shear free rotation, the translating matter case if translating rigidly  can always be transformed to a frame where the system is static  \cite{GriffithsB}.

The field of a cylinder of matter that is in translational motion along its length has not been  {studied in detail,} and any differences with the static case are unknown. It is therefore of interest to determine whether or not frames are dragged by motion along the cylinder in a way similar to that in which they are dragged around it.


%
\section{Lewis vacuum spacetime coupled with a cosmological constant} 
%
%
\renewcommand{\theequation}{7.\arabic{equation}} \setcounter{equation}{0}

\subsection{Stationary cylindrical vacuum spacetime with cosmological constant}

Santos \cite{Santos} in 1993 gave a set of solutions for stationary cylindrically symmetric spacetimes of the Lewis \cite{Lewis} form, published in 1932, with a cosmological constant.
Krasinski \cite{Krasinski} had previously given these solutions, but in a different form by using coordinates in which the static limit is hard to obtain. The solutions obtained by Santos are reexamined by MacCallum and Santos \cite{MacCallum} in 1998 where they make some remarks on the problem of the definition of cylindrical symmetry and they find too the essential parameters of the solutions.

We consider the cylindrically symmetric spacetime metric in the form
\begin{equation}
ds^2=fdt^2-2kdtd\phi-dr^2-e^{\mu}dz^2-ld\phi^2, \label{lc1}
\end{equation}
where $f$, $k$, $\mu$ and $l$ are functions of $r$ only. The ranges of the coordinates are $-\infty<t<\infty$   {for} the time coordinate, $0\leq r<\infty$ for the radial coordinate, $-\infty<z<\infty$ for the axial coordinate and $0\leq\phi\leq 2\pi$ for the angular coordinate with the hypersurfaces $\phi=0$ and $\phi=2\pi$ being identified. The general solution for    Einstein's equations  
for (\ref{lc1}), in the notation of \cite{MacCallum}, is
\begin{eqnarray}
\label{lc2}
f &=& G_0^{2/3}\Big\{T_1^2\exp[2(p_0-1/3)U]\nb\\
&& -X_1^2\exp[2(p_3-1/3)U]\Big\}, \nb\\ 
k &=& G_0^{2/3}\Big\{X_1X_2\exp[2(p_3-1/3)U]\nb\\ && -T_1T_2\exp[2(p_0-1/3)U]\Big\}, \nb\\ 
e^{\mu} &=& G_0^{2/3}\Big\{Z^2\exp[2(p_2-1/3)U]\Big\}, \nb\\ 
l&=& G_0^{2/3}\Big\{X_2^2\exp[2(p_3-1/3)U]\nb\\
&& -T_2^2\exp[2(p_0-1/3)U]\Big\}, 
\end{eqnarray}
where the distinct metric solutions are given by
\begin{eqnarray}
G_0=\frac{\cosh({\sqrt{3|\Lambda|}r)}}{\sqrt{3|\Lambda|}}, \;\; \Lambda<0, \label{lc5}\\
G_0=\frac{\exp(\sqrt{3|\Lambda|}r)}{\sqrt{3|\Lambda|}}, \;\; \Lambda<0, \label{lc6}\\
G_0=\frac{\sinh({\sqrt{3|\Lambda|}r)}}{\sqrt{3|\Lambda|}}, \;\; \Lambda<0, \label{lc7}\\
G_0=r, \;\; \Lambda=0, \label{lc8}\\
G_0=\frac{\sin({\sqrt{3\Lambda}r)}}{\sqrt{3\Lambda}}, \;\; \Lambda>0, \label{lc9}
\end{eqnarray}
with, respectively,
\begin{eqnarray}
U=\arctan|\sinh({\sqrt{3|\Lambda|}}r)|, \;\; \Lambda<0, \label{lc10}\\
U=\exp(-{\sqrt{3|\Lambda|}}r), \;\; \Lambda<0, \label{lc11}\\
U=\ln\tanh[(\sqrt{3|\Lambda|}r)/2], \;\; \Lambda<0, \label{lc12}\\
U=\ln r, \;\; \Lambda=0, \label{lc13}\\
U=\ln\tan[({\sqrt{3\Lambda}})/2], \;\; \Lambda>0. \label{lc14}
\end{eqnarray}
The constants $p_j$ with $j=0,2,3$ satisfy
\begin{equation}
\sum_jp_j=1, \;\; \sum_jp_j^2=\frac{2\eta+1}{3}, \label{lc15}
\end{equation}
with $\eta=-1,0,1,1,1$, respectively, for (\ref{lc5}-\ref{lc9}). Note that for $\eta=0$ the only real solution is $p_j=1/3$ for all $j$; other cases with $\eta=0$ and all cases with $\eta=-1$, will have complex values of $p_j$. The resulting general metric for a stationary cylindrically symmetris Einstein space can be assumed with, say, $p_2$ to be real, which leaves with the possibility of complex $p_0$ and $p_3$. The periodic   {character} of $\phi$ can be taken to be specified by the parameters $T_2$ and $X_2$. Hence the essential parameters describing the metric (\ref{lc2}) are $T_2$, $X_2$, $p_2$ and the cosmological constant $\Lambda$. The inessential free parameters $T_1$, $X_1$, $R$ and $Z$ may be required in order to match the metric to an interior by using the appropriate matching conditions. The case $k=0$ in (\ref{lc2}) is reducible to the static cylindrically symmetric   {LT solutions} studied in Section  4. For further analysis of the limits and properties of the metric    {(\ref{lc2})} see \cite{MacCallum}.

 {Before proceeding further, we note that in   \cite{Santos} the above solutions were presented in  a slightly different form. In particularly, the function $G$ was given by,
\bqun
\lb{7.14}
G &=& \begin{cases}
C_1 \cosh\left(\sqrt{3|\Lambda|} \; r \right) + C_2 \sinh\left(\sqrt{3|\Lambda|} \; r \right), & \Lambda < 0,\cr
C_1  + C_2  r, & \Lambda = 0,\cr
C_1 \cos\left(\sqrt{3\Lambda} \; r \right) + C_2 \sin\left(\sqrt{3\Lambda} \; r \right), & \Lambda > 0,\cr
\end{cases}\nb\\
\equn
where $C_{1}$ and $C_2$ are two arbitrary  constants. However, using the transformation $r' = r - r_0$, where $r_0$ is a constant, it can be seen that such given $G$ can be always brought 
to the forms given by  (\ref{lc5})-(\ref{lc9}) by properly choosing $r_0$ \cite{MacCallum}. }

\subsection{A source producing stationary cylindrical vacuum spacetime with cosmological constant}

Bonnor et al.  {\cite{BonnorC}} and Griffiths and Santos  {\cite{GriffithsA}} have identified a family of compound spacetimes for which the interior of an infinite cylinder of finite radius is described by the G\"odel solution \cite{Godel}. They showed that this can be matched to a vacuum exterior obtained by Santos \cite{Santos}. The resulting spacetime is cylindrically symmetric and stationary.

The perfect fluid solution obtained by G\"odel \cite{Godel} can be expressed in comoving coordinates in the cylindrical form
\begin{eqnarray}
ds^2=dt^2+\frac{2\sqrt 2}{\omega}\sinh^2(\omega r)dtd\phi-dr^2-dz^2 \nonumber\\
-\frac{1}{\omega^2}[\sinh^2(\omega r)-\sinh^4(\omega r)]d\phi^2, \label{lc16}
\end{eqnarray}
where $\omega$ is an arbitrary constant and the   {coordinates} $r$, $\phi$ and $z$ have the same range as in (\ref{lc1}). The coordinate $r$ may be interpreted as the proper distance from a regular axis at $r=0$, and the rotational symmetry of the metric (\ref{lc16}) about the axis can be clearly seen. The fluid four velocity is $u_{\alpha}=(u_t,0,0,0)$, and the constant mass density $\rho$ (the pressure $p=0$) and cosmological constant $\Lambda$ are given by
\begin{equation}
8\pi\rho=4\omega^2, \;\; \Lambda=-2\omega^2. \label{lc17}
\end{equation}
This implies that the fluid density is positive and the cosmological constant negative. The metric has no curvature singularities and its form explicitly reduces to a vacuum Minkowski spacetime in the   {slow rotation } limit, $\omega\rightarrow 0$. This solution may be interpreted as describing a homogeneous spacetime with a fluid source that is rigidly rotating. In the Newtonian analogue of such a situation, relative to any point, there will exist a cylinder of finite radius around the axis of rotation on which the particles of the fluid would move at the speed of light. For all points outside this cylinder, the particles of the classical fluid will be moving faster than light. Such a situation is not possible in a relativistic theory, but this feature is reflected in the appearance of   {CTCs}.

In fact, it can be seen from (\ref{lc16}) that curves on which $t$, $r$ and $z$ are constant are either spacelike or timelike according to whether $\sinh^2(\omega r)$ is, respectively, less than or greater than 1. They are null when $\omega r_1=2\sqrt 2$. Moreover, since $\phi=2\pi$ is identified with $\phi=0$, these are   {CTCs} whenever $r$ is larger  than this value, but they are not geodesics. Hence we consider that for the interior cylinder $\omega r<2\sqrt 2$.

Now we consider the matching of an interior cylinder filled with a G\"odel fluid with its cylindrical surface described by its proper radius $0<r_S<r_1$ and for its exterior spacetime (\ref{lc1}) with $r_S<r<\infty$. Both proper radii in (\ref{lc1}) and (\ref{lc16}) are taken to be equal. Then, according to the Lichnerowicz matching conditions, the metric functions and their derivatives must be continuous across the junction $r_S$. By performing these long calculations one obtains the different parameters of (\ref{lc1}) expressed in terms of the radius $r_S$ and the cosmological constant $\Lambda$. It can be shown that asymptotically the spacetime becomes the anti-de Sitter one,   { which is what one might expect}. Furthermore, this stationary system is expected to be qualitatively similar to the van Stockum solution studied in Section  6. For the details of the calculations concerning this model see  \cite{BonnorC} and \cite{GriffithsA}.

  {
\subsection{Geodesic motion, confinement and extragalactic jet formation}
}
 {Ubiquitous extragalactic jets appear in active galaxies and various models have been suggested
for their origin \cite{Ferrari98}. A classic model is that of Blandford and Rees \cite{BR74,BR75}, in which a hot
plasma is assumed to be steadily generated at the center of the galaxy. The central object is
surrounded by a gravitationally confined rotating gas cloud. The thermal pressure of this gas cloud
constrains the outflowing relativistic plasma within two oppositely directed channels. An equilibrium
flow is possible only if the channel shape adjusts so that a nozzle forms where the external
pressure has dropped to one-half its central value. The shape of the channel is assumed to adjust
itself to a de Laval nozzle. This configuration, with a light fluid supporting a heavier fluid in a
gravitational field, is well known to be unstable.  }

 { In the studies of the geodesic motion of the van Stockum solution \cite{Stockum},
Opher, Santos and Wang found that the test particles are always confined inside a finite cylindrical shell, while their 
motion along the symmetry axis is free  \cite{Opher}. This motivated them to  argue  that the confinement of 
the test particles in the radial direction might
provide another source for the extragalactic jet formation. Their arguments go as follows: The
gravitational field produced by jets usually is negligible compared with the one produced by the
matter at the center of the galaxy. Thus, to the first-order approximation, it is sufficient to model
those jets as made of test particles. Also, almost all the galaxies are rotating, as a first step, one can
model the center of a galaxy by a rotating cylinder. This approximation seems reasonable as long
as the gravitational field in the middle of the rotating galaxy is concerned, though they admitted that it
is indeed highly simplified. Assuming that the above simplified model can capture some essence
of physics, one can see that the confinement can be related to the jets. }

 { Certainly, to have the above idea really work, more realistic models should be built. Recently, such studies were extended
to the Kerr solution  \cite{Pacheco,Gariel1,Gariel2,Gariel4,Gariel3}, and found  that particles can be ejected from the ergosphere surface through  the repulsive gravitational fields,
which are  created by the spin of the Kerr black hole. Again, these ejected particles are highly collimated and might provide  a mechanism for the observed extragalactic jets.}

  {In the following, using  the five-parameter class of  Wright solutions  \cite{Wri65} as an example, we shall show the generic properties of the confinements of test particles in
rotating spacetimes. For this purpose, 
 let us first note that the G\"odel solution given by (\ref{lc16}) belongs to one of the two classes of solutions found by Wright in \cite{Wri65}. The other class of Wright
solutions contains five free parameters, and can be cast in the form \cite{PSW96}, 
\beq
\lb{7.3.1}
ds^2 = dt^2  +  2\beta dt d\phi - \delta^2 dr^2 - \gamma^2d\phi^2 - \alpha^2dz^2,
\eeq
where
\bqun
\lb{7.3.2}
\alpha &=& e^{-r^2}, \nb\\
\beta &=& \pm \left(\frac{A}{B}\right) r^2 + C, \nb\\
\gamma^2 &=& \frac{A^2}{2B^2}r^2 - \frac{\Lambda}{B^2}e^{-2r^2} + \frac{D}{B^2}  - \left(\frac{A}{B}r^2 \pm C\right)^2,\nb\\
\delta^2 &=& \frac{4r^2e^{-2r^2}}{D +\frac{1}{2}A^2r^2 - \Lambda e^{-2r^2}},
\equn
where $A, B, C$ and $D$ are  four integration constants, and $\Lambda$ is the cosmological constant. The energy density and four-velocity of the dust fluid are given by
\beq
\lb{7.3.3}
\rho = \frac{1}{\kappa} \left(A^2e^{2r^2}+ 2\Lambda\right), \quad u^{\mu} = \delta^{\mu}_{t},
\eeq
where $\kappa$ is the Einstein coupling constant.  In \cite{PSW96}, it was shown that three of these five parameters are completely fixed by the  elementary flatness condition  (\ref{1.4})
and the existence of  the symmetry axis (\ref{1.5}),  which lead to 
 \bqu
 \lb{7.3.4}
 B = a^2 + \Lambda, \quad C = 0, \quad D = \Lambda,
 \equ
 where $a \equiv A/2$. Setting
 \beq
 \lb{7.3.5}
 u = 2 r^2, \quad \varphi = \frac{\phi}{a^2 + \Lambda},
 \eeq
 it can be shown that the above Wright solutions with the conditions given by (\ref{7.3.4}) reduce precisely to the Lanczos ones presented  in \cite{Lanczos}, 
 which represents the gravitational field of a rigidly rotating dust cylinder coupled with a cosmological constant, $\Lambda$. }

  { Then, the corresponding Lagrangian for test particles reads,
\bqun
\lb{7.3.6}
{\cal{L}} \equiv \frac{1}{2}\left(\dot{t}^2  +  2\beta \dot{t} \dot{\phi} - \delta^2 \dot{r}^2 - \gamma^2\dot{\phi}^2 - \alpha^2\dot{z}^2\right),
\equn
where $\dot{t} \equiv dt/ds$, etc., with $s$ being the affine parameter. Hence,  the geodesic motions along $z$- and $r$-axes are given by
 \cite{PSW96}, 
\bqun
\lb{7.3.7a}
\dot{z} &=& P_z e^{x^2}, \\
\lb{7.3.7b}
\dot{x}^2 &=& a^2e^{x^2}\left[V_0 - V(x)\right],  
\equn
where $x \equiv \sqrt{2} \; r$, and 
\bqun
\lb{7.3.8a}
V(x) &\equiv& E^2 x^2 + P_z^2 e^{x^2} + \frac{\left(a^2+\Lambda\right)^2}{a^2x^2} P_{\phi}^2 \nb\\
&& -   \frac{\Lambda\left(1 - e^{-x^2}\right)} {a^2x^2}\left(E^2 - \epsilon - P_{z}^2e^{x^2}\right), \\
\lb{7.3.8b}
V_0 &\equiv& E^2 - \epsilon - 2EP_{\phi} \frac{a^2+\Lambda}{a},
\equn
where $\epsilon = 0, 1, -1$, respectively for null, timelike and spacelike geodesics, and $E$, $p_{\phi}$ and $P_z$ are the constants of motion, defined by
\bqun
\lb{7.3.9}
E &\equiv& \frac{\partial {\cal{L}}}{\partial \dot{t}}, \quad P_{z} \equiv \frac{\partial {\cal{L}}}{\partial \dot{z}},
\quad P_{\phi} \equiv \frac{\partial {\cal{L}}}{\partial \dot{\phi}}.
\equn
It is interesting to note that $V(x)$ takes the form,
\beq
\lb{7.3.10}
V(x) \simeq E^2 x^2 + P_{z}^2 \left(1 + \frac{\Lambda}{a^2x^2}\right) e^{x^2},
\eeq
as $x \gg 1$. Thus, it is monotonically increasing when $x$ becomes sufficiently large.  Since $\dot{x}^2 \ge 0$, from (\ref{7.3.7b}) and (\ref{7.3.10}) we can see that there must exist a maximal value, say, $x_{max}$, so that 
(\ref{7.3.7b}) has solution for $x \le x_{max}$. That is, the geodesics are always confined inside a finite radius $r \le r_{max} \equiv x_{max}/\sqrt{2}$. However, 
the motion along the symmetry axis is free, as it can be seen from (\ref{7.3.7a}). }

  { The above properties were noticed when studying the G\"odel space-time,  independently by Kundt \cite{Kundt56}  and 
Chandrasekhar and Wright \cite{CW61}, and later by Opher, Santos and Wang \cite{Opher}  in the study of   the van Stockum solution \cite{Stockum}.
Combining these studies with   the ones carried out above, as well as  the ones carried out  in the Kerr spacetime  \cite{Pacheco,Gariel1,Gariel2,Gariel4,Gariel3}, one can see  that in rotating 
spacetimes confinement occurs  generically in the radial direction,
whereas the motion in the axial direction is free. Therefore, it is quite plausible to argue that such   confinements  might provide a mechanism for  extragalactic jet formation.  }

  {As a matter of fact, in \cite{Gariel3} the kinematic behavior of extragalactic jets was investigated. By numerical calculations, it was found that a transition from
non-relativistic to ultrarelativistic speeds at subparsec scales exists. This transition
comes sooner and more abruptly than in models based on magnetic paradigm,
which indicates that one needs a weaker magnetic field to explain observed synchrotron
radiation. This ejection phenomenon  was attributed  to the repulsive effect of the gravitomagnetic Kerr field.}

%
\section{Cylindrical wormholes} 
%
\renewcommand{\theequation}{8.\arabic{equation}} \setcounter{equation}{0}

 The possibility of unusual spacetime geometries was   {noticed} by researchers 
  as soon as the gravitational field was identified with curvature in Einstein's 
  theory. Flamm \cite{wh1} already in 1916 noticed that 
  the spatial part of Schwarzschild's spacetime describes a kind of tunnel between 
  two flat asymptotic regions, a configuration now called a wormhole after 
  the suggestion of Wheeler \cite{wh2} who also introduced
  the concept of spacetime foam on extremely small length scales.
  Nowadays, wormholes, although not yet discovered in Nature or constructed
  artificially, are one of actively discussed subjects in gravitational physics 
  both due to their properties of interest (e.g., as possible distance shortcuts, 
  time machines, agents in quantum entanglements, etc.) and in connection 
  with such fundamental issues as energy conditions and topological censorship
  \cite{wh3,wh4,wh5,wh5a}. Possible observable effects of  \whs\ in astronomy are 
  studied in  a number of papers \cite{wh6,wh7,wh8,wh9}, see also references 
  therein. More general reviews can be found in \cite{wh4,wh10,wh11,wh12}.

  One of the main problems in \wh\ physsics in GR is the necessity of 
  ``exotic'', or phantom matter for \wh\ existence, violating the Weak Energy 
  Condition (WEC) and its part, the Null Energy Condition (NEC), at least 
  near a \wh\ throat (the narrowest place). This result is known for a long time 
    {\cite{wh3,wh12,wh12a}}, at least for static \whs\ with compact throats of finite area
  and for \asflat\ spacetimes.\cite{wh5}.

  In theories extending GR, it is possible to find \wh\ solutions 
  without exotic matter, replacing it with geometric or field quantities 
  that are absent in GR, e.g., in Einstein-Cartan theory \cite{wh13,wh14},  
  theories involving the Gauss-Bonnet invariant \cite{wh15}, multidimensional 
  models  {\cite{wh16,wh17,ABK18,KLS15,SK16},} etc. However, if we adhere to GR, which is 
  justified by its excellent experimental status, then a question of interest is 
  whether or not it is possible to obtain phantom-free \whs\ by abandoning 
  some conditions of the firmly established theorems. Meanwhile, 
  asymptotic flatness is a necessary requirement if a \wh\ is expected 
  to be observable from distant regions with small curvature.

  Some phantom-free \wh\ solutions in GR can be recalled, beginning with 
  Kerr's metric with a large angular momentum and Zipoy's static vacuum 
  solution \cite{wh20} as well as their electromagnetic and scalar 
  extensions \cite{wh21, wh22}, but all of them possess a naked ring singularity 
  that bounds a disk playing the role of a throat.  Therefore, despite their 
  asymptotic flatness, they can hardly be welcomed. Others \cite{wh23,wh24}, 
  being phantom-free, lack asymptotic flatness. All such shortcomings may 
  probably be interpreted as signs of the topological censorship. 
  
  Cylindrical symmetry seems to be a way to avoid topological censorship 
  because, even assuming asymptotic flatness in the radial ($r$) direction, we 
  have no such flatness in the longitudinal one, $z$, since at large $|z|$ spacetime   
  at given $r$ remains equally curved as it was at small $r$. And, provided 
  the system is $t$-independent (stationary), the same is true for the null 
  ($z,t$) direction. So, there is hope to obtain phantom-free \whs, \asflat\ in 
  the radial direction.   
  
  Among numerous known \cy\ \wh\ solutions in GR (see, among others,
  \cite{wh25, wh26, wh27, wh28, wh29, wh30} and references therein) 
  many are phantom-free but none are \asflat, for the same reason for which the 
    {LC} vacuum solution is only \asflat\, if it is   {completely flat, that is, only when the entire spacetime is Minkowiski.} To solve the problem, 
  the following construction was suggested \cite{wh26}: a central \wh\
  region with a throat, surrounded on both sides by Minkowski regions, matched 
  through thin cylindrical shells $\Sigma_-$ and $\Sigma_+$. 
  These shells will possess nonzero densities and pressures, their    {SETs}
   being determined by jumps of the extrinsic curvature according 
  to the well-known Darmois-Israel formalism. A phantom-free model will be obtained 
  if in the internal region and on the two surfaces all SETs respect the WEC.
  
  In this  {section} we will describe the conditions on the geometry and SETs of 
  matter for obtaining static or stationary (rotating) \cyl\ \wh\ configurations and 
  present some particular configurations. Among them are two examples of rotating
  phantom-free \wh\ models, \asflat\ on both sides of the throat.

\bigskip
\subsection{Wormhole definitions and existence conditions}

  We begin with   {the} stationary \cyl\ metric
\bearr                                                  \label{wh-1}
         ds^2 = \e^{2\gamma(x)}[ dt - E(x)\e^{-2\gamma(x)}\, d\varphi ]^2
       - \e^{2\alpha(x)}dx^2 
\nnn \cm\cm	- \e^{2\mu(x)}dz^2 - \e^{2\beta(x)}d\varphi^2,
\ear
  where $x$, $z\in \R$ and $\varphi\in [0, 2\pi)$ are the radial, longitudinal and 
  angular coordinates. The coordinate $x$ is specified up to a reparametrization
  $x \to f(x)$, so its range depends on both its choice (the ``gauge'') and the
  geometry itself. It differs from (6.1) in notations ($f \mapsto \e^{2\gamma}$,
  $k \mapsto E$, $\mu \mapsto 2\mu$, $l \mapsto \e^{2\beta} - E^2 \e^{-2\gamma}$)
  and in that we here preserve gauge freedom whereas in (6.1) the gauge 
  (in the present notations) is chosen as $\alpha =\mu$. The quantity $r = \e^\beta$
  has the meaning of circular radius. The off-diagonal component $E$ corresponds
  to rotation, unless $E\e^{-2\gamma} = \const$, in which case one can make $E=0$
  by redefining (resynchronizing) the time coordinate. If $E=0$, the metric \rf{wh-1}
  is static and coincides with (A.1).
  
  The metric \rf{wh-1} is said to describe a {\it \wh}\ if either (i) the circular radius 
  $r(x) = \e^{\beta(x)}$ has a regular minimum (the corresponding surface 
  $x=\const$ is called an {\it $r$-throat}) and is large or infinite far from this minimum or 
  (ii) the same is true for the area function $a(x) = \e^{\mu+\beta}$ 
  (its minimum defines an $a$-throat)  \cite{wh25,wh26}. For certainty, we choose
   the definition connected with the radius $r(x)$, looking more evident: while
   moving to smaller $r$, we approach a would-be symmetry axis $r=0$, 
   but, instead, reach a minimum of $r$ and observe its subsequent growth,
   paying no attention to what happens to the $z$ scale. 
  
   Asymptotic flatness in some limit of $x$ implies $r \to \infty$, finite limits of $\gamma$, 
   $\mu$ and $E$ (thus no rotation) and $\e^{-\alpha}r' \to 1$ to provide a Euclidean
   shape of large circles. It is clear that if a \wh\ is \asflat\ on both extremes of the $x$ range,
  it contains both $r$- and $a$-throats.
  
   In the static case, assuming the most general SET of matter compatible with the 
   symmetry,
\beq                  \label{wh-2}
		T\mN = \diag (\rho, - p_x, -p_z, -p_\phi)		
\eeq      
   and using the expressions (A.2) for the Ricci tensor components, it is easy to show
   \cite{wh25} that at an $a$-throat near a minimum of $\e^{\beta+\mu}$ one should 
   inevitably have
\beq  		\label{wh-3}
		\rho < p_x \leq 0,
\eeq      
  which means the necessity of a negative density, violating the WEC. Thus static, twice
  \asflat\ \whs\ are impossible. On the other hand, an $r$-throat only implies  
\beq  		\label{wh-4}  
		\rho - p_x - p_z + p_\phi < 0, 
\eeq 
  which does not necessarily violate the WEP. Though, if $p_z = p_\phi$ (as, for 
  example, in isotriopic fluids), \rf{wh-4} reads $\rho - p_x < 0$, violating the dominant energy
  condition if $\rho > 0$. 
  
  Thus in the static case we can have phantom-free \wh\ solutions in the above sense 
  but without   {two-side} asymptotic flatness.
    
  Apart from asymptotic flatness, other restrictions also follow from combinations of 
  \rf{wh-8}, for example, the following no-go theorem \cite{wh25}:
    
 {\sl   
   A static, cylindrically symmetric wormhole, mirror-symmetric with respect to its throat, 
   cannot exist in GR with matter whose SET satisfies the conditions $T^2_2 = T^3_3$ and 
   $T^0_0 \geq 0$.   }

  This concerns, in particular, isotropic fluids with any equations of state and scalar fields
  $\Phi(x)$ with a variety of Lagrangians, such as $L_s = (\d \Phi)^2 - 2V(\Phi)$ with any
   potentials $V$ and more general $k$-essences  with $L_s = f(X, \Phi)$ where        
  $X = (\d \Phi)^2$ and $f$ is an arbitrary function.
  
  What changes if there is rotation? To make it clear, we note that rotation 
  in spacetime with the metric \rf{wh-1}    
  can be characterized by the angular velocity $\omega(x)$ 
  of a congruence of timelike curves defined as \cite{wh26, wh31, wh32}
\beq                                                  \label{wh-5}
          \omega = \half (E\e^{-2\gamma})' \e^{\gamma-\beta-\alpha},
\eeq
    {where a prime stands for $d/dx$}. Furthermore, in our reference frame is comoving   {with}
  matter in its rotation, then the SET component $T^3_0 = 0$, hence (due to 
  Einstein's equations) the Ricci tensor component 
  $R_0^3 \sim (\omega \e^{2\gamma+\mu})' = 0$, and \cite{wh26}
\beq       	      					\label{wh-6}
	\omega = \omega_0 \e^{-\mu-2\gamma}, \cm \omega_0 = \const.
\eeq
  Consequently, from \rf{wh-5} we find, 
\beq                          \label{wh-7}
	E(x) = 2\omega_0 \e^{2\gamma(x)} \int \e^{\alpha+\beta-\mu-3\gamma}dx.
\eeq
  The nontrivial components of    {Einstein's equations 
  in the form  $R\mN = \kappa \tT\mN = \kappa(T\mN -\half \delta\mN T)$}
   are
      {
\bearr                 \label{wh-8}
      R^0_0 = \e^{-2\alpha}\left[\gamma'' + \gamma'(\sigma' -\alpha')\right] + 2\omega^2 
      	= \kappa \tT^0_0,
\nnnv      
      R^1_1 = \e^{-2\alpha}\left[\sigma'' + \sigma'{}^2 - 2U - \alpha'\sigma'\right]- 2\omega^2
      		= \kappa \tT^1_1,
\nnnv      
      R^2_2 = \e^{-2\alpha}\left[\mu'' + \mu'(\sigma' -\alpha')\right] = \kappa \tT^2_2,  
\nnnv      
      R^3_3 = \e^{-2\alpha}\left[\beta'' + \beta'(\sigma' -\alpha')\right] - 2\omega^2 = \kappa \tT^3_3,  
\nnnv
      R^0_3 = G^0_3 = E \e^{-2\gamma}\left(R^3_3 - R^0_0\right) = \kappa \tT^0_3, 
\ear
}
 where we use the notations
\beq               \label{wh-9}
            \sigma = \beta + \gamma + \mu, \qquad
            U = \beta'\gamma'  + \beta'\mu' + \gamma' \mu'.
\eeq     

  From \rf{wh-8} it follows \cite{wh26} that the diagonal components of the Ricci
  ($R\mN$) and Einstein ($G\mN$) tensors split into those for the static metric 
  \rf{wh-1} with $E=0$ plus contributions involving $\omega$: 
\bearr 		\label{wh-10}
    		R\mN = {}_s R\mN + {}_\omega R\mN, \quad\
	  {{}_\omega R\mN = \omega^2 \diag (2, -2, 0, -2)},  
\nnnv     		
		G\mN = {}_s G\mN + {}_\omega G\mN,  \quad\
   	  {{}_\omega G\mN = \omega^2 \diag (3, -1, 1, -1)},  
\nnn   	
\ear
  where ${}_s R\mN$ and ${}_s G\mN$ are the static parts (see (A.2)). The tensors 
  ${}_s G\mN$ and ${}_\omega G\mN$ (each separately) obey the conservation law 
  $\nabla_\alpha G^\alpha_\mu =0$ in terms of the static metric. Therefore,
  ${}_\omega G\mN/\kappa$ behaves as one more SET with exotic properties 
  (the effective energy density is $ -3\omega^2/\kappa <  0$) that favors the  
  existence of wormholes, as confirmed by a number of examples \cite{wh26,wh27,wh32}.
  
  By \rf{wh-8}, if the diagonal components of Einstein's equations are solved, 
  their only off-diagonal component is automatically fulfilled.  

\subsection {Examples of static \whs}

  A number of such examples are presented in \cite{wh25}. Let us here mention two of them.
  The first is with an azimuthal magnetic field, see (3.24), (3.25) in the case $b > k$. 
  The solution is neither \asflat\ nor regular, with the magnetic induction $B$ and 
  curvature invariants blowing up as $u \to -\infty$.   
  
  Wormhole solutions have also been obtained with nonlinear electromagnetic fields
  with Lagrangians $L_e = L(F), \ F= F\mn F\MN$ \cite{wh33}. Their properties can be 
  more diverse, but there are restrictions following from the field equations.   
  For example, a flat asymptotic region can appear only on one side of the throat
  and, moreover, only  at the expense of a non-Maxwell behavior of $L(F)$ at small $F$ 
  \cite{wh33}.
  
  Another example concerns a massless scalar field $\Phi$ with the Lagrangian 
  $L_s = \eps g\MN \Phi_{,\mu} \Phi_{,\nu}$ in the presence of a cosmological constant
  $\Lambda = - 3\lambda^2 < 0$ ($\lambda >0$), where $\eps =1$ corresponds to a
   canonical scalar and $\eps = -1$ to a phantom one. Then the field equations for the 
   metric \rf{wh-1} with $E=0$, in the harmonic gauge $\alpha = \beta+\gamma+\mu$,
   have the family of solutions \cite{wh25}
\bearr                                \label{wh-11}
		\Phi = Cx,\quad\	       \gamma = \frac 13 \alpha -Ax,   
\nnn
               \beta = \frac 13 \alpha +Ax + Bx,  \quad\    \mu = \frac 13 \alpha - Bx,    
\nnn
		\e^{-\alpha} = \vars{ (\lambda/h) \sinh (hx), &  h>0,\yy
				   \lambda x, & h = 0,\yy
				  (\lambda/h) \sin (hx), &   h < 0,			}  
\ear            
   where $A, B, C, h$ are integration constants related by 
\beq                                 \label{wh-12}
	    h^2 \sign h = 3 (A^2 + AB + B^2) + 3 \eps \kappa C^2.         
\eeq
   Among them, \wh\ solutions are those where $\beta(x) \to \infty$ on both ends of the 
   range of $x$. Assuming $x > 0$, we see that as $u\to 0$ we have $\beta\to \infty$ 
   for any values of the constants. Let us look when $\beta\to\infty$ at large $x$
   
   With a normal scalar field, $\eps =1$, we have $h > 0, C \ne 0$, $x \in \R_+$,
   and  $\alpha \approx -hx \ \then \ 3\beta \approx (3A + 3B - h)x$,
   and we obtain a \wh\ if $A + B > h/3$. The metric can be written as
\bearr                                \label{wh-13}
             ds^2 = Y \Big(\e^{-2Bx} dt^2 - Y^2 dx^2 -\e^{-2Ax} dz^2
\nnn   \      
    			- \e^{2Ax + 2Bx} d\phi^2\Big), \quad\ 
				    Y := \bigg(\frac {h}{\lambda \sinh(hx)}\bigg)^{2/3}.     
\ear         

  Other \wh\ solutions are obtained with $h \leq 0$. Thus, if $h <0$, then 
  $x \in (0, \pi/|h|)$, and all metric functions $\alpha,\beta,\gamma,\mu$, being governed by 
  $\sin (|h|x)$, tend to infinity on both ends of this range, where  repulsive singularities occur. 
   
  Among these solutions, there is a symmetric subfamily: it corresponds to
  $\eps =-1$, $C \ne 0$, $A = B = 0$, $h < 0$. In accord with the above said,
  it has negative energy density, $T^0_0 = - \Phi'{}^2\e^{-2\alpha} + \Lambda/\kappa$. 
  In terms of the Gaussian coordinate $l = (1/\lambda)\log \tan (hx/2)$ ($l \in \R$ 
  is a length in the radial direction), the metric in this case can be written as
\beq                                    \label{wh-14}
    ds^2 = -dl^2
         + \biggl(\frac{|h|}{\lambda}\cosh\frac{l}{\lambda}\biggr)^{2/3}
	     (dt^2 - dz^2 - d\phi^2).
\eeq
               
  \subsection {Examples of rotating \whs}
  
   Assuming $E \ne 0$ and $\omega \ne 0$, we return to \rf{wh-6}--\rf{wh-8} and
   see, in particular, that \rf{wh-6} is incompatible with asymptotic flatness 
   since finite values of  $\gamma(x)$ and $\mu(x)$ imply finite $\omega$.
   Accordingly, among numerous known rotating \cy\ \wh\ solutions, none are 
   \asflat, and to obtain configurations potentially observable by astronomers,
   one has to use matching with external regions on some surfaces $\Sigma_\pm$
   as described above. 
   
   Furthermore, in a phantom-free model, all involved SETs should respect the 
   WEC, and it is a separate challenge to find such solutions. For example, 
   it has been shown  \cite{wh26,wh28} that a construction with two external 
   Min\-kow\-ski regions around a \wh\ region does not lead to phantom-free models:
   the SET on at least one of the shells $\Sigma_\pm$ violates the WEC, 
   if the SET of matter filling the internal \wh\ region has the property
   $T^0_0 = T^3_3$. This no-go theorem \cite{wh28} was proved 
   by analyzing the matching conditions using the solution of the equation 
   $R^0_0 = R^3_3$. Therefore neither of \wh\ solutions known for scalar fields
   (including those with exponential and other potentials \cite{wh27}) are suitable 
   for our purpose. 

 \subsubsection{Weak energy condition on a junction surface}   
         
   In   {the following}, we will present two examples of models that solve the problem 
   designated above. But before that we formulate a convenient criterion of WEC 
   fulfillment on a junction cylinder $\Sigma$ ($x = x_s$) separating two regions  $\D_-$ 
   ($x \leq x_s$) and $\D_+$ ($x \geq x_s$) with two different metrics of the form \rf{wh-1}. 
   The metric on $\Sigma$ should be the same when looking from $\D_-$ or $\D_+$,
   therefore
\beq                                                                                  \label{wh-15}
      [\beta] = 0, \quad [\mu] = 0, \quad     [\gamma] = 0, \quad [E] =0,
\eeq
  where, for any $f(x)$, $[f] = f(x_s+0) - f(x_s -0)$. The coordinates $t, z, \phi$ 
  are identified in the whole space, while the radial coordinates in $\D_+$ and $\D_-$ 
  may be different since all quantities involved in the matching conditions are
  insensitive to this choice.

  The material content of $\Sigma$ is found using the Darmois-Israel formalism 
  \cite{wh34,wh35} in terms of discontinuities of the extrinsic curvature of $\Sigma$.
\bearr                                                                                      \label{wh-16}
	S_a^b =  (8\pi G)^{-1} [\tK_a^b], \quad
                  		\tK_a^b := K_a^b - \delta_a^b K^c_c, 		
\ear
  where $a, b, c = 0, 2, 3$, and $K_{ab} = \half \e^{-\alpha} g'_{ab}$. 
  The question is whether $S_a^b$ satisfies the WEC requirements 
\beq                                                                                          \label{wh-17}
	S_{00}/g_{00} = \sigma_s \geq 0, \qquad\
				S_{ab}\xi^a \xi^b \geq 0,
\eeq
  where $\sigma_s$ is the surface energy density and $\xi^a$ an arbitrary null
  vector on $\Sigma$. The second inequality makes the content of the NEC as a 
  part of the WEC, and taken together, they provide $\sigma_s \geq 0$ in any 
  reference frame on $\Sigma$.
  
  The condition $\sigma_s \geq 0$ is easily found and reads
\beq                                                               \label{wh-18}
 		 [\e^{-\alpha} (\beta' + \mu')] \leq 0.
\eeq 		 
  The radial coordinates $x$ can be chosen in different ways in $\D_+$ and $\D_-$, 
  and the prime denotes a coordinate derivative defined for each region separately.  
  
  The NEC should hold for {\it any\/} null vector $\xi^a$ on $\Sigma$. However, instead
  of directly verifying this, we can note that the WEC holds if in a comoving reference 
  frame the density $\sigma_s$ and the  pressures $p_i$ in all directions satisfy the 
  inequalities 
\beq   				                               \label{wh-19}
  	\sigma_s \geq 0, \cm  \sigma_s + p_i \geq 0. 
 \eeq 	
  A difficulty is that $\Sigma$ is in general not described in a comoving frame, but
  the needed quantities can be found as eigenvalues of $S_{ab}$ written in an 
  orthonormal frame. In this way we obtain the following WEC criterion \cite{wh29,wh30}: 
\bearr                                      \label{wh-20}
	 a \geq 0, \qquad  a+c \geq 0, 
\nnnv
	a+c+\sqrt{(a-c)^2 +4 d^2} + 2b \geq 0,
\ear                                 
   where 
\bearr                                                 \label{wh-21}
        a = -[\e^{-\alpha}(\beta'+\mu')] \qquad
	b = [\e^{-\alpha}(\beta' + \gamma')],
\nnn    
        c= [\e^{-\alpha}(\gamma'+ \mu')], \qquad\    
        d = - [\omega]
 \ear
  
\subsubsection{Minkowski regions}

   If we place Minkowski regions $\M_{\pm}$ around ``internal'' rotating \wh\ regions,
  we must take the Minkowski  metric in a rotating reference frame. To do so, 
  in the inertial-frame metric $ds_{\rm M}^2 = dt^2 - dX^2 - dz^2 - X^2 d\varphi^2$  
  we substitute $\varphi \to \varphi + \Omega t$ ($\Omega = \const$ is the angular 
  velocity) to obtain
\beq                                                                                         \label{wh-22}
	      ds_{\rm M}^2 = dt^2 - dX^2 - dz^2 - X^2 (d\varphi + \Omega dt)^2.
\eeq
  The relevant quantities in the notations of \rf{wh-1} are
\bearr                                                                                   \label{wh-23}
      \e^\alpha = \e^\mu = 1, \quad   \e^{2\gamma} =  1 - \Omega^2 X^2,\quad
      \e^{2\beta} = \frac{X^2}{1 - \Omega^2 X^2},
\nnn
      E = \Omega X^2, \qquad  \omega = \frac{\Omega}{1 - \Omega^2 X^2}.
\nnn 
      \gamma' = -\frac{\Omega^ X}{1 - \Omega^2 X^2}, \qquad  
      \beta'  = \frac 1X  + \frac{\Omega^2 X}{1 - \Omega^2 X^2}.
\ear
  This metric is stationary and suitable for matching if $|X| <  1/|\Omega|$, 
  so that the linear rotational velocity is smaller than $c$.

 \subsubsection{A model with stiff matter, $p = \rho$}   
 
   The perfect fluid with $p = \rho$ is a maximally stiff matter compatible with causality, 
   where the velocities of sound and light coincide. This circumstance makes easier 
   finding exact solutions, including wave and inhomogeneous ones \cite{wh36}; 
   stiff matter also finds various applications in cosmology, see, e.g., \cite{wh37,wh38}
   and references therein.
   
   A rotating \wh\ solution with such a fluid reads \cite{wh30}
\bearr                                      \label{wh-24}  
  	   ds^2 = \bigg(P dt -  \frac{y}{\omega_0} d\varphi\bigg)^2 
  	   - \frac{dy^2}{2\omega_0^2(k^2+y^2)} - dz^2 
\nnn \quad  	   
  	   - (k^2 + y^2) \frac{d\varphi^2}{2\omega_0^2},\qquad
  	   			\kappa \rho_0 = \omega_0^2 = \const.
\ear  
    The solution contains an arbitrary time scale factor $P >0$ and integration constants 
    $k$ and $\omega_0$.  It is regular in the whole range $y\in \R$, but at 
    $y^2 > k^2$ it has $g_{33} > 0$, so that $\varphi$-circles are   {CTCs}.
  
    Consider matching with the Minkowski metric \rf{wh-22} on  
     $\Sigma_+$ ($y=y_0 >0$, $X=X_0 > 0$). The matching conditions yield
\beq                \label{wh-25}
                    P^2 = 1 - \Omega^2 X^2, \quad\
                    \Omega X^2 = \frac{P y}{\omega_0}, \quad\
                    \frac{k^2+y^2}{2\omega_0^2} = \frac{X^2}{P^2},
\eeq       
     where the index ``zero'' near $X$ and $y$ is omitted without risk of confusion.
     We have six system parameters: four ($P,  k, \omega_0, y$) from the internal
     metric \rf{wh-24} and two ($\Omega, X$) from the external one. Choosing three of them 
     as independent ones: $X = X_0$ (having the dimension of length),
     $y =y_0$ and $P$, we obtain from \rf{wh-25} 
\beq            \label{wh-26}  
		  \Omega = \frac{\sqrt{1-P^2}}{X} , \quad 
		  \omega_0 = \frac{P y}{\sqrt{1-P^2}X}, \quad
		  k^2 = y^2 \frac{1+P^2}{1-P^2}.
\eeq        
  As a result, the quantities $a,b,c,d$ from \rf{wh-21} are expressed as 
\bearr                      \label{wh-27} 
	a = \frac{P^3 y  -1}{P^2X}, \quad\
	b = \frac{1 - Py}{X}, 
\nnnv
	c = \frac{P^2-1}{P^2X}, \quad\
	d = \frac{P y - 1 + P^2}{P^2 X \sqrt{1-P^2}}.  
\ear      
  The factor $1/X$ is common and irrelevant for the inequalities of interest. 
  An inspection shows that under the condition 
\beq                       \label{wh-28} 
		y \geq \frac{2 - P^2}{P^3}
\eeq    
  the inequalities \rf{wh-20} are satisfied.
  
  What changes on $\Sigma_-$ specified by $X = -X_0 < 0$ and $y= -y_0 <0$? 
  The common factor $1/X$ becomes negative. But the signs of all discontinuities also change: 
  while on $\Sigma_+$ we had $[f] = f_{\rm out} - f_{\rm in}$ for any $f$, on 
  $\Sigma_-$ we have the opposite. Therefore, the WEC requirements do not change and 
  hold with \rf{wh-28} applied to $|y|$ instead of $y$. Thus we have obtained a completely 
  phantom-free \wh\ model.
  
   We also notice that \rf{wh-26} implies $y_0^2 < k^2$, hence $y^2 < k^2$
   in the whole internal region, and there are no   {CTCs}.

\subsubsection {A model with an anisotropic fluid}    
    
  One more model was obtained in \cite{wh29} with the SET
\beq  \label{SET-2}
	   T\mN = \rho \diag (1, -1, 1, -1)  \ \oplus \   T^0_3 = -2\rho E \e^{-2\gamma},
\eeq
   describing an anisotropic fluid chosen by analogy with that of a longitudinal
   magnetic field in the static case (see Section 3), which cannot be directly extended 
   to involve rotation. (A full description of the anisotropic fluid 
   formalism in  the metric \rf{wh-1} may be found, e.g., in \cite{Debbasch}.)
   
   The solution, in the gauge $\alpha = \beta +  \gamma + \mu$, is written as
\bearr
       \rho = \rho_0 \e^{-2\gamma - 2\mu},  \quad\ \rho_0 = \const > 0,
\nnnv 
      r^2 \equiv \e^{2\beta}= \frac{r_0^2}{Q^2 (x_0^2-x^2)}, \quad\
		\e^{2\gamma} = Q^2 (x_0^2-x^2), 
\nnnv 
		\e^{2\mu} =  (x_0 -x)^{1-x/x_0} (x_0 +x)^{1+ x/x_0},	
\nnnv			
	E = \frac {r_0} {2x_0^2} \bigg[ 2x_0 x 	+ (x_0^2 - x^2) \ln\frac{x_0+x}{x_0 -x} \bigg],
\nnn	 \cm		
	x_0 := \frac{|\omega_0|}{\kappa\rho_0 r_0}, \qquad\ Q^2 := \kappa \rho_0 r_0^2,      
\ear
   The solution contains integration constants $\omega_0$, $\rho_0$ and $r_0$,
   and  dimensionless constants $x_0$ and $Q$ introduced for convenience.  
   The coordinate $x$ ranges from $-x_0$ to $x_0$. Since $r \to \infty$ as $x\to \pm x_0$, 
   it is a \wh\ solution, it is symmetric with respect to the throat $x=0$. However,
   the extremes $x = \pm x_0$ are curvature singularities, with the Kretschmann 
   invariant $\sim |x_0-x|^{-4}$. 
  
   To construct an \asflat\ configuration, we join it at some $x= \pm x_s < x_0$ 
  to Minkowski regions at $X = \pm X_s$ and obtain
\bearr             \label{jun-2}
		X = r_0, \cm  
		Q^2 (x_0^2 -x^2) = 1 - \Omega^2 X^2 =: P^2, 
\nnn
		2x_0^2 \sqrt{1-P^2} = 2xx_0 + (x_0^2 -x^2) \ln \frac {x_0+x}{x_0-x},		
\ear    
  where the index $s$ near $x$ is omitted. To satisfy $[\mu] =0$,
  we change the $z$ scale in $\M_\pm$, obtaining $-g_{zz} = M^2 := \e^{2\mu(x_s)}$
  taken from the internal metric.

  It is convenient to use the ratio $y = x_s/x_0$ and to introduce the notation 
  $L = L(y) = \ln[(1+y)/(1-y)]$. So we have   
\bearr
	 M = M(y) = \big(1-y\big)^{-(1-y)/2}\big(1+y\big)^{-(1+y)/2},
\nnnv                 
	P^2(y) = (1-y^2)\bigg[1 - y L(y) - \frac 14 (1-y^2) L^2(y)\bigg].
\ear  
   Calculation of the quantities $a,b,c,d$ used in the WEC requirements \rf{wh-20} 
   on $\Sigma_\pm$ gives   
\bearr
	a = - \frac 1{P^2(y)} + \frac {M(y)}{x_0^2}\bigg(\frac{y}{1-y^2} + \Half L(y)\bigg),
\nnn
	b = 1,
\nnn     
        c = - \frac 1{P^2(y)} + 1 + \frac {M(y)}{x_0^2}\bigg(\frac{y}{1-y^2} - \Half L(y)\bigg),
\nnn
	\pm d =  -\frac{\sqrt{1-P^2(y)}}{P^2(y)} + \frac{M(y)}{x_0^2(1-y^2)}, 
\ear      
  where the insignificant factor $1/r_0$ at each of them has been ignored.
  
  The expressions for $a,b,c$ and $d^2$ depend on two parameters, $x_0$ and $y$, 
  and coincide on $\Sigma_+$ and $\Sigma_-$, hence the conditions \rf{wh-20}
  for the two surfaces also coincide. A numerical analysis shows that \cite{wh29}:

\medskip  
  \noi$\bullet$ The condition $0< P^2(y) < 1$, required by construction, holds for 
  $0< y \lesssim 0.564$;
  
\medskip    
  \noi$\bullet$ The conditions $a > 0$ and $a >c$ hold in a significant range of $x_0$ 
  and $y$, for example, for $x_0 = 0.5,\ y \in (0.15, 0.47)$ and for 
  $x_0 = 0.3,\ y \in (0.05, 0.53)$.

\medskip    
  \noi$\bullet$ The condition $a+c >0$ also holds in a ccertain range of the 
  parameters, e.g, for $x_0 = 0.5,\ y \in (0.15, 0.38)$ and for $x_0 = 0.3,\ y \in (0.05, 0.51)$.
  
  Thus in the parameter space ($x_0, y$) there is a significant range in which this
  \asflat\ \wh\ model completely satisfies the WEC. One can also verify that $g_{33} <0$
  in the internal region between $\Sigma_-$ and $\Sigma_+$, hence the model does not 
  contain   {CTCs}.

  This section demonstrates the possible existence of \cy\ \whs\ \asflat\ on both 
  sides of the throat without WEC violation. It should be noted that the study of such systems 
  is at an early stage, and many questions are yet to be answered, including, above all, 
  possible observational signatures of such objects if they can really exist.  
  
  In conclusion, we would like to mention that \cy\ thin shells in many studies were used 
  not as auxiliary objects needed to reach asymptotic regions but as the main sources 
  of gravity in \wh\ solutions, see \cite{wh40,wh41,wh42} and references therein.

%
\section{Gravitational collapse and emission of gravitational waves} 
%
\renewcommand{\theequation}{9.\arabic{equation}} \setcounter{equation}{0}

 Gravitational collapse of a realistic body has been one of the most  important problems in Einstein's theory, and due to the 
complexity of   {Einstein's} field equations, the problem even in simple cases, such as, spacetimes with spherical symmetry, is still 
not well understood, and new phenomena keep emerging \cite{Josh1994}. Particularly,  in 1991 Shapiro and Teukolsky  studied  numerically the
problem of a dust spheroid, and found that only if the spheroid is compact enough, a black hole can be formed  \cite{ST1991,ST1992}.  Otherwise, the collapse most
likely ends with a naked singularity. Soon,  Barrab\'es, Israel and Letelier constructed an analytical model of a collapsing convex thin shell and found
that in certain cases no apparent horizons are formed, too \cite{BIL1991}. Similar results were also found in more general cases \cite{BGL1992}.
However, since in all the cases considered by them, the external gravitational field of the collapsing    {spheroid/shell} is not known,  one cannot exclude  the formation of an 
event horizon outside of the spheroid/shell. 

  {In addition, Wald and Iyer proved that even in the Schwarzschild spacetime there exists a family of Cauchy surfaces, 
which come arbitrarily close to the black-hole singularity at the center,  but are such that there do not exist any trapped surfaces lying within the past of any of these Cauchy surfaces
\cite{WI91}. Then, they argued that, in any spherically symmetric spacetime describing gravitational collapse to a Schwarzschild black hole, it should be possible to choose a 
spaceiike slicing by Cauchy surfaces which terminate when a singularity (outside the matter) is reached, such that the region of the spacetime covered by the slicing contains no 
outer trapped surfaces. As a result, the evidence that no outer trapped surfaces exist does not always mean that a naked singularity must be resulted.
In order to determine whether a black hole or naked singularity occurs in the examples of Shapiro and Teukolsky, it will be necessary to continue the evolution of their spacetimes 
considerably further into the future.}

To answer the above question, a simplified model was considered analytically by Apostolatos and Thorne (AT), that is,  a cylindrical shell that is made of
counter-rotating particles,   and AT showed  that the centrifugal forces associated with an arbitrarily small amount of
rotation, by themselves, without the aid of any pressure, can halt the collapse at some non-zero, minimum radius, and the shell will then oscillate
until it settles down at some final, finite radius, whereby a spacetime singularity is prevented from forming on the symmetry axis \cite{AT1992}. 
It should be noted that    {AT  considered} only the case where the shell has zero total angular momentum  and is momentarily
 static and radiation-free. In    {more realistic situations}, the spacetime has neither cylindrical symmetry nor zero angular momentum, and gravitational and
particle radiations are always expected to occur. In particular, when radiation is taken into account, modeled by an outgoing radiation fluid  \cite{LW1994}, 
   {Pereira and Wang}    found  \cite{PW00b} that similar  results to a rotating spheroid   obtained in  \cite{ST1992} also hold for a collapsing cylindrical shell made of
 counterrotating dust particles: {\em in some cases the angular momentum of the dust particles 
is strong enough to halt the collapse,  so that  a spacetime singularity is prevented from forming, while in other cases it is not, and  a line-like 
spacetime singularity is finally formed on the symmetry axis}.

 {Recently, East revisited  the same problem studied by Shapiro and Teukolsky in \cite{ST1991,ST1992} but using different coordinates, and  found
that the final state in all cases studied is a black hole plus gravitational radiation \cite{East19}. Though initially
distorted, the proper circumferences of the apparent horizons that are found do not significantly
exceed the hoop conjecture bound  \cite{Thorne72}.  }

\subsection{ The Hoop conjecture and critical collapse}

    {The} studies of gravitational collapse with cylindrical symmetry have been generalized to various cases, including
 the collapse of cylindrical shells with or without finite thickness \cite{Ech93,KM94,LW94,PW02,HS05,NIK08,PHMS09,Kha11,SA12,GB14}, 
 and matter fields filled the whole collapsing spacetimes \cite{Nol02,WSSW03,HN04,NM04,NM05,KN06,NKMH07,SA07,NHKM09,SF11,SA11,SA11b,CC14,MSA15,COR16,BSMS17,CC17,SA17,GM18}.  
   {In particular,  the general matching conditions of two arbitrary   ER  spacetimes  with a thin shell were given in \cite{PW02},
 and without a thin shell in \cite{PHMS09}. It is remarkable that in the latter the junction  conditions can be explicitly solved. In addition, in this
 case   for a collapsing shear free isotropic cylindrical fluid, only a Robertson-Walker dust interior is possible, as shown in \cite{PHMS09}.}

   {All the results obtained so far} are consistent with the hoop conjecture \cite{Thorne72}: {\em black holes with horizons form when and only when a mass $M$ gets compacted into a region whose circumference is ${\cal{C}}
 \lesssim 4\pi M$ in every direction} \footnote{It should be noted that the circumference of a given region is not gauge-invariant, and a well-defined physical quantity of it
 is still absent in the general case, although it is not a problem in cylindrical spacetimes.}.

 In most of these studies, the spacetimes are described by the  KJEK metric (\ref{1.2}).  Then,  the   no-go theorem regarding the existence of black holes in cylindrical spacetimes is applicable \cite{Wang05}.
 However,   the naked singularities formed in cylindrical collapse may not be considered as counterexamples of
 the cosmic censorship conjecture (CCC) \cite{Penrose69}, as one can always argue that such spacetimes are not realistic, while the CCC is often referred to   collapse of realistic matter \cite{Josh1994}.
  Nevertheless, cylindrical spacetimes can be used to understand  other important properties of realistic collapse, such as the role that rotation can play,    {and the nonlinear interaction between 
  gravitational and matter fields.}
  
 Critical phenomena in gravitational collapse of  cylindrically symmetric spacetimes are an example of such kind of studies. In 1993, Choptuik studied the non-linearity of the Einstein field equations near
the threshold of black hole formation and revealed very rich phenomena  \cite{Chop93}, which are quite similar to critical phenomena in Statistical
Mechanics and Quantum Field Theory \cite{Golden,Golden2}. In particular, by numerically studying the gravitational collapse of a massless scalar field in $3+1$-dimensional spherically symmetric
spacetimes, Choptuik found that the mass of such formed black holes takes the form,
\bqu
\lb{9.1}
  M_{BH} = C(p)\left(p -p^{*}\right)^{\gamma},
\equ
where $C(p)$ is a constant and depends on the initial data, and $p$ parameterizes a family of initial data in such a way that when
$p > p^{*}$  black holes are formed, and when $p < p^{*}$ no black holes are formed. It was shown that, in contrast to $C(p)$, the
exponent $\gamma$   is {\em universal} to all the families of initial data studied, and was numerically  determined as $\gamma \sim 0.37$.
The solution with $p = p^{*}$, usually called the critical solution, is found also {\em universal}. Moreover, for the massless
scalar field it is periodic, too. Universality  of the critical solution and the exponent $\gamma$, as well as the power-law scaling of the
black hole mass all have given rise to the name {\em Critical Phenomena in Gravitational Collapse}. Choptuik's studies were soon generalized to other matter fields, 
   {see, for example, the review articles} \cite{Wang01,GG07} (For more recent studies of  critical phenomena, see, for example, \cite{ZL15,CY16,GB17}, and references therein.).

\begin{figure}
\centering
\includegraphics[width=8.cm]{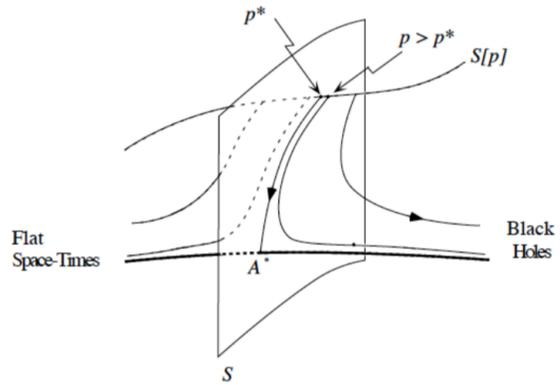}
\caption{\small  The phase space of the dynamic system of the Einstein-scalar field equations. The hypersurface $S$ is the
critical surface of co-dimension one, which separates the basin of black holes from the basin of flat spacetimes. A
generic smooth family of initial data $S[p]$ always passes the two basins at the critical point $p= p^{*}$ on the critical hypersurface.
All the initial data on the hypersurface will collapse to the critical solution $A^{*}$, which is a fixed point on the hypersurface when it has continuous self-similarity,  and a fixed cycle when
it has discrete self-similarity. All details of initial data are soon washed out during the collapsing process, and the collapse with initial
data near the critical point will be very similar to the critical collapse. This similarity can  last almost to the fixed point $A^{*}$, whereby the one unstable mode suddenly draws
the collapse either to form black hole or    {to a flat} spacetime, depending on whether $p > p^{*}$ or $p < p^{*}$ \cite{Wang01}.  {The bottom bold horizontal line containing the point $A^*$
denotes the final states of the initial data $S[p]$.}} 
\label{Fig9.1}
\end{figure}

 To understand the phenomena deeply, analytical studies have been put forwards \cite{Gar01,GG02,CF01,HW02}. However, due to the nonlinearity of    {Einstein's} field equations, even in spherically symmetric spacetimnes,
 the problem is still too complicated to be studied analytically.  In 2003,  cylindrically symmetric spacetimes with homothetic  self-similarity were studied  and a class of exact solutions to the Einstein-massless scalar 
field equations was found \cite{Wang03}. Among other things, from the analysis of their linear perturbations, it was found that there exists one solution that has  one and only unstable 
mode. By definition \cite{Wang01,GG07},  this is a critical solution that   may  sit on a boundary that separates two different basins of attraction in the phase space [see Fig.\ref{Fig9.1}].

Recently, a self-similar cylindrical scalar field with non-minimal coupling was studied \cite{CN14a,CN14b}, and  a 2-parameter family of solutions with a regular axis was found, and their global structure
was    {studied}. 

In addition, gravitational radiation from cosmic strings and their instabilities were also studied in cylindrically symmetric spacetimes \cite{BB91,CA92,WS96,BS96,BBS96,Gre96,WN97,NW97,BSWW99}.

\subsection{Polarizations  and Faraday rotation}

 As is well known, due to Birhkoff's theorem \cite{HE73},  {the spherically symmetric vacuum spacetimes are uniquely  described by  the Schwarzschild black hole solution, 
 and it is not possible to have gravitational radiation in such spacetimes.} So, one   of the next simplest cases is spacetimes with 
cylindrical symmetry. Studying cylindrical GWs, Piran, Safier and Stark (PSS) first discovered that, due to the nonlinear interaction between ingoing and outgoing cylindrical GWs, a phenomenon similar to
the electromagnetic Faraday rotation exists, but now one  GW serves as the medium for the other, so the polarizations of these waves get rotated  \cite{PS85,PSS85}. The spacetimes considered by PSS
are the ones described by  the KJEK metric (\ref{1.2}) with $W = r$, which is possible when spacetimes are vacuum, 
 \bqu
 \lb{9.2}
 ds^2 =  e^{2(\gamma - \psi)}\left(dt^2 - dr^2\right) - e^{2\psi}\left(dz + \omega d\phi\right)^2 - r^2e^{-2\psi}d\phi^2.
 \equ
 Introducing the quantities, 
  \bqun
 \lb{9.3}
 I_{+} &\equiv& 2\left(\psi_{,t} + \psi_{,r}\right), \;\; O_{+} \equiv 2\left(\psi_{,t} - \psi_{,r}\right), \nb\\
 I_{\times} &\equiv& \frac{e^{2\psi}}{r} \left(\omega_{,t} + \omega_{,r}\right), \;\;
 O_{\times} \equiv \frac{e^{2\psi}}{r} \left(\omega_{,t} - \omega_{,r}\right),
 \equn
PSS interpreted $I_{+}, \; I_{\times}$ and  $O_{+}, \; O_{\times}$ as representing ingoing and outgoing cylindrical GWs, respectively. The indices $+$ and $\times$ denote the two independent polarizations. 
With the boundary conditions at the symmetry axis ($r = 0$), $I_{+} = O_{+},\; I_{\times} = - O_{\times}$, PSS found numerically that the propagation of these waves displays a reflection of ingoing to outgoing waves 
(and vice versa),  in combination of a rotation of the polarization vector between the $+$ and $\times$ modes.   

\begin{figure}
\centering
\includegraphics[width=8.cm]{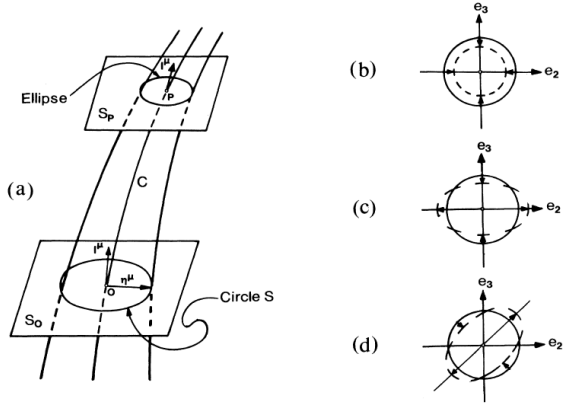} 
\caption{\small   {(a)} A null geodesic congruence meets the $S_O$ plane in the circle $S$. Because of the force generated by the $\Psi_0$ GW component, the image of the circle $S$ on the $S_P$ plane is a deformed into an ellipse \cite{Wang91}.  {(b) When a matter wave passes through the circle $S$ located on the $S_O$ plane, its image on the $S_P$ plane is a uniformly-contracted circle, due to the attractive forces of matter.
(c) When a $\Psi_0$ GW with only the ``+" polarization passes through the circle $S$,  its image is distorted into an ellipse with its major axis along the $e_2$-direction.
(d) When a $\Psi_0$ GW with only the ``$\times$" polarization passes through the circle $S$,  its image is distorted into an ellipse but now with its major axis along the direction
with $45^0$ with respect to the $e_2$-direction.}} 
\label{Fig9.2}
\end{figure}

 Note that the definition of the $+$ and $\times$ modes given in  (\ref{9.3}) is coordinate-dependent, and the polarizations are defined with respect to the chosen coordinates. In \cite{Wang91,Wang91th}, a 
 coordinate-independent definition was given for plane GWs. However, the difference between   {metric} (\ref{9.2}) and plane symmetric spacetimes only lies in the existence of a symmetry axis and the closed orbit of the Killing vector
 $\xi = \partial_{\phi}$. Therefore, it can be easily generalized to the current case. In the following, we shall give a brief outline, and for details we refer readers to  \cite{Wang91,Wang91th}. 
 Let us first introduce two null coordinates $u \equiv (t + r)/\sqrt{2}, \; v \equiv (t - r)/\sqrt{2}$, so the metric (\ref{1.2}) takes the form,
  \bqun
 \lb{9.4}
 ds^2 &=&  2e^{-M}dudv - e^{-U}\left[e^{V}\cosh{\cal{W}}d\phi^2 - 2\sinh{\cal{W}}d\phi dz \right.\nb\\
 && \left. + e^{-V}\cosh{\cal{W}}dz^2\right],
 \equn
 where 
  \bqun
 \lb{9.5}
 \psi &=& - \frac{1}{2}\left(U + V - \ln\cosh{\cal{W}}\right), \nb\\
     \gamma &=& - \frac{1}{2}\left(M + U + V - \ln\cosh{\cal{W}}\right), \nb\\
 \omega &=& - e^{V}\tanh{\cal{W}},\;\;\; W = e^{-U}\frac{\cosh^{1/2}{\cal{W}}}{\sinh{\cal{W}}}. 
 \equn
Introducing    a null tetrad $\left(l^{\mu}, \; n^{\mu},\; m^{\mu},\; {\bar{m}}^{\mu}\right)$  \cite{NP62},   via the relations, 
  \bqun
 \lb{9.6}
 l^{\mu} &=& B \delta^{\mu}_{v}, \;\;  n^{\mu}=  A \delta^{\mu}_{u}, \nb\\
  m^{\mu} &=& \zeta^2 \delta^{\mu}_{2} + \zeta^3 \delta^{\mu}_{3}, \;\;
 {\bar{m}}^{\mu} =   \overline{\zeta^2} \delta^{\mu}_{2} + \overline{\zeta^3} \delta^{\mu}_{3},\nb\\
 \zeta^2 &=& \frac{e^{(U-V)/2}}{\sqrt{2}}\left(\cosh\frac{\cal{W}}{2} + i \sinh\frac{\cal{W}}{2}\right),\nb\\
  \zeta^3 &=&  \frac{e^{(U+V)/2}}{\sqrt{2}}\left(\sinh\frac{\cal{W}}{2} + i \cosh\frac{\cal{W}}{2}\right),
  \equn
$M = \ln(A B)$, one can show that each of the two null vectors $l^{\mu}$ and $n^{\mu}$ defines a null geodesic congruence, 
\bqu
\lb{NGs}
l^{\nu}\nabla_{\nu}l^{\mu} = - B(\ln A)_{,v} l^{\mu}, \;\;\; n^{\nu}\nabla_{\nu}n^{\mu} =  A(\ln B)_{,u} n^{\mu}. 
\equ
Thus, when $A = 1$ ($B = 1$) the null geodesics defined by $l^{\nu}$ ($n^{\mu}$) are affinely parameterized. 

Without loss of generality, let us first consider the null geodesics defined by  $l^{\nu}$. Let $\eta^{\mu}$ be the deviation vector between two neighbor geodesics and $\eta^{\nu}l_{\nu} = 0$. Then, it can be shown that 
the geodesic deviation is given by
 \bqun
 \lb{9.7}
 \frac{D^{2}\eta^{\mu}}{D\lambda^2} &=& - R^{\mu}_{\;\;\; \nu\lambda \rho} l^{nu}l^{\rho} \eta^{\lambda}  = - e^{M}\left[\Phi_{00} e^{\mu\nu}_{o} - 2 {\mbox{Re}}\left(\Psi_0\right)e_{+}^{\mu\nu} \right. \nb\\
 &&\left.
 - 2 {\mbox{Im}}\left(\Psi_0\right)e_{\times}^{\mu\nu}\right]\eta_{\nu},
  \equn
where  ${\mbox{Re}}\left(\Psi_0\right)$ $\left({\mbox{Im}}\left(\Psi_0\right)\right)$ denotes  the real (imaginary) part of $\Psi_0 \; (\equiv - C_{\mu\nu\lambda\rho}l^{\mu}m^{\nu}l^{\lambda}m^{\rho})$, 
$2\Phi_{00} \equiv - R_{\mu\nu}l^{\mu}l^{\nu}$ \cite{NP62},  and
 \bqun
 \lb{9.8}
 e_{o}^{\mu\nu} &\equiv& e_{2}^{\mu} e_{2}^{\nu}  + e_{3}^{\mu} e_{3}^{\nu}, \;\; e_{+}^{\mu\nu} \equiv e_{2}^{\mu} e_{2}^{\nu}  - e_{3}^{\mu} e_{3}^{\nu},\nb\\
 e_{\times}^{\mu\nu} &\equiv& e_{2}^{\mu} e_{3}^{\nu}  + e_{3}^{\mu} e_{2}^{\nu},
  \equn
with $e_2 \equiv (m + \bar{m})/\sqrt{2},\; e_3 \equiv - i (m - \bar{m})/\sqrt{2}$. The Weyl scalar $\Psi_0$ is interpreted as representing the transverse GW component propagating alone the $n^{\nu}$ direction \cite{Sze65,Wang91}. 

For physically realistic matter, we have $\Phi_{00} \ge 0$ \cite{HE73}. Equation (\ref{9.7}) allows the following physical interpretation.
Let $S_0$ and $S_P$ be infinitesimal    {2-dimensional} elements spanned by the two spacelike vectors $e_2$ and $e_3$, and orthogonal to a null geodesic $C$
at neighbor points $O$ and $P$ of $C$,  and let $S$ be an infinitesimal circle with center $O$, lying in $S_0$ [see Fig.\ref{Fig9.2}]. Suppose that a beam of light rays meets $S_0$ in the
circle $S$,  then let us observe the image of these light rays on $S_P$. The first term on the right-hand side of (\ref{9.7})
shows that matter fields always make the circle $S$ contracted [see Fig.\ref{Fig9.2} (b)]. The second term, corresponding
to the contribution of the real part of $\Psi_0$, makes the circle elliptic with the main major axis along $e_2$  [see Fig.\ref{Fig9.2} (c)].
The last term on the right-hand side of   (\ref{9.7}), corresponding to the contribution of the imaginary part of $\Psi_0$,
makes the circle also elliptic but with the main major axis tilted at 45 to $e_2$ and $e_3$   [see Fig.\ref{Fig9.2} (d)]. Thus, the image
of these light rays on $S_P$ is     {an   ellipse}.   
Making the rotation in the $(e_2, e_3)$-plane with an angle $\varphi_0$ defined by
  \bqu
 \lb{9.9}
 \tan 2\varphi_0 = - \frac{{\mbox{Im}}\left(\Psi_0\right)}{{\mbox{Re}}\left(\Psi_0\right)},  
 \equ
  {and then} in terms of $(e_2', e_3')$,   (\ref{9.7}) takes the form, 
 \bqun
 \lb{9.10}
 \frac{D^{2}\eta^{\mu}}{D\lambda^2} &=& - R^{\mu}_{\;\;\; \nu\lambda \rho} l^{nu}l^{\rho} \eta^{\lambda}  \nb\\
 &=& - e^{M}\left[\Phi_{00} {e'}^{\mu\nu}_{o} - 2 \left(\Psi_0\overline{\Psi}_0\right)^{1/2}{e'}_{+}^{\mu\nu}\right]\eta_{\nu}. ~~
  \equn
 It follows that the main major axis of the ellipse is along
$e'_2$-direction. We call $e'_2$   the direction of the polarization of the $\Psi_0$ wave.  The angle $\varphi_0$  is the polarization angle of
this   wave component  with respect to the $(e_2, e_3)$-basis. The relative accelerations of {neighboring} geodesics
are proportional to $\left(\Psi_0\overline{\Psi}_0\right)^{1/2}$, which does not relate to any observer. Thus, $\left(\Psi_0\overline{\Psi}_0\right)^{1/2}$ represents the absolute strength
of the relative accelerations of neighboring rays.

Similarly,  we can consider  the geodesic deviation of the null congruence defined by $n^{\mu}$, and the geodesic deviation will take the form,  
 \bqun
 \lb{9.11}
 \frac{D^{2}\eta^{\mu}}{D\lambda^2} &=& - R^{\mu}_{\;\;\; \nu\lambda \rho} n^{\nu}n^{\rho} \eta^{\lambda}  = - e^{M}\left[\Phi_{22} e^{\mu\nu}_{o} - 2 {\mbox{Re}}\left(\Psi_4\right)e_{+}^{\mu\nu}\right. \nb\\
 &&   \left. + 2 {\mbox{Im}}\left(\Psi_4\right)e_{\times}^{\mu\nu}\right]\eta_{\nu},
  \equn
where    $\Psi_4 \equiv - C_{\mu\nu\lambda\rho}n^{\mu}\bar{m}^{\nu}n^{\lambda}\bar{m}^{\rho}$, 
and $2\Phi_{22} \equiv - R_{\mu\nu}n^{\mu}n^{\nu}$ \cite{NP62}. The Weyl scalar $\Psi_4$ is interpreted as representing the transverse GW component propagating alone the $l^{\nu}$ direction \cite{Sze65,Wang91}. 
Rotating the basis $(e_2, e_3)$ now with an angle,
  \bqu
 \lb{9.12}
 \tan 2\varphi_4 =  \frac{{\mbox{Im}}\left(\Psi_4\right)}{{\mbox{Re}}\left(\Psi_4\right)},  
 \equ
 the geodesic deviation equation (\ref{9.11}) takes a similar form of (\ref{9.10}). Thus, the angle $\varphi_4$ denotes  the polarization angle of the $\psi_4$ wave component  with respect to the $(e_2, e_3)$-basis.

It is important to note that the basis $(e_2, e_3)$ defined above is not parallel-transported (freely falling) along the null geodesic congruence. As a result, the angles $\varphi_0$ and $\varphi_4$ defined above have physical meaning only
locally. Thus, in order to compare the polarizations of a given GW at different locations along the wave path, we need to find a parallel-transported basis, and then define the polarization angle with respective    {to} 
this parallel-transported basis.  In \cite{Wang91}, it was shown that if  $(e_2, e_3)$ is rotated by an angle $\varphi_0^{(0)}$, 
  \bqu
 \lb{9.13}
2  \varphi_{0, u} ^{(0)} - \sinh{\cal{W}} V_{,u} = 0,  
 \equ
the resulted basis is  parallel-transported along the path of the $\Psi_0$    {wave}. Thus, the angle 
  \bqu
 \lb{9.14}
\theta_0 \equiv \varphi_0 -  \varphi_0^{(0)},  
 \equ
 determines the polarization direction of the $\Psi_0$ wave with respect to this parallel-transported basis. Similarly, the angle 
   \bqu
 \lb{9.15}
\theta_4 \equiv \varphi_4 -  \varphi_4^{(0)},  
 \equ
 determines the polarization direction of the $\Psi_4$ wave with respect to the parallel-transported basis along the path of the $\Psi_4$ wave,
 where
  \bqu
 \lb{9.16}
2  \varphi_{4, v} ^{(0)} - \sinh{\cal{W}} V_{,v} = 0.
 \equ
 
 Note that in the derivations of the polarizations of the cylindrical GWs, we used the null geodesic deviations. One can equally use timelike geodesic deviations as done in \cite{Wang91th} 
   {(See also \cite{Sze65} for the general case)}, and the same results will be  obtained, as it is expected.  
 
 Moreover, for the rotating cylindrical GWs described by     {metric} (\ref{1.3}),  choosing the null tetrad 
 \bqun
 \lb{9.17}
 l^{\mu} &=& \frac{e^{\psi-\gamma}}{\sqrt{2}\; A} \left(1, 1, 0, -\omega\right), \;\;
  n^{\mu} = \frac{A e^{\psi-\gamma}}{\sqrt{2}} \left(1, -1, 0, -\omega\right), \nb\\
  m^{\mu} &=& \frac{1}{\sqrt{2}}\left(0, 0, e^{-\psi}, i  \frac{e^{\psi}}{\cal{W}}\right), \nb\\ 
  \bar{m}^{\mu} &=& \frac{1}{\sqrt{2}}\left(0, 0, e^{-\psi}, - i \frac{e^{\psi}}{\cal{W}}\right),     
\equn
Pereira and Wang studied the polarizations of the rotating GWs, and found that the geodesic deviation equation (\ref{9.7}) is replaced by \cite{PW00},
\bqun
\lb{9.18}
\frac{D^{2}\eta^{\mu}}{D\lambda^{2}} &=& - R^{\mu}_{\nu\alpha\beta}l^{\nu}l^{\beta}\eta^{\alpha}
= \left[\Phi_{00}e^{\mu\nu}_{o} + \Psi_{0}e^{\mu\nu}_{+} \right.\nb\\
&& \left. + i \left(\Psi_{1}+ \Phi_{01}\right)e_{03}^{\mu\nu} + i \left(\Psi_{1}+ \Phi_{01}\right) e_{13}^{\mu\nu}\right]\eta_{\nu},\nb\\
\equn
where $\Psi_{1} \equiv   - C_{\mu\nu\lambda\rho}l^{\mu}n^{\nu}l^{\lambda}{m}^{\rho}$, 
and $2\Phi_{01} \equiv \left(R_{\mu\nu} - Rg_{\mu\nu}/4\right)l^{\mu}m^{\nu}$ \cite{PW00},
$A = e^{\gamma - \psi}$, 
\bqun
\lb{9.19}
e^{\mu\nu}_{03} &\equiv& e^{\mu}_{0}e^{\nu}_{3} 
+ e^{\mu}_{3}e^{\nu}_{0},\;\;
e^{\mu\nu}_{13} \equiv e^{\mu}_{1}e^{\nu}_{3} 
+ e^{\mu}_{3}e^{\nu}_{1},\nb\\
e^{\mu}_{0} &\equiv& \frac{l^{\mu} + n^{\mu}}{\sqrt{2}},\quad
e^{\mu}_{1} \equiv \frac{l^{\mu} - n^{\mu}}{\sqrt{2}},
\equn
and $e_2, \; e_3$ and $e^{\mu\nu}_{+}$ are as defined above.  It is remarkable to note that now the $\Psi_0$ wave has only one polarization, the ``+" mode, along $e_2$ direction, and the ``$\times$" mode now is absent. 

\begin{figure}
\centering
\includegraphics[width=8.cm]{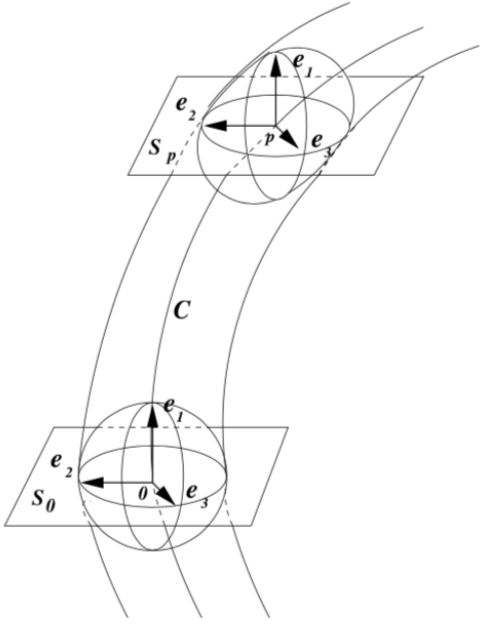}
\caption{\small  A spherical ball consisting of photons
 cuts $S_{O}$ in the circle $S$ with the point $O$ as its center. The image
 of the ball at the point $P$ is turned into a
spheroid with the main major axis along a line at $45^{0}$ 
with respect to $e_{1}$ in the plane spanned by $e_{1}$ 
and $e_{3}$ because of the interaction of $\Psi_{1}$
and $\Phi_{01}$, while the rays are left undeflected  in the 
$e_{2}$-direction \cite{PW00}.} 
\label{Fig9.3}
\end{figure}

To see the physical interpretation of the last two terms, let us consider a tube along the null geodesic $C$.  Consider a sphere consisting of photons, 
which will cut $S_{O}$ in the circle $S$ with the point $O$ as its center,
as shown in Fig.\ref{Fig9.3}. Then, the last term in  (\ref{9.18})
will make the image of the sphere at the point $P$ as a 
spheroid with the main major axis along a line at $45^{0}$ 
with respect to $e^{\mu}_{1}$ in the plane spanned by $e^{\mu}_{1}$ 
and $e^{\mu}_{3}$, while the rays are left undeflected  in the 
$e^{\mu}_{2}$-direction. This can be seen clearly by performing 
a rotation in the plane spanned by $e^{\mu}_{1}$ 
and $e^{\mu}_{3}$,
$e^{\mu}_{1} = \cos\alpha {e'}_{1}^{\mu} - \sin\alpha {e'}_{3}^{\mu}$, 
$e^{\mu}_{3} = \sin\alpha {e'}_{1}^{\mu} + \cos\alpha {e'}_{3}^{\mu}$,
which leads to
\bqu
\lb{eq10}
e^{\mu\nu}_{13} = \sin(2\alpha)({e'}_{1}^{\mu}{e'}_{1}^{\nu} 
- {e'}_{3}^{\mu}{e'}_{3}^{\nu})
 + \cos(2\alpha)({e'}_{1}^{\mu}{e'}_{3}^{\nu} 
+ {e'}_{3}^{\mu}{e'}_{1}^{\nu}).
\equ
Thus, choosing $\alpha = \pi/4$, we have
 $e^{\mu\nu}_{13} = {e'}_{1}^{\mu}{e'}_{1}^{\nu} - {e'}_{3}^{\mu}{e'}_{3}^{\nu},\; (\alpha = \pi/4)$.
Combining it with  (\ref{9.18}), we can see that the last term will make a circle in the $({e'}_{1}, {e'}_{3})$-plane into
an ellipse with its main major axis along the ${e'}_{1}$-axis, which is at $45^{0}$ with respect to the $e_{1}$-axis [cf. Fig.\ref{Fig9.3}]. 

Similar physical interpretation can be given to the second term, but note that $e_0$ defined by (\ref{9.19}) now is timelike. 

It should be noted that in the case of timelike geodesics the $\Psi_{1}$ term  deflects the sphere into an ellipsoid
\cite{Sze65}.  
Moreover, the second term in  (\ref{9.18}) is absent in the timelike geodesic case. The effect of this term 
will make a clock ``flying" with the photons slow down, in addition to the effect of deflecting the photons in the $e^{\mu}_{3}$-direction.
It is interesting to note that there is a fundamental difference between the time delay caused by this term and the one caused by a Lorentz 
boost. The latter, in particular, has no contribution to geodesic deviations, timelike or null \cite{PW00}. 

From the above analysis we can also see that for a pure Petrov type $N$
GW propagating along the null geodesic congruence, in which only
the component $\Psi_{0}$ is different from zero, the
GW has only one polarization state,
similar to the KJEK metric (\ref{9.4}) with ${\cal{W}}  = 0$. The difference between
these two cases is that in the case
${\cal{W}}  = 0$, the polarization angle remains the same even at
different points along the wave path {  {for the KJEK metric (\ref{9.4})}, 
while in the rotating case   { (\ref{1.3})} with  $\omega\not= 0$, in general
this is no longer true. In fact, it is easy to 
show that 
\bqu
\lb{eq12}
e^{\mu}_{2;\nu}l^{\nu}  = 0,\;\;
e^{\mu}_{3;\nu}l^{\nu} 
= - W e^{\psi - \chi}\frac{\omega,_{r}}{2\sqrt{2}}
 \left(e^{\mu}_{0} + e^{\mu}_{1}\right).
\equ
Thus,   {for the spacetimes described by metric (\ref{1.3})},  although $e^{\mu}_{2}$ is parallel-transported along the null geodesic congruence, $e^{\mu}_{3}$ in general is not, and is rotating
with respect to a parallel-transported basis.  Since the polarization angle of the $\Psi_{0}$ wave remains the same with respect to $e^{\mu}_{3}$,
the polarization direction is also rotating with respect to the parallel-transported basis.

Note that the effects of polarizations on the formation of spacetime singularities were studied for plane GWs \cite{Wang91a}, and one can generalize such studies to cylindrical GWs, and 
similar results  should be   {expected to hold here}, too.

%
\section{Conclusions} 
%
%
\renewcommand{\theequation}{10.\arabic{equation}} \setcounter{equation}{0}

With the beginning of the era of  gravitational wave astronomy \cite{GWs18},  {the strong gravitational field regime will be soon explored   {observationally} in various aspects}. 
Such   {theoretical} studies can be carried out analytically and/or numerically. In the former, due to the complexity of   {Einstein's} field equations, symmetries of spacetimes are
often imposed, such as spherical, plane and cylindrical \cite{StephaniA}. Although they are all ideal models, and in realistic situations any of these symmetries may 
not exist, they do provide attackable  {examples}, from which some fundamental issues of physics can be addressed. One of the typical  {cases} is the Schwarzschild  
solution, which plays the central role in the studies of black hole physics. Another  {case}  is the LC solution, discovered by LC in 1919 \cite{Levi},   marked
as the second  { exact solution to} the Einstein's field equations discovered historically, and has been studied extensively in the past decades, in the  {efforts} to understand the nonlinearity of Einstein's theory. 

In this  article, we have first defined cylindrically symmetric spacetimes \cite{CSV99}, and shown that the most general form of the metric can be cast in the form (\ref{1.6}) \cite{Wain81},
with the conditions  (\ref{1.4}) and (\ref{1.5}) as the existence of a symmetry axis. A  {spacetime} without   a  {symmetry} axis is referred to as a cyclically symmetric spacetime \cite{Barnes00},  
as the axis may be singular and so not part of the manifold, or the topology of the manifold may be such that no axis exists.  {Depending on additional assumptions, the metric can reduce to the 
KJEK   metric (\ref{1.2}), the rotating MQ metric (\ref{1.3}), or the ER metric (\ref{1.6e})}. 

Then,  we have provided a general review on the main  properties and possible sources of the LC solution (Sec. 2),     {when} it is coupled with an electromagnetic field (Sec. 3),
 or with  the cosmological constant (the LT solution, Sec. 4), or  with a perfect fluid (Sec. 5). 

The Lewis vacuum solution is the generalization of the static LC solution to the case with rotation, and has been reviewed in Sec. 6.  
When it is coupled with a cosmological constant, the solution was first studied by Lanczos  \cite{Lanczos}, then by Wright  \cite{Wri65}, Krasimski  \cite{Krasinski}, Santos  \cite{Santos}, 
and Pereira, Santos  and Wang  \cite{PSW96}, among many others. One of the remarkable features   is that the confinement of particles  near  the symmetry axis. Such particles  are highly 
 collimated and could provide an alternative   mechanism for the formation of the observed extragalactic jets \cite{Pacheco,Gariel1,Gariel2,Gariel3,PSW96,Opher}.

In Sec. 8, the studies of cylindrically symmetric wormholes  {have been reviewed, and various models have been presented, while}
in Sec. 9, the gravitational collapse of cylindrically symmetric sources has been summarized, and particular attention has been paid to the understanding of the Hoop conjecture
\cite{Thorne72} and critical phenomena  at the threshold of black hole formation  \cite{Chop93}.  In this section, a gauge-invariant definition of the polarizations of cylindrically 
symmetric gravitational waves has been also given \cite{Wang91th,Wang91},  and  the interesting features of gravitational Faraday rotation \cite{PS85,PSS85}  of such waves  
have been reviewed.   {Thus, now it would be extremely interesting to find some observational effects from  
 polarizations of gravitational waves and gravitational Faraday rotation. }

 With the promise of increasing duration, observational sensitivity and the number of detectors, many more gravitational wave events are expected to be detected.  
 In particular, LISA  \cite{LISA} is expected to observe tens of thousands of compact galactic binaries during its nominal four year  mission lifetime  \cite{CR17}.
 Then, the properties of the strong field regime of gravitational fields will be explored in various directions, including the (nonlinear) interaction 
of gravitational waves and matter fields. 

 { In addition to GWs produced by these putative sources, another important target of GW experiments 
 is the stochastic gravitational wave background (SGWB) of cosmological origin \cite{CCD08}. One of such cosmological sources is 
 the cosmic superstring network \cite{CF18}, which emits GWs throughout the history of the Universe, and generates a SGWB from the superposition of many uncorrelated sources,
 as mentioned in Introduction.
 Recently, it was found that LISA will be able to detect such a background with $G\mu/c^2 \gtrsim {\cal{O}}\left(10^{-17}\right)$ \cite{Auclair19}, while currently 
 the most stringent bounds come from pulsar timing arrays \cite{BOS18}, which yield the upper bound  $G\mu/c^2 \lesssim {\cal{O}}\left(10^{-11}\right)$. On the other hand, depending on the string
 network model, LIGO-Virgo observations could constrain it to  $G\mu/c^2 \lesssim {\cal{O}}\left(10^{-14}\right)$ \cite{Abbott19,Abbott18}.}

So far, cylindrical spacetimes have been studied mainly in the  {form} of the KJEK   metric (\ref{1.2}). As shown in Introduction, this is only a 
particular case of the general metric (\ref{1.6}).  So, a natural question is to which extent the properties obtained so far hold in more general cases? What are the generic features 
of Einstein's theory of gravity? which are independent of the symmetry assumed but obtained from the studies of cylindrical spacetimes. 
Which kind of  guidelines can one draw for the future observations of  gravitational waves from such studies? Therefore, despite of the fact that cylindrical spacetimes have been 
extensively studied in the past several decades, and various remarkable results were obtained, there are still many important questions that are still waiting for answers.

\section*{Acknowledgements}

A.W. would like to thank B. Mashhoon for valuable discussions and suggestions. He
 is supported in part by the National Natural Science Foundation of China (NNSCF) with the Grants Nos. 11675145 and 11975203.
 The work of K.B. was partly supported by the RUDN University program 5-100
  and was also performed within the framework of the Center FRPP 
  supported by MEPhI Academic Excellence Project (contract No. 02.a03.21.0005,
  27.08.2013).  

%

%
\section*{Appendix A: Static cylindrical spacetimes } 
\renewcommand{\theequation}{A.\arabic{equation}} \setcounter{equation}{0}
%

In this Appendix,  we shall  present some general statements and relations about the static cylindrical spacetimes. 
 
 The most general static metric admitting translational symmetry along two mutually 
 orthogonal axes has the form [Cf. (\ref{1.6e})], 
\beq                                                        					 \label{ds-cy}
    ds^2 = \e^{2\gamma(u)} dt^2 - \e^{2\alpha(u)} du^2 -
						\e^{2\mu(u)} dz^2 - \e^{2\beta(u)} d\phi^2,
\eeq
 where $u$ is an arbitrary admissible ``radial'' coordinate, with a freedom of 
 choosing it for one's convenience. It is important that the coordinates $z$ and $\phi$ appear in 
 \rf{ds-cy} (and hence in the field equations) on equal grounds and admit different interpretations.
 Thus, assuming $z\in \R$ and $\phi\in \R$, we obtain a symmetry called pseudoplanar, 
 being an extension of planar symmetry (the latter requires additional isometry under
 rotations in the ($z,\phi$) plane). The assumption of both $z\in \S^1$ and $\phi\in \S^1$ leads 
 to toroidal symmetry. We here choose to deal with cylindrical symmetry, assuming that 
 $z\in \R$ is the longitudinal coordinate, and $\phi \in \S^1$ is the angular one.  

 Let us present for further calculations the nonzero components of the Ricci tensor for the metric
 \rf{ds-cy} 
 without specifying the choice of the coordinate $u$, 
   {
\bear                                   \label{Ric}
    R^0_0 &=&  \e^{-2\alpha}\left[\gamma'' + \gamma' (\gamma'-\alpha'+\beta'+\mu')\right],
\nn
    R^1_1 &=&  \e^{-2\alpha}
		\left[\gamma''+\mu''+\beta''+\gamma'{}^2+\mu'{}^2\right.\nb\\
		&& \left.+\beta'{}^2 -\alpha'(\gamma'+\mu'+\beta')\right],
\nn
    R^2_2 &=&  \e^{-2\alpha}\left[\mu'' + \mu'(\gamma'-\alpha'+\beta'+\mu')\right],
\nn
    R^3_3 &=&  \e^{-2\alpha}\left[\beta'' + \beta'(\gamma'-\alpha'+\beta'+\mu')\right],
\ear
} where the prime denotes $d/du$, and the coordinates are numbered according
  to the scheme $(0,1,2,3) = (t,u,z,\phi)$. It is also   {useful} to write the component
  $G^1_1$ of the Einstein tensor $G\mN = R\mN - \half \delta\mN R$ that contains
  only first-order derivatives:
\beq                                                            \label{G11}
      G^1_1 =   {-} \e^{-2\alpha}\left(\gamma'\mu' + \beta'\gamma' + \beta'\mu'\right).
\eeq

  The coordinate $u$ can be chosen in different ways, depending on the purpose of a study,
  convenience and taste, and is usually fixed by some relation between the functions
  $\alpha,\beta,\gamma,\mu$.   For example, the LC solution in the previous section was 
  written under the conditions $\alpha = \mu$ [Cf.\eq (2.8)] and $\alpha =0$ [Cf.\eq (2.14)].
     
  However, the components $R\mN$ take the simplest form if we choose the 
  harmonic radial coordinate defined by the condition, 
\beq                        						\label{harm}
		\alpha = \beta + \gamma + \mu.
\eeq 
 
  
  Let us here, for illustration,  
  show how the LC vacuum solution is obtained using \rf{harm}. The equations $R\mN =0$
  then immediately lead to $\beta'' = \gamma'' = \mu'' =0$, whence 
\beq                                                           \label{vac1}
      \beta(u) = bu + b_0, \;\; \gamma(u) = cu + c_0,\;\;  \mu(u) = hu + h_0,
\eeq
  with six integration constants, among which $b_0,\ c_0,\ h_0$ can be
  turned to zero by changing scales along the $t$ and $z$ axes and by choosing
  the zero point of the $u$ coordinate. Moreover, the first-order equation
  $G^1_1 =0$ leads to a relation between the remaining constants $b,\ c,\ h$:
\beq
		bc + bh + ch =0,                                        \label{vac2}
\eeq
  hence there are precisely two significant constants, in accord with what was said previously.
  The metric takes the form
\beq                                                                            \label{vac3}
      		ds^2 = \e^{2cu}dt^2 - \e^{2(b+c+h)u} du^2
			      					- \e^{2hu} dz^2 - \e^{2bu}d\phi^2.
\eeq
  It is quite suitable for a further study, and its form (2.8), preferred for historical reasons,  
  is obtained from (\ref{vac3}) using \rf{vac2} and the following relations and notations:
\bearr                                                                               \label{vac5}
	\e^{(b+c) u} = k\rho, \cm\ \ \ k = (b+c)^{(b+c)/(b+c+h)}, 
\nnn
	 2\sigma = c/(b+c), \cm a^2 = (b+c)^{-2b/(b+c+h)}.
\ear
  This transformation is valid for $b+c \ne 0$ and also requires some finite rescaling 
  along the $t$ and $z$ axes. 


}

\end{document}